\documentclass[12pt]{article}

\usepackage[english]{babel}
\usepackage[utf8x]{inputenc}
\usepackage[T1]{fontenc}

\pdfoutput=1

\usepackage[a4paper,top=3cm,bottom=3cm,left=3.1cm,right=3.1cm,marginparwidth=1.75cm]{geometry}

\usepackage{bm}
\usepackage{amsmath}
\usepackage{empheq}
\usepackage{amssymb}
\usepackage{amsthm}
\usepackage{graphicx}
\usepackage{subfig}
\usepackage[colorinlistoftodos]{todonotes}
\usepackage{hyperref}
\hypersetup{colorlinks,citecolor=blue,urlcolor=blue,bookmarks=false,hypertexnames=true}

\numberwithin{equation}{section}

\usepackage{fouriernc}
\usepackage{Royal}
\usepackage{lettrine}

\usepackage[font=footnotesize]{caption}

\usepackage{mathrsfs}
\usepackage{authblk}

\usepackage{appendix}
\usepackage[final]{showlabels}
\usepackage{breakcites}

\theoremstyle{plain}
\newtheorem{theorem}{Theorem}[section] 

\theoremstyle{definition}
\newtheorem{definition}[theorem]{Definition} 
\newtheorem{example}[theorem]{Example} 
\newtheorem{lemma}[theorem]{Lemma}

\newtheorem{observation}[theorem]{Observation}
\newtheorem{remark}[theorem]{Remark}
\newtheorem{corollary}[theorem]{Corollary}

\makeatletter
\providecommand*{\diff}%
{\@ifnextchar^{\DIfF}{\DIfF^{}}}
\def\DIfF^#1{%
	\mathop{\mathrm{\mathstrut d}}%
	\nolimits^{#1}\gobblespace}
\def\gobblespace{%
	\futurelet\diffarg\opspace}
\def\opspace{%
	\let\DiffSpace\!%
	\ifx\diffarg(%
	\let\DiffSpace\relax
	\else
	\ifx\diffarg[%
	\let\DiffSpace\relax
	\else
	\ifx\diffarg\{%
	\let\DiffSpace\relax
	\fi\fi\fi\DiffSpace}

\DeclareMathOperator{\sign}{sign}
\DeclareMathOperator{\ind}{ind}
\DeclareMathOperator{\spanspan}{span}

\newcommand{\R}{\mathbb{R}}

\newcommand{\C}{\mathbb{C}}

\newcommand{\Sch}{\mathbb{S}}
\newcommand{\D}{\mathbb{D}}

\DeclareMathAlphabet{\mathcal}{OMS}{cmsy}{m}{n}

\usepackage{xcolor}
\usepackage{framed}
\colorlet{shadecolor}{blue!15}

\title{Analytic signal in many dimensions \vspace*{1cm}}

\author{Mikhail Tsitsvero\footnote{This work was supported by the LABEX MILYON (ANR-10-LABX-0070) of Universit{\'e} de Lyon, within the program "Investissements d'Avenir" (ANR-11-IDEX- 0007) operated by the French National Research Agency (ANR).\\ e-mail: tsitsvero@gmail.com}, Pierre Borgnat and Paulo Gon\c{c}alves}

\begin{document}

\maketitle

\begin{abstract}
In this work we extend analytic signal theory to the multidimensional case when oscillations are observed in the $d$ orthogonal directions. First it is shown how to obtain separate phase-shifted components and how to combine them into instantaneous amplitude and phases. Second, the proper hypercomplex analytic signal is defined as holomorphic hypercomplex function on the boundary of certain upper half-space. Next it is shown that correct phase-shifted components can be obtained by positive frequency restriction of hypercomplex Fourier transform. Necessary and sufficient conditions for analytic extension of the hypercomplex analytic signal into the upper hypercomplex half-space by means of holomorphic Fourier transform are given by the corresponding Paley-Wiener theorem. Moreover it is demonstrated that for $d>2$ there is no corresponding non-commutative hypercomplex Fourier transform (including Clifford and Cayley-Dickson based) that allows to recover phase-shifted components correctly. 
\end{abstract}

\section{Introduction}

\lettrine[lraise=0.015, lines=3]{A}{nalytic signal} is a holomorphic/analytic complex-valued function defined on the boundary of upper complex half-plane. The boundary of upper half-plane coincides with $\R$ and therefore analytic signal is given by the mapping $f_a: \R \rightarrow \C$. The problem of extending a complex-valued function defined on the boundary of a complex region was originally posed by Riemann and later by Hilbert \cite{pandey2011hilbert}. In \cite{hilbert1912grundzuge} Hilbert introduced transformation that relates real and imaginary parts of analytic function on the boundary of an open disk/upper half-plane, that is now known as Hilbert transform, and is used to define analytic signal. As from the beginning of XX-th century the concepts of amplitude and phase of a complex function became ubiquitous in quantum mechanics. Additionally in 1946 Gabor \cite{gabor1946theory} proposed to use analytic signal to encode instantaneous amplitude and phase for communication signals. Analytic signal representation of oscillation processes is peculiar because it allows to define the concepts of instantaneous amplitude, phase and frequency in unique and convenient way \cite{vakman1977amplitude}. During the last few decades an interest emerged towards extension of analytic signal concept to multidimensionsional domains, mainly motivated by the problems coming from the fields ranging from image/video processing to multidimensional oscillating processes in physics like seismic, electromagnetic and gravitational waves. In other words by all the domains that focus on oscillating processess in more than one spatial/temporal direction. 

Typically one is not able to fully describe oscillating process in several dimensions by using just complex numbers due to insufficient degrees of freedom of complex numbers. In this case one hopes to rely on algebraic constructions that extend ordinary complex numbers in a convenient manner. Such constructions are usually referred to as hypercomplex numbers \cite{SKETCHINGHYPERCOMPLEX}. 
While complex numbers allow to relate oscillations of the form
\begin{equation}
f(x) = a(x) \cos  \phi(x) 
\end{equation}
to the complex Euler form
\begin{equation}
f_a(x) = a(x) e^{i \phi(x)},
\end{equation}
the goal of this paper is to relate oscillations of the form 
\begin{equation}
f\left(x_1, \dotsc, x_d  \right) = a(x_1, \dotsc, x_d) \cos \phi_1 \left( x_1 \right) \cos \phi_2 \left( x_2 \right) \cdot \dotsc \cdot \cos \phi_d \left( x_d \right)
\end{equation}
to the corresponding hypercomplex Euler form 
\begin{equation}
f_S \left( x_1, \dotsc, x_d \right) = a_S \left( x_1, \dotsc, x_d \right) e^{e_1 \phi_1 \left( x_1 \right)} \cdot e^{e_1 \phi_2 \left( x_2 \right)} \cdot \dotsc \cdot e^{e_d \phi_d \left( x_d \right)}.
\end{equation}

 A number of works addressed various issues related to the proper choice of hypercomplex number system, definition of hypercomplex Fourier transform and partial Hilbert transforms for the sake of studying instantaneous amplitude and phases. Mainly these works were based on properties of various spaces such as $\C^d$, quaternions, Clifford algebras and Cayley-Dickson constructions. In the following we list just some of the works dedicated to the studies of analytic signal in many dimensions. To the best of our knowledge, the first works on multidimensional analytic signal appeared in the early 1990-s including the work of Ell on hypercomplex transforms \cite{ell1992hypercomplex}, the work of B{\"u}low \cite{bulow2001hypercomplex} and the work of Felsberg and Sommer on monogenic signals \cite{felsberg2001monogenic}. Since then there was a huge number of works that studied various aspects of hypercomplex signals and their properties. Studies of Clifford-Fourier transform and transforms based on Caley-Dickson construction applied to the multidimensional analytic signal method include \cite{sangwine2007hypercomplex}, \cite{hahn2011unified}, \cite{le2014instantaneous}, \cite{hahn2013quasi}, \cite{ell2014quaternion}, \cite{ell2014quaternion}, \cite{le2008h}, \cite{bihan2010hyperanalytic}, \cite{alfsmann2007hypercomplex}. Partial Hilbert transforms were studied by \cite{bulow2001hypercomplex}, \cite{yu2008bedrosian}, \cite{Zhang2014} and others. Recent studies on Clifford-Fourier transform include \cite{brackx2006two}, \cite{mawardi2006clifford}, \cite{de2011class}, \cite{de2012clifford}, \cite{de2011class}, \cite{ell2007hypercomplex}, \cite{bahri2008uncertainty}, \cite{felsberg2001fast} and many others. Monogenic signal method was recently studied and reviewed in the works \cite{bernstein2014fractional}, \cite{bridge2017introduction} and others. Unfortunately we cannot provide the complete list of references, however we hope that reader can find all the relevant references for example in the review works \cite{hahn2016complex} and \cite{bernstein2013generalized,le2017foreword}. 

The hypercomplex analytic signal $f_S$ is expected to extend all useful properties that we have in 1-D case. First of all one should be able to extract and generalize instantaneous amplitude and phases to $d$ dimensions. Second, we expect that hypercomplex Fourier transform of analytic signal will be supported only over a positive quadrant of hypercomplex space. Third, conjugated parts of complex analytic signal are related by Hilbert transform, so we can expect that conjugated components in hypercomplex space should be related also by some combination of Hilbert transforms. And finally it should be possible to extend hypercomplex analytic signal to some uniquely defined holomorphic function inside a region of hypercomplex space. 

We address these issues in sequential order. First of all we start without anything that is hyper- or complex- by considering the Fourier integral formula and showing that Hilbert transform is related to the modified Fourier integral formula. This fact allows definition of instantaneous amplitude, phase and frequency {\it without any reference to hypercomplex numbers and holomorphic functions}. We proceed by generalizing the modified Fourier integral formula to several dimensions and defining the phase-shifted components that can be combined into instantaneous amplitude and phases. Second, we address the question of existence of holomorphic functions of several hypercomplex variables. It appears that commutative and associative hypercomplex algebra generated by the set of elliptic ($e^2_i = -1$) generators is a suitable space for hypercomplex analytic signal to live in. We refer to such a hypercomplex algebra as {\it Scheffers space} and denote it by $\Sch_d$. Hypercomplex analytic signal is defined as an extension of a hypercomlex holomorphic function of several hypercomplex variables to the boundary of certain upper half-space that is denoted by $\Sch^d_+$. We then observe the validity of Cauchy integral formula for functions $\Sch^d_+ \rightarrow \Sch_d$, where the integral is calculated over tori-like hypersurfaces inside $\Sch^d_+$ and deduce the corresponding partial Hilbert transforms that relate hypercomplex conjugated components. Finally it appears that hypercomplex Fourier transform of analytic signal with values in Scheffers space is supported only over non-negative frequencies, while one can use holomorphic Fourier transform for extension of analytic signal inside the upper half-space $\Sch_+^d$.

The paper is organised as follows. In Section \ref{section:in one dimension}, we present general theory of analytic signal, establish modified Fourier integral formula and relate it to the Hilbert transform. In Section \ref{section:in many dimensions}, it is shown how to obtain phase-shifted components in several dimensions and combine them to define the instantaneous amplitude, phases and frequencies. Then, in Section \ref{section:analytic functions of hypercomplex variable} we describe general theory of hypercomplex holomorphic functions of several hypercomplex variables and introduce Cauchy integral formula as well as Hilbert transform for polydisk and upper half-space. Next, in Section \ref{section:commutative hypercomplex Fourier transform}, we review the commutative hypercomplex Fourier transform and show that analytic signal can be reconstructed by positive frequency restriction of the hypercomplex Fourier transform of the original real-valued oscillating process. We conclude the section by proving Bedrosian's theorem for hypercomplex analytic signal. In Section \ref{section:analytic extension into upper space}, necessary and sufficient conditions for extension of analytic signal by means of holomorphic Fourier trasform are given by the Paley-Wiener theorem for functions with positively supported hypercomplex Fourier transform. In Section \ref{section:why clifford- transform doesnt fit} we prove that positive frequency restriction of the class of non-commutative hypercomplex Fourier transforms (including Clifford and Cayley-Dickson) {\it does not} provide correct phase-shifted components for $d > 2$. The paper is concluded by Section \ref{section: application examples} where we provide several illustrative examples of amplitude computation as well as observe limitations of the approach in the case when oscillations are not aligned with the chosen orthogonal directions of the ambient space.

%
%
%
%

\section{In one dimension}
\label{section:in one dimension}
  Let us assume that we have some oscillating function $f: \mathbb{R} \rightarrow \mathbb{R}$ and we want to obtain its envelope function, that ``forgets'' its local oscillatory behavior, and instantaneous phase, that shows how this oscillatory behavior evolves. In one dimension one gets analytic signal by combining the original signal with its phase-shifted version. The phase-shifted version is given by the Hilbert transform $\tilde{f}$ of the original function $f$
 \begin{equation}
 \label{eq:Hilbert transform 1D}
 	\tilde{f} = H \left[ f \right] (x) = \text{p.v.} \frac{1}{\pi} \int\limits_{\mathbb{R}} \frac{f(y)}{x-y} \, \diff y =  \lim_{\varepsilon \rightarrow 0} \frac{1}{\pi} \int\limits_{|x-y|>\varepsilon} \frac{f(y)}{x-y} \, \diff y, \ \ x \in \mathbb{R}.
 \end{equation}

 The analytic signal $f_a : \mathbb{R} \rightarrow \mathbb{C}$ is obtained by combining $f$ with $\tilde{f}$, i.e. $f_a = f + i \tilde{f}$, where $i^2 = -1$. This definition comes from the fact that analytic signal is extension of holomorphic (analytic) function in the upper half-plane $\mathbb{C}_+$ to its boundary $\partial \mathbb{C}_+ = \mathbb{R}$, i.e. $f$ and $\tilde{f}$ are harmonic conjugates on the boundary $\partial \mathbb{C}_+$. Basic theory of Hilbert transform is concisely given in Appendix \ref{appendix:Hilbert transform in 1D}. We will need several standard definitions.
 \begin{definition}
 The Fourier transform of a function $f: \R \rightarrow \C$ is defined as 
 \begin{align}
 \label{eq:Fourier transform}
 \begin{split}
 	&\hat{f}(\omega) = F \left[ f \right](\omega) = \int\limits_{\R} f(x) e^{-i \omega x} \, \diff x, \\
 	&f(x) = F^{-1} \left[ \hat{f} \right] (x) = \frac{1}{2 \pi} \int\limits_{\R} \hat{f}(\omega) e^{i \omega x} \, \diff \omega.
 \end{split}
 \end{align}
\end{definition}
\begin{definition}
 The sign function is defined as 
 \begin{align}
 \sign(x) = \begin{cases}
 -1 & \text{if } x < 0, \\
 0 & \text{if } x = 0, \\
 1 & \text{if } x > 0. \end{cases}
 \end{align}
 \end{definition}
 It turns out that $F[f_a](\omega)$ is supported only on $[0,\infty)$ and therefore spectrum of $f_a$ does not have negative frequency components which frequently are redundant for applications. This is due to the following relation.
  \begin{lemma}
 	Suppose $f: \R \rightarrow \R$ is continuous and continuously differentiable, then
 	\begin{equation}
 	\label{eq:Ff1 spectrum}
 	F \left[ \tilde{f} \right] (\omega) = \int\limits_{-\infty}^{\infty} H[f](x) \, e^{-i \omega x} \, \diff x = -i \operatorname{sign}(\omega) F[f](\omega)
 	\end{equation}
 \end{lemma}
 Combination of the function $f$ and its Hilbert transform $\tilde{f}$ provides all the necessary information about the envelope function, instantaneous phase and frequency.
 \begin{definition}[Instantaneous amplitude, phase and frequency]
 	\label{def:instantaneous amplitude, phase and frequency 1-D}
 	\ \\
 \begin{itemize}
 	\item The absolute value of complex valued analytic signal is called {\it instantaneous amplitude} or {\it envelope} $a(x)$
 	\begin{equation}
 	a(x) = |f_a(x)| = \sqrt{f(x)^2+\tilde{f}(x)^2}.
 	\end{equation}
 	\item The argument $\phi(x)$ of analytic signal $f_a(x)$ is called {\it instantaneous phase}
 	\begin{equation}
 	\phi(x) = \arg [ f_a ](x) = \arctan \left( \frac{\tilde{f}(x)}{f(x)}\right).
 	\end{equation}
 	\item The {\it instantaneous frequency} $\nu(x)$ is defined as derivative of instantaneous phase
 	\begin{equation}
 	\nu(x) = \frac{\diff \phi}{\diff x}.
 	\end{equation}
 \end{itemize}
 \end{definition}
 \begin{remark}
 	\label{obs:Ff=1+sgnOmega}
 	 Analytic signal $f_a(x)$ could be defined in terms of Fourier transform, by discarding the negative frequency components 
 	 \begin{equation}
 	 \label{eq:Ff=1+sgnOmega}
 	 \hat{f}_a(\omega) = \left[ 1 + \sign (\omega) \right] \hat{f}(\omega).
 	 \end{equation} 
 \end{remark}
 
 Quite complementary perspective on the relationship between $f$ and $\tilde{f}$ may be obtained by observing that 
 \begin{equation}
 H \left[ \cos (\omega x' + \phi) \right](x) = \sin(\omega x + \phi).
 \end{equation}
 Therefore we observe how Hilbert transform phase-shifts the cosine by $\pi/2$ turning it into sine. Next we give definitions of Fourier sine and cosine transforms that will appear useful in defining the phase-shifted components. 
 \begin{definition}
 The Fourier cosine and sine transforms of continuous and absolutely integrable $f:\R \rightarrow \R$ are defined by
 \begin{align}
 F_c [f](\omega) =  \int\limits_{-\infty}^\infty f(x) \cos(\omega x) \, \diff x, \\
 F_s [f](\omega) =  \int\limits_{-\infty}^\infty f(x) \sin(\omega x) \, \diff x.
 \end{align}
 \end{definition}
 
 In the following we will rely on the Fourier integral formula that follows directly from the definition of Fourier transform.
 \begin{lemma}[Fourier integral formula]
 	Any continuous and absolutely integrable on its domain $f: \R \rightarrow \R$ may be represented as
 \begin{equation}
 \label{eq:Fourier integral formula}
 f(x)= \frac{1}{\pi} \int\limits_{0}^{\infty} \int\limits_{-\infty}^{\infty} f(x') \cos \left[ (x-x') \omega \right] \, \diff x' \diff \omega. 
 \end{equation}
 \begin{proof}
 	From the definition of Fourier transform we can write the decomposition
 	\begin{equation}
 	f(x)= \frac{1}{2 \pi} \int\limits_{-\infty}^{\infty} \int\limits_{-\infty}^{\infty} f(x') e^{ (x-x') \omega } \, \diff x' \diff \omega.
 	\end{equation}
 	Taking the real part of both sides and observing that cosine is even in $\omega$ we arrive to the result.
 \end{proof}
 \end{lemma}
 
 In the following theorem we deduce the modified Fourier integral formula and relate it to the Hilbert transform.
 \begin{theorem}
 	\label{theorem:Hf = f1}
 	Let $f: \R \rightarrow \R$ be continuous and absolutely integrable and let $H[f]$ be the Hilbert transform of $f$. Then we have
 	\begin{equation}
 	\label{eq:1 dim phase}
 		H[f](x) =  \frac{1}{\pi} \int\limits_{0}^{\infty} \int\limits_{-\infty}^{\infty} f(x') \sin \left[ (x-x') \omega \right] \, \diff x' \, \diff \omega.
 	\end{equation}
 	\begin{proof}
		Let us denote for convenience the left-hand side of \eqref{eq:1 dim phase} by $\tilde{f}$ and right-hand side by $f_1$. In the proof we treat the generalized function $\delta(x)$ as an infinitely localized measure or a Gaussian function in the limit of zero variance. Under this assumption, we can write the following cumulative distribution function
		\begin{equation}
		H(x) = \int\limits_{-\infty}^x \delta(s) \, \diff s,
		\end{equation}
 		which is the Heaviside step function $H(\omega) = \frac{1}{2}(1+\sign(\omega))$, and, in particular, we have
 		\begin{equation}
 		\label{eq: H(0)= int delta}
 		H(0)= \int\limits_{-\infty}^{0} \delta(s) \, \diff s = \int\limits_{0}^{\infty} \delta(s) \, \diff s = \frac{1}{2}.
 		\end{equation}
 		Then we write the Fourier transform of the right hand side of \eqref{eq:1 dim phase} as
 		\begin{align}
 		\label{eq:Ff1 expansion}
 		\begin{split}
 		F [f_1] (\omega') & =
 		\frac{1}{\pi} \int\limits_{-\infty}^{\infty} \left( \int\limits_{0}^{\infty} \int\limits_{-\infty}^{\infty} f(x') \sin \left[ \omega (x-x') \right] \, \diff x' \diff \omega \right) \ e^{-i \omega' x} \, \diff x \\
 		& = \frac{1}{2 \pi i} \int\limits_{-\infty}^{\infty} \int\limits_{0}^{\infty} \int\limits_{-\infty}^{\infty} f(x') \left[ e^{i \omega (x-x')} - e^{-i \omega (x-x')} \right] e^{-i \omega' x} \, \diff x' \diff \omega \diff x \\
 		& = \frac{1}{2 \pi i} \int\limits_{-\infty}^{\infty} \int\limits_{0}^{\infty} F[f](\omega) e^{i(\omega-\omega')x} \, \diff \omega \diff x - \frac{1}{2 \pi i} \int\limits_{-\infty}^{\infty} \int\limits_{0}^{\infty} F[f](-\omega) e^{i(-\omega-\omega')x} \, \diff \omega \diff x  \\
 		& = -i \int\limits_{0}^{\infty} F[f](\omega) \delta(\omega - \omega') \, \diff \omega + i \int\limits_{0}^{\infty} F[f](-\omega) \delta(-\omega - \omega') \, \diff \omega.
 		\end{split}
 		\end{align}
 		In the derivations above we used the relation $\int\limits_{-\infty}^{\infty} e^{i(\omega-\omega')x} \, \diff x = 2 \pi \delta(\omega - \omega') $. 
 		Now we see, by taking into account \eqref{eq: H(0)= int delta}, that
 		\begin{align}
 		\begin{split}
 		&\int\limits_{0}^{\infty} F[f](\omega) \delta(\omega - \omega') \, \diff \omega = H(\omega') F[f](\omega'),  \\
 		&\int\limits_{0}^{\infty} F[f](-\omega) \delta(-\omega - \omega') \, \diff \omega = H(-\omega') F[f](\omega').
 		\end{split}
 		\end{align}
 		Finally, from the last line in \eqref{eq:Ff1 expansion} we observe that
 		\begin{align}
 		F[f_1](\omega) = \begin{cases}
 		-i F[f](\omega), & \text{if } \omega > 0, \\
 		- \frac{i}{2} F[f](\omega) +  \frac{i}{2} F[f](-\omega), & \text{if } \omega = 0, \\
 		i F[f](\omega), & \text{if } \omega < 0, \end{cases}
 		\end{align}
 		which is exactly in accordance with \eqref{eq:Ff1 spectrum} and therefore $f_1 = \tilde{f}$.  		
 	\end{proof}
 \end{theorem}

 Now by expanding sine of difference in the integrand of \eqref{eq:1 dim phase}, we can write
 \begin{align}
 \label{eq: f1 expansion}
 \begin{split}
 f_1(x) =& \ \frac{1}{\pi} \int\limits_{0}^{\infty} \int\limits_{-\infty}^{\infty} f(x') \left[ \sin \left( x \omega \right) \cos \left( x' \omega \right) - \cos \left(x \omega \right) \sin \left( x' \omega \right) \right] \, \diff x' \, \diff \omega \\
 =& \ \frac{1}{\pi} \int\limits_{0}^{\infty} \left( \int\limits_{-\infty}^{\infty} f(x') \cos \left( x' \omega \right) \diff x' \right) \sin \left( x \omega \right) \, \diff \omega \\
 & \ - \frac{1}{\pi} \int\limits_{0}^{\infty} \left( \int\limits_{-\infty}^{\infty} f(x') \sin \left( x' \omega \right) \diff x' \right) \cos \left( x \omega \right) \, \diff \omega.
 \end{split}
 \end{align}
 For the sake of brevity we introduce the following notation
 \begin{align}
 \label{eq:notation indices 1D}
 \begin{split}
 &\alpha_0 = \alpha_0 (x, \omega) =\cos \left( \omega x \right), \\
 &\alpha_1 = \alpha_1 (x, \omega) = \sin \left( \omega x \right), \\
 &\alpha^0 = \alpha^0 (\omega) = \int\limits_{-\infty}^{\infty} f(x') \cos \left( x' \omega \right) \, \diff x' \\
 &\alpha^1 = \alpha^1 (\omega) = \int\limits_{-\infty}^{\infty} f(x') \sin \left( x' \omega \right) \, \diff x', \\
 &\hat{\alpha}^0 = \hat{\alpha}^0 (x) = \frac{1}{2\pi} \int\limits_{-\infty}^{\infty} \hat{f}(\omega) \cos \left( \omega x \right) \, \diff \omega, \\
 &\hat{\alpha}^1 = \hat{\alpha}^1 (x) = \frac{1}{2\pi} \int\limits_{-\infty}^{\infty} \hat{f}(\omega) \sin \left( \omega x \right) \, \diff \omega.
 \end{split}
 \end{align}
 For the integration over $\omega$ of a pair of functions $a(\omega)$ and $b(x,\omega)$ we will write
 \begin{align}
 \label{eq: notation brackets 1D}
 \begin{split}
 \left\langle a, b \right\rangle_+ &= \frac{1}{\pi} \int\limits_{0}^{\infty} a(\omega) b(x,\omega) \, \diff \omega, \\
 \left\langle a, b \right\rangle &= \frac{1}{2\pi} \int\limits_{-\infty}^{\infty} a(\omega) b(x,\omega) \, \diff \omega.
 \end{split}
 \end{align}

 \begin{observation}
 	\label{observation: 1D phases}
 	Equipped with the notation introduced in \eqref{eq:notation indices 1D} and \eqref{eq: notation brackets 1D} and using Fourier integral formula \eqref{eq:Fourier integral formula} we can expand $f_0 = f(x)$ in the following form
 	\begin{equation}
 	\label{eq:f0 with brackets}
 	f_0 = \left\langle \alpha^0, \alpha_0 \right\rangle_+ + \left\langle \alpha^1, \alpha_1 \right\rangle_+.
 	\end{equation}
 	Similarly using the results of Theorem \ref{theorem:Hf = f1} we can write for $f_1 = \tilde{f}$
 	\begin{equation}
 	\label{eq:f1 with brackets}
 	f_1 = \left\langle \alpha^0, \alpha_1 \right\rangle_+ - \left\langle \alpha^1, \alpha_0 \right\rangle_+.
 	\end{equation}
 \end{observation}
 Above observation shows that the original signal $f$ is obtained by first projecting $f$ onto sine and cosine harmonics and then by reconstructing using the same harmonics correspondingly. Hilbert transform $f_1$, or phase-shifted version of $f$, is obtained differently though. First we project initial function $f$ on sine and cosine harmonics to obtain $\alpha^1$ and $\alpha^0$, however then for reconstruction the phase-shifted harmonics are used, i.e. $\alpha_0$ and $\alpha_1$.
 \begin{observation}
 	The Fourier transform of a general complex-valued function $f: \R \rightarrow \C$, with $f(x) = f_0(x) + i f_1(x)$, may be written in terms of projection on phase-shifted harmonics as
 	\begin{align}
 	\hat{f}(\omega) = F[f](\omega) = \alpha^0(\omega) - i \alpha^1(\omega),
 	\end{align}
 	which follows if we apply Euler's formula $e^{ix} = \cos x + i \sin x$ to the definition \eqref{eq:Fourier transform}. The inverse transform may be written as
 	\begin{equation}
 	F^{-1} [ \hat{f} ](x) = \hat{\alpha}^0(x) + i \hat{\alpha}^1(x).
 	\end{equation}
 	The original function is recovered by
 	\begin{equation}
 	f(x) = F^{-1} \left[ \hat{f} \right] (x) = \left\langle \alpha^0, \alpha_0 \right\rangle + \left\langle \alpha^1, \alpha_1 \right\rangle + i \left( \left\langle \alpha^0, \alpha_1 \right\rangle - \left\langle \alpha^1, \alpha_0 \right\rangle \right).
 	\end{equation}
 \end{observation}
 A variety of approaches for generalization of the analytic signal method to many dimensions has been employed. For example quaternionic-valued generalization of analytic signal has been realized \cite{bulow2001hypercomplex} and works well in case $d=2$. Later on, in the Example \ref{example:quaternionic example}, we show why and how the quaternionic based approach is consistent with presented theory. As a next step we are going to generalize formulas \eqref{eq:f0 with brackets} and \eqref{eq:f1 with brackets} to multidimensional setting, first, without relying on any additional hypercomplex structure, while later in Sections \ref{section:analytic functions of hypercomplex variable} and \ref{section:commutative hypercomplex Fourier transform} we choose a convenient hypercomplex algebra and Fourier transform.

\section{In many dimensions}
\label{section:in many dimensions}
In several dimensions we are puzzled with the same questions on what are the adequate definitions of instantaneous amplitude and phase. Again we start by considering some function $f : \mathbb{R}^d \rightarrow \mathbb{R}$ having some local oscillatory behavior. We start by extending the Fourier integral formula to many dimensions.

For convenience and brevity we introduce the following notation. For a given binary $\{0,1\}$-vector $\bm{j} \in \{0,1\}^d$ we define the functions $\alpha_{\bm{j}}: \R^d \times \R^d \rightarrow \R$ by
\begin{equation}
\label{eq:definition of alpha functions}
\alpha_{\bm{j}} = \alpha_{\bm{j}} (\bm{x}, \bm{\omega}) = \prod_{l=1}^{d} \cos \left(\omega_l x_l - j_l \frac{\pi}{2} \right).
\end{equation}

 \begin{theorem}
	Any continuous and absolutely integrable over its domain $f: \R \rightarrow \R$ may be decomposed as
	\begin{align}
	\label{eq:Fourier integral formula many dimensions}
	\begin{split}
	f(x_1, \dotsc, x_d)= \ &\frac{1}{\pi^d} \int\limits_{0}^{\infty} \int\limits_{-\infty}^{\infty} \dotsi \int\limits_{0}^{\infty} \int\limits_{-\infty}^{\infty} f(x'_1,x'_2,\dotsc,x'_d) \cos \left[ (x_1 - x'_1) \omega_1 \right] \cdot \dotsc \\ &\cdot \cos \left[ (x_d - x'_d) \omega_d \right] \, \diff x'_1 \diff \omega_1 \dotsc \diff x'_d \diff \omega_d. 
	\end{split}
	\end{align}
	or by employing the notation \eqref{eq:definition of alpha functions},
	\begin{equation}
	\label{eq:Fourier integral formula multi-d short-form}
	f(\bm{x}) = \frac{1}{\pi^d} \bm{\int\limits_0^\infty \int\limits_{-\infty}^{\infty} } f(\bm{x'}) \alpha_{\bm{0}} \left(\bm{x-x'}, \bm{\omega} \right) \,  \diff \bm{x'} \diff \bm{\omega}
	\end{equation} 
	with $\bm{0} \in \{ 0,1 \}^d$ being the all-zeros vector.
	\begin{proof}
		Proof is the same as for 1-D case.
	\end{proof}
 \end{theorem}
 Similarly to the result of Theorem \ref{theorem:Hf = f1} next we define the phase-shifted copies of the oscillating function $f$.
 \begin{definition}
 	The phase-shifted version $f_{\bm{j}}: \R \rightarrow \R$, in the direction $\bm{j} \in \{0,1\}^d$, of the function $f: \R \rightarrow \R$ is given by
 	\begin{equation}
 	\label{eq:fj definition}
 	f_{\bm{j}}(\bm{x}) = \frac{1}{\pi^d} \bm{\int\limits_0^\infty \int\limits_{-\infty}^{\infty} } f(\bm{x'}) \alpha_{\bm{j}} \left(\bm{x-x'}, \bm{\omega} \right) \,  \diff \bm{x'} \diff \bm{\omega}.
 	\end{equation}
 \end{definition}
\noindent Next we give the definitions of instantaneous amplitude, phase and frequency. 
 \begin{definition}
 	\label{def:instantaneous amplitude, phase and frequency} \ \\ \vspace{-0.5cm}
 	\begin{itemize}
 		\item The square root of sum of all phase-shifted copies of signal is called {\it instantaneous amplitude} or {\it envelope} $a(\bm{x})$
 		\begin{equation}
 		\label{eq:envelope in many dim}
 		a(\bm{x}) =  \sqrt{\sum_{\bm{j}\in \{0,1\}^d} f_{\bm{j}} (\bm{x})^2}.
 		\end{equation}
 		\item The {\it instantaneous phase} $\phi_{\bm{j}}(\bm{x})$ in the direction $\bm{j} \in \{0,1\}^d$ is defined by
 		\begin{equation}
 		\phi_{\bm{j}}(\bm{x}) = \arctan \left( \frac{f_{\bm{j}}(\bm{x})}{f(\bm{x})}\right).
 		\end{equation}
 		\item The {\it instantaneous frequency} $\nu_{\bm{j}}(\bm{x})$ in the direction $\bm{j} \in \{0,1\}^d$ is defined as partial derivatives of the corresponding instantaneous phase in the directions given by $\bm{j}$
 		\begin{equation}
 		\label{eq:definition of instantaneous freq many-dim}
 		\nu_{\bm{j}}(\bm{x}) = \partial^{\bm{j}} \phi_{\bm{j}} = \partial_{1}^{j_1} \circ \partial_{2}^{j_2} \circ ... \circ \partial_{d}^{j_d} \left( \phi_{\bm{j}} \right) (\bm{x}),
 		\end{equation}
 		where we used multi-index notation for derivatives, $\partial_i^1 = \frac{\partial}{\partial x_i}$ and $\partial_i^0 = \operatorname{id}$. Note that no Einstein summation convention was employed in \eqref{eq:definition of instantaneous freq many-dim}.
 	\end{itemize}
 \end{definition}
 We can generalize Theorem \ref{theorem:Hf = f1} by relating phase-shifts $f_{\bm{j}}$ to the corresponding multidimensional Hilbert transforms. However first let us define Hilbert transform in several dimensions.
 \begin{definition}
 	\label{def:Hj}
 	The Hilbert transform in the direction $\bm{j} \in \{0,1\}^d$ is defined by
 	\begin{equation}
 	\label{eq:Hj}
 	H_{\bm{j}} \left[ f \right] (\bm{x}) =  \text{p.v.} \frac{1}{\pi^{|\bm{j}|}} \bm{\int\limits_{\R^{|\bm{j}|}}} \frac{f(\bm{y})}{(\bm{x}-\bm{y})^{\bm{j}}} \, \diff \bm{y}^{\bm{j}},
 	\end{equation}
 	where $|\bm{j}|$ gives the number of $1$-s in $\bm{j}$, $(\bm{x}-\bm{y})^{\bm{j}} = \prod_{i=1}^d (x_i - y_i)^{j_i}$ and $\diff \bm{y}^{\bm{j}} = \diff y_1^{j_1} ... \diff y_d^{j_d} $, i.e. integration is performed only over the variables indicated by the vector $\bm{j}$.
 \end{definition} 
 \begin{theorem}
 	\label{fj=Hj}
 	For a continuous $f: \R \rightarrow \R$ that is absolutely integrable in its domain, we have
 	\begin{equation}
 		f_{\bm{j}} = H_{\bm{j}} \left[ f \right].
 	\end{equation}
 	\begin{proof}
 		proof follows the lines of the proof of Theorem \ref{theorem:Hf = f1}.
 	\end{proof}
 \end{theorem}

It is useful to express the phase-shifted functions $f_{\bm{j}}$ as a combination of harmonics $\alpha_{\bm{j}}$ and corresponding projection coefficients $\alpha^{\bm{i}}$. We define projection coefficients similarly to \eqref{eq:notation indices 1D},
 \begin{align}
 \label{eq:notation alpha^j many dim}
 \begin{split}
 &\alpha^{\bm{j}} = \alpha^{\bm{j}}(\bm{\omega}) = \bm{\int\limits_{-\infty}^{\infty}} f(\bm{x}) \alpha_{\bm{j}} \left( \bm{x}, \bm{\omega} \right) \, \diff \bm{x}, \\
 &\hat{\alpha}^{\bm{j}} = \hat{\alpha}^{\bm{j}}(\bm{x}) = \frac{1}{(2 \pi)^d} \bm{\int\limits_{-\infty}^{\infty}} \hat{f}(\bm{\omega}) \alpha_{\bm{j}} \left( \bm{x}, \bm{\omega} \right) \, \diff \bm{\omega}, \\
 &\hat{\alpha}_{+}^{\bm{j}} = \hat{\alpha}_{+}^{\bm{j}}(\bm{x}) = \frac{1}{\pi^d} \bm{\int\limits_{0}^{\infty}} \hat{f}(\bm{\omega}) \alpha_{\bm{j}} \left( \bm{x}, \bm{\omega} \right) \, \diff \bm{\omega}
 \end{split}
 \end{align}
 and brackets, similarly to \eqref{eq: notation brackets 1D}, as
 \begin{align}
 \label{eq:bracket notation multidimensional}
 \begin{split}
 	 \left\langle a, b \right\rangle_+ = \left\langle a, b \right\rangle_+ (\bm{x})= \frac{1}{\pi^d} \bm{\int\limits_{0}^{\infty}} a(\bm{\omega}) b(\bm{x},\bm{\omega}) \, \diff \bm{\omega}, \\
 	 \left\langle a, b \right\rangle = \left\langle a, b \right\rangle (\bm{x})= \frac{1}{(2\pi)^d} \bm{\int\limits_{-\infty}^{\infty}} a(\bm{\omega}) b(\bm{x},\bm{\omega}) \, \diff \bm{\omega},
 \end{split}
 \end{align}
 for any two functions $a(\bm{\omega}) = a(\omega_1,\dots,\omega_d)$ and $b(\bm{x},\bm{\omega}) = b(x_1,\dots,x_d, \omega_1, \dots, \omega_d)$.
 It is useful to observe how phase-shifted functions $f_{\bm{j}}$ could be obtained from $\alpha^{\bm{j}}$. 
 \begin{theorem}
 	For a given function $f: \R^d \rightarrow \R$, one obtains phase-shifted copy $f_{\bm{j}}$, in the direction $\bm{j} \in \{0,1\}^d$, by 
 	\begin{equation}
 	\label{eq:fj and alphas}
 	f_{\bm{j}} = \sum_{\bm{i} \in \{0,1\}^d} (-1)^{\left| \left( \bm{i} \oplus \bm{j} \right) \ominus \bm{i} \right|} \left\langle \alpha^{\bm{i} \oplus \bm{j}}, \alpha_{\bm{i}} \right\rangle_+,
 	\end{equation}
 	where $\oplus$ is a binary exclusive OR operation acting elementwise on its arguments and $\ominus$ is defined as following: $1 \ominus 0 = 1$ and the result is $0$ otherwise.
 	\begin{proof}
 		The proof follows immediately after we expand all the cosines and sines of difference in \eqref{eq:fj definition} and make all the substitutions from \eqref{eq:definition of alpha functions} and \eqref{eq:notation alpha^j many dim}.
 	\end{proof}
 \end{theorem}

\begin{example}
Short illustration may be helpful to understand the rule \eqref{eq:fj and alphas}. Let us suppose that we have $\bm{j} = (1,0,1,1,0)$ and $\bm{i} = (1,1,0,0,0)$. In this case $\bm{i} \oplus \bm{j} = (1 \oplus 1, 0 \oplus 1, 1 \oplus 0, 1 \oplus 0, 0 \oplus 0) = (0,1,1,1,0)$. Then we have $\left( \bm{i} \oplus \bm{j} \right) \ominus \bm{i} = (0 \ominus 1, 1 \ominus 1, 1 \ominus 0, 1 \ominus 0, 0 \ominus 0) = (0,0,1,1,0)$ therefore we will have ``$+$'' sign in front of $\left\langle \alpha^{01110}, \alpha_{11000} \right\rangle_+$ for $f_{10110}$. 
\end{example}

\begin{remark}
	Before moving to the theory of hypercomplex holomorphic functions and analytic signals, it will be useful to note that from the definition \eqref{eq:fj definition} there will be in total $2^d$ different components $f_{\bm{j}}$. It says us that most likely the dimensionality of hypercomplex algebra in which analytic signal should take its values should be also $2^d$.
\end{remark}

 \section{Holomorphic functions of several variables}
 \label{section:analytic functions of hypercomplex variable}
 \subsection{Commutative hypercomplex algebra}
 \label{section:Commutative hypercomplex algebra}
 The development of hypercomplex algebraic systems started with the works of Gauss, Hamilton \cite{hamilton1844ii}, Clifford \cite{clifford1871}, Cockle \cite{cockle1849iii} and many others. Generally a hypercomplex variable $w$ is given by the linear combination of $n$ hypercomplex units $\{ e_i \}$
 \begin{equation}
 w = \sum_{i=0}^n w_i e_i,
 \end{equation}
 with coefficients $w_i \in \R$. Product rule for the units $e_i$ is given by the structure coefficients $\gamma^s_{ij} \in \R$
 \begin{equation}
 \label{eq:multiplication gamma constants}
 e_i e_j = \sum_{s=0}^n \gamma^s_{ij} e_s.
 \end{equation}
 Only algebras with unital element or module are of interest to us, i.e. those having element $\epsilon = 1$ such that $\epsilon x = x \epsilon = x$. The unital element have expansion
 \begin{equation}
 \epsilon = \sum_{i=0}^n \epsilon_i e_i.
 \end{equation}
 For simplicity we will always assume that the unital element $\epsilon = e_0 = 1$.
 
 The units $e_i, \ i \neq 0$, could be subdivided into three groups \cite{Catoni2008}.
 \begin{enumerate}
 	\item $e_i$ is {\it elliptic unit } if $e_i^2 = -e_0$
 	\item $e_i$ is {\it parabolic unit } if $e_i^2 = 0$
 	\item $e_i$ is {\it hyperbolic unit } if $e_i^2 = e_0$.
 \end{enumerate}
 
 \begin{example}
 	General properties of some abstract hypercomplex algebra $\Sch$ that is defined by its structure constants $\gamma^s_{ij}$ may be quite complicated. We will focus mainly on the commutative and associative algebras constructed from the set of elliptic-type generators. For example suppose we have two elliptic generators $e_1^2 = -1$ and $e_2 = -1$, then one can construct the simplest associative and commutative elliptic algebra as $\Sch_2 = \operatorname{span}_{\R} \left\{ e_0 \equiv 1,e_1,e_2, e_3 \equiv e_1 e_2 \right\}$. The algebra have $2^2 = 4$ units. For $\Sch_2$ we have the following multiplication table
 	\begin{center}
 	\begin{tabular}{|c|c c c|}
 		\hline 
 		$e_0$ & $e_1$ & $e_2$ & $e_3$ \\ 
 		\hline 
 		$e_1$ & $-e_0$ & $e_3$  & $-e_2$ \\ 
 		$e_2$ & $e_3$  & $-e_0$ & $-e_1$  \\ 
 		$e_3$ & $-e_2$  & $-e_1$  & $e_0$ \\ 
 		\hline
 	\end{tabular} 
  	\captionof{table}{Multiplication table for $\Sch_2$ algebra}
 \end{center}
 \end{example}
 
 One could assemble various types of hypercomplex algebras depending on the problem. All of the above unit systems have their own applications. For example, numbers with elliptic unit relate group of rotations and translations of $2$-dimensional Euclidean space to the complex numbers along with their central role in harmonic analysis. Algebra with parabolic units may represent Galileo's transformations, while algebra with hyperbolic units could be used to represent Lorentz group in special relativity \cite{Catoni2008}. In this work we mainly focus on the concepts of instantaneous amplitude, phase and frequency of some oscillating process. Elliptic units are of great interest to those studying oscillating phenomena due to the famous Euler's formula. We briefly remind how it comes. Taylor series of exponential is given by
 \begin{equation}
 e^z = 1 + \frac{z}{1!} + \frac{z^2}{2!} + \frac{z^3}{3!} + \cdots = \sum_{n=0}^{\infty} \frac{z^n}{n!}.
 \end{equation}
 We could have various Euler's formulae. In elliptic case when $z = e_i x$, $e_i^2 = -1$, we can write from the above Taylor's expansion
 \begin{equation}
 \label{eq:eulers formula}
 e^{e_i x} = \cos (x) + e_i \sin (x).
 \end{equation}
 When $e_i^2 = 0$ we have
 \begin{equation}
 e^{e_i x} = 1 + e_i x.
 \end{equation}
 And finally when $e_i^2 = 1$ we get
 \begin{equation}
 e^{e_i x} = \cosh (x) + e_i \sinh (x).
 \end{equation}
 It is due to the relation \eqref{eq:eulers formula} that one relies on the complex exponentials for analysis of oscillating processes. One may well expect that to obtain amplitude and phase information, one will need a Fourier transform based on the algebra containing some {\it elliptic} numbers.
  
 It was a question whether the chosen hypercomplex algebra for multidimensional analytic signal should be commutative, anticommutative, associative or neither. Whether it is allowed to have zero divisors or not. In Section \ref{section:why clifford- transform doesnt fit} we show that commutative and associative algebra not only suffices but also is essentially a {\it necessary} condition to define Fourier transform coherently with \eqref{eq:fj and alphas}. Based on these general considerations we define the simplest associative and commutative algebra for a set of $d$ elliptic units. We call it {\it elliptic Scheffers\footnote{Due to the work \cite{scheffers1893generalisation} of G. W. Scheffers (1866-1945)} algebra} and denote it by $\Sch_d$.
 
 \begin{definition}
 	The elliptic Scheffers algebra $\Sch_d$ over a field $\R$ is an algebra of dimension $2^d$ with unit $\epsilon = e_0 = 1$ and generators $\{ e_1, \dots, e_d\}$ satisfying the conditions $e^2_i = -1, \ e_i e_j = e_j e_i, \ i,j = 1, \dots, d$. The basis of the algebra $\Sch_d$ consists of the elements of the form $e_0 = 1, \ e_{\beta} = e_{\beta_1} e_{\beta_2} \dots e_{\beta_d}, \ \beta = \left( \beta_1, \dots , \beta_s \right), \ \beta_1 < \beta_2 < \dots < \beta_s, \ 1 \leq s \leq d$. Each element $w \in \Sch_d$ has the form
 	\begin{equation}
 	\label{eq:hypercomplex variable}
 	w = \sum_{\beta = 0}^{2^d - 1} w_{\beta} e_{\beta} = w_0 \epsilon + \sum_{s = 1}^d \sum_{\beta_1 < \dots < \beta_s} w_{\beta} e_{\beta}, \ \ w_{\beta} \in \R.
 	\end{equation}
 \end{definition}
 \noindent If otherwise not stated explicitly we will consider mainly algebras over $\R$.
 To illustrate the definition above we can write any $w \in \Sch_2$ in the form
 \begin{equation}
 w = w_0 + w_1 e_1 + w_2 e_2 + w_{12} e_{1} e_{2}.
 \end{equation}
 \begin{remark}
 	We will use three different notations for indices of hypercomplex units and their coefficients. The dimension of Scheffers algebra $\Sch_d$ is $2^d$ therefore there always will be $2^d$ indices. However one can choose how to label them. First notation is given by natural numbering of hypercomplex units. For example for $\Sch_2$ we will have the following generators $\{ e_0 \equiv 1, e_1, e_2, e_3 \equiv e_1 e_2 \}$. Second notation is given by the set of indices in subscript, i.e. for $\Sch_2$ we have the generators $\{ e_0 \equiv 1, e_1, e_2, e_{1,2} \}$. Third notation uses binary representation $\bm{j} \in \{0,1\}^d $. For $\Sch_2$ example generators will be labeled as $\{ e_{00} \equiv 1, e_{10}, e_{01}, e_{11} \}$. To map between naturally ordered and binary representations we use index function $\ind(\cdot)$, for example in case of $\Sch_2$ we have $\ind(11) = \{1,2\} = 3$ as well as we have $\ind(3) = 11 \in \{ 0,1\}^2$ or $\ind (1,2) = 3 \in [0,1,...,2^d-1]$. Even though the function $\ind(\cdot)$ is overloaded, in practice it is clear how to apply it.
 \end{remark}
 
 \begin{remark}
 	\label{remark:norm and Banach algebra}
 The space $\Sch_d$ is a Banach space with norm of $w \in \Sch_d$ defined by
 \begin{equation}
 \label{eq:norm on Scheffers space}
 | w |_{\Sch_d} = \sqrt{\sum_{\beta =0}^{2^d-1} | w_\beta |^2}, \ w_\beta \in \R,
 \end{equation}
 while sometimes we will write just $| w |$ instead of $| w |_{\Sch_d}$.
 \end{remark}
 One can check that $\Sch_d$ is a unital commutative ring. The main difference of the elliptic Scheffers algebra from the algebra of complex numbers is that factor law does not hold in general, i.e. if we have vanishing product of some non zero $a,b \in \Sch_d$, it does not necessarily follow that either $a$ or $b$ vanish. In case $ab=0, \ a,b \neq 0$,  $a$ and $b$ are called zero divisors. {\it Zero divisors are not invertible}. However what is of actual importance for us is that $\Sch_d$ has subspaces spanned by elements $\{e_0, e_i\}$, each of which has the structure of the field of complex numbers and therefore factor law holds inside these subspaces. We will need this result to define properly the Cauchy integral formula later \cite{pedersen1997cauchy}.
 
 \begin{definition}
 	The space $\Sch(i), \ i = 1,...,d$, is defined by 
 	\begin{equation}
 	\Sch(i) = \left\{ a + be_i | a,b \in \R \right\}.
 	\end{equation}
 \end{definition}
 Real and imaginary parts of an element $\zeta = a + b e_i$ are defined in an obvious way by $\Re (\zeta) = a$ and $\Im (\zeta) = b$, while norm of $\zeta$ is given by $| \zeta |_{\Sch(i)} = \sqrt{a^2 + b^2}$. 
 Similarly to the complex analysis of several variables we define {\it open disk, polydisk} and {\it upper/lower half-planes}.
 \begin{definition}
 	The unit disk $\D(i) \subset \Sch(i)$ of radius $1$ centered at zero is defined as
 	\begin{equation}
 	\D(i) = \left\{ x \in \Sch(i): |x| < 1 \right\}.
 	\end{equation}
 	Sometimes we will write $\D(w_i, r)$ for a disk of radius $r$ centered at $w_i \in \Sch(i)$. 
 \end{definition}
 
 \begin{definition}
 	The upper half-plane $\Sch_+(i)$ is given by
 	\begin{equation}
 	\Sch_+(i) = \left\{ a + be_i | a,b \in \R; b>0 \right\},
 	\end{equation}
 	while the lower half-plane will be denoted by $\Sch_-(i)$
 	\begin{equation}
 	\Sch_-(i) = \{ a + be_i | a,b \in \R; b<0 \}.
 	\end{equation}
 \end{definition}
 Finally we combine these spaces to construct the domain for hypercomplex functions of several variables.
 \begin{definition}
 	\label{def:direct sum of S}
 	The {\it total Scheffers space} $\Sch^d$ is a direct sum of $d$ subalgebras $\Sch(i)$
 	\begin{equation}
 	\Sch^d = \bigoplus_{i=1}^d \Sch(i).
 	\end{equation}
 \end{definition}
  \begin{definition}
 	\label{def:direct sum of S plus}
 	The {\it upper Scheffers space} $\Sch_+^d$ is a direct sum of $d$ subalgebras $\Sch_+(i)$
 	\begin{equation}
 	\Sch_+^d = \bigoplus_{i=1}^d \Sch_+(i).
 	\end{equation}
 \end{definition}

\begin{definition}
	The {\it mixed Scheffers space}, denoted by $\Sch_{+ \bm{j}}$, is the direct sum of upper spaces marked by $1$-s in a vector $\bm{j} \in \{0,1\}^d$ and lower spaces marked by $0$-s
	\begin{equation}
	\label{eq: mixed Scheffers space}
	\Sch_{+ \bm{j}} = \left( \bigoplus_{i: \bm{j}_i = 1} \Sch_+ (i) \right) \oplus \left( \bigoplus_{i: \bm{j}_i = 0} \Sch_- (i) \right).
	\end{equation}
\end{definition}

  \begin{remark}
 	\label{remark:norm and Banach algebra for S^d}
 	The space $\Sch^d$ is a Banach space with norm of a vector $\bm{v} \in \Sch^d$, $\bm{v} = \left( v_1, \dotsc, v_d \right)$, each $v_i \in \Sch(i)$, given by
 	\begin{equation}
 	\label{eq:norm on domain Scheffers space}
 	| \bm{v} |_{\Sch^d} = \sqrt{\sum_{i} | v_i |_{\Sch(i)}^2}, \ v_i \in \Sch(i).
 	\end{equation}
 	Sometimes we will write for simplicity $| \bm{v} |$ instead of $| \bm{v} |_{\Sch^d}$.
 \end{remark}

  \begin{definition}
 	For $\bm{j} \in \{0,1\}^d$ we define $\bm{j}$-polydisk $\D_{\bm{j}} \subset \Sch^d$ as a product of $|\bm{j}|$ disks $\D(i)$ 
 	\begin{equation}
 	\D_{\bm{j}} = \prod_{i \in \ind(\bm{j})} \D(i)
 	\end{equation}
 	and we will write for $d$-polydisk $\D^{d} = \prod_{i = 1}^d \D(i)$ when we take product of all $d$ disks. Sometimes we will use polydisk of radius $r$ centered at vector $\bm{w} \in \Sch^d$ which is denoted by $\D_{\bm{j}}(\bm{w},r) = \prod_{i \in \ind(\bm{j})} \D(w_i, r)$.
 \end{definition}
 
 \begin{observation}
 	The boundary of $\Sch^d_+$ coincides with $\R^d$, i.e.
 	\begin{equation}
 	\partial \Sch_+^d = \overline{\Sch^d_+} \setminus \Sch^d_+ \simeq \R^d.
 	\end{equation}
 \end{observation}
 While each $\Sch(i)$ is equivalent to the complex plane $\C$, the combination $\Sch(i) \oplus \Sch(j)$ is not equivalent to $\C^2$ because imaginary units have different labels -- in a sense $\Sch(i) \oplus \Sch(j)$ contains more information than $\C^2$.
 	 
 \subsection{Holomorphic functions of hypercomplex variable}
 \label{subsection: Holomorphic (analytic) functions of hypercomplex variable}
The scope of this section is to define hypercomplex holomorphic/analytic functions of type $f:\Sch^d \rightarrow \Sch_d$. Indeed we are interested in rather restricted case of mappings between hypercomplex spaces which finally will be consistent with our definition of analytic signal as a function on the boundary of the polydisc in $\Sch^d$ in full correspondence with $1$-dimensional analytic signal. There are some similarities and differences with theory of functions of several complex variables. We refer the reader to the book \cite{krantz2001function} for details on the function theory of several complex variables and to the concise review \cite{gong2007concise} to refresh in memory basic facts on complex analysis. We start by providing several equivalent definitions of a holomorphic function of a complex variable by following Krantz \cite{krantz2001function}. Then we define holomorphic function of several hypercomplex variables in a pretty similar fashion. Starting from the very basics will be instructive for the hypercomplex case.

\begin{definition}
The derivative of a complex-valued function $f: \C \rightarrow \C$ is defined as a limit
\begin{equation}
\frac{\partial f}{\partial z} \left( z_0 \right) = \lim_{z \rightarrow z_0} \frac{f(z) - f(z_0)}{z-z_0}.
\end{equation}
If the limit exists one says that $f$ is {\it complex differentiable} at point $z_0$.
\end{definition}

\begin{definition}
	If $f: \Omega \rightarrow \C$, defined on some open subset $\Omega \subseteq \C$, is complex differentiable at every point $z_0 \in \Omega$, then $f$ is called {\it holomorphic} on $\Omega$.
\end{definition}

Complex differentiability means that the derivative at a point does not depend on the way sequence approaches the point. If one writes two limits, one approaching in the direction parallel to the real axis and one parallel to the imaginary axis, then after equating the corresponding real and imaginary terms one obtains the Cauchy-Riemann equations and therefore equivalent definition of a holomorphic function.

\begin{definition}
	A function $f: \Omega \rightarrow \C$, explicitly written as a combination of two real-valued functions $u$ and $v$ as $f(x+iy) = u(x,y) + i v(x,y)$, is called {\it holomorphic} in some open domain $\Omega \subseteq \C$ if $u$ and $v$ satisfy the Cauchy-Riemann equations
	\begin{align}
	\label{eq:cauchy-riemann equations}
	\begin{split}
	&\frac{\partial u}{\partial x} = \frac{\partial v}{\partial y}, \\
	&\frac{\partial u}{\partial y} = -\frac{\partial v}{\partial x}.
	\end{split}
	\end{align}
\end{definition}

A holomorphic function of a complex variable $z \in \C$ can be compactly introduced using the derivatives with respect to $z$ and conjugated variable $\bar{z}$
\begin{align}
\label{eq:complex differentiantion z and z bar}
\begin{split}
\frac{\partial}{\partial z} = \frac{1}{2} \left( \frac{\partial}{\partial x} - i \frac{\partial}{\partial y} \right), \\
\frac{\partial}{\partial \bar{z}} = \frac{1}{2} \left( \frac{\partial}{\partial x} + i \frac{\partial}{\partial y} \right).
\end{split}
\end{align}

\begin{definition}
	A function $f: \Omega \rightarrow \C$ is called {\it holomorphic} in some open domain $\Omega \in \C$ if for each $z \in \Omega$
	\begin{equation}
	\frac{\partial f}{\partial \bar{z}} = 0,
	\end{equation}
	which means that holomorphic function is a proper function of a single complex variable $z$ and not of the conjugated variable $\bar{z}$.
\end{definition}
 Contour integration in the complex space is yet another viewpoint on the holomorphic functions. Complex integration plays the central role in our work due to our interest in Hilbert transform that is defined as a limiting case of contour integral over the boundary of an open disk (see Appendix \ref{appendix:Hilbert transform in 1D}).

\begin{definition}
	A continuous function $f: \Omega \rightarrow \C$ is called {\it holomorphic} if for each $\omega \in \Omega$ there is an $r = r(w)>0$ such that $\overline{\D}(w,r) \subseteq \Omega$ and
	\begin{equation}
	\label{eq:cauchy formula 1d - definition}
	f(z) = \frac{1}{2 \pi i} \oint\limits_{|\zeta - w| = r} \frac{f(\zeta)}{\zeta - z} \, \diff \zeta
	\end{equation}
	for all $z \in \D(w,r)$.
\end{definition}

\begin{remark}
	\label{remark:zero divisors}
	Structure of $\C$ as a field is important for the definition of the integral in \eqref{eq:cauchy formula 1d - definition}.  Only because every element of a field has a multiplicative inverse, we can write the integration of the kernel $\frac{1}{z}$. However in imaginary situation when the contour of integration passes through some points where $(\zeta-z)$ is not invertible, we will be unable to write the Cauchy formula \eqref{eq:cauchy formula 1d - definition}.
\end{remark}

Finally we can justify the word ``analytic'' in the title of the paper by giving the following definition. Note that analytic and holomorphic means essentially the same.
\begin{definition}
	A function $f: \Omega \rightarrow \C$ is called {\it analytic} (holomorphic) in some open domain $\Omega \in \C$ if for each $z_0 \in \Omega$ there is an $r = r(z_0) >0$ such that $\D (z_0, r) \subseteq \Omega$ and $f$ can be written as an absolutely and uniformly convergent power series
	\begin{equation}
	f(z) = \sum_{k} a_k \left( z - z_0 \right)^k
	\end{equation}
	for all $z \in \D \left( z_0,r \right)$.
\end{definition}

Having introduced basic notions of holomorphic functions now we are ready to review the holomorphic function theory in the hypercomplex space. Theory of holomorphic functions was generalized to the functions of commutative hypercomplex variables by G.W. Scheffers in 1893 \cite{scheffers1893generalisation}, then his work was extended in 1928 by P.W. Ketchum \cite{ketchum1928analytic} and later by V.S. Vladimirov and I.V. Volovich \cite{vladimirov1984superanalysis1}, \cite{vladimirov1984superanalysis2} to the theory of superdifferentiable functions in superspace with commuting and anti-commuting parts. One important distinction to previous works is that we present different Cauchy formula from one given by Ketchum and Vladimirov. Their approach is based on complexification (i.e. $w_\beta \in \C$) of the underlying hypercomplex algebra, while we are working in the hypercomplex algebra over reals. We briefly describe their formula in Appendix \ref{appendix: Cauchy formula of Ketchum and Vladimirov}.

Let us recall, having in mind Remarks \ref{remark:norm and Banach algebra} and \ref{remark:norm and Banach algebra for S^d}, the definition of continuity for a function $f: A \rightarrow \Sch_d$ defined on some open subset $A \subseteq \Sch^d$. We say the limit of $f$ as $\bm{\zeta}$ approaches $\bm{\zeta}_0$ equals $a$, denoted by $\lim_{\bm{\zeta} \rightarrow \bm{\zeta}_0} = a$ if for all $\varepsilon > 0$ there exists a $\delta > 0$ such that if $\bm{\zeta} \in A$ and $\left| \bm{\zeta} - \bm{\zeta}_0 \right| < \delta$ then $\left| f \left( \bm{\zeta} \right) - a \right| < \varepsilon$. Then $f$ is said to be continuous at $\bm{\zeta}_0$ if $\lim_{\bm{\zeta} \rightarrow \bm{\zeta}_0} = f(\bm{\zeta}_0)$. If $f$ is continuous for every point $\bm{\zeta}_0 \in A$ then $f$ is said to be continuous on $A$.

First by following \cite{scheffers1893generalisation} we give definition of a holomorphic function in some general hypercomplex space $\Gamma$ that is defined by its structure constants $\gamma^s_{ik}$, see \eqref{eq:multiplication gamma constants}. This definition generalizes ordinary Cauchy-Riemann equations and follows from the same argument that hypercomplex derivative at a point should be independent on the way one approaches the point in a hypercomplex space. Finally we will restrict attention to the functions $\Sch^d \rightarrow \Sch_d$ and will see that we are consistent with our previous derivations.
\begin{definition}[\cite{scheffers1893generalisation}]
	A function $f: \Gamma \rightarrow \Gamma, \ w \mapsto f(w)$	is \textit{holomorphic} in an open domain $\Omega \subseteq \Gamma$ if for each point $w \in \Omega$ it satisfies the generalized Cauchy-Riemann-Scheffers equations
	\begin{equation}
	\label{eq:scheffers equations}
	\frac{\partial f_s}{\partial w_k} = \sum_{i=0}^{n} \gamma^s_{ik} \frac{\partial f_i}{\partial w_0}, \ s,k = 0, \dots, n.
	\end{equation}
\end{definition}
General equations could be simplified if we consider the functions that have only subsets of $\Gamma = \Sch_d$ as their domain, in our case we consider functions $f: \Sch(i) \rightarrow \Sch_d$ that map plane $\Sch(i)$ in $\Sch_d$ to the total space $\Sch_d$. If we restrict our domain only to the plane $\Sch(i)$ and insert values of structure constants for $\Sch_d$ we obtain simple Cauchy-Riemann equations (see for example eqq. \eqref{eq:scheffers equations 2d z1} and \eqref{eq:scheffers equations 2d z2} for 2-D case). We will arrive to the explicit equations in a while. Mappings from restricted domains, i.e. in our case $\Sch(i) \subseteq \Sch_d$, are described in details in \cite{ketchum1928analytic}.

\begin{remark}
	To avoid confusion the comment on the relationship between the spaces $\Sch(i)$, $\Sch^d$ and $\Sch_d$ will be useful. While we can embed trivially $\Sch(i) \hookrightarrow \Sch^d$ as well as $\Sch(i) \hookrightarrow \Sch_d$, one may tend to think that it will also be natural to embed $\Sch^d \hookrightarrow \Sch_d$. Last embedding always exists because dimensionality of $\Sch_d$ is $2^d$ and is non smaller that the dimensionality of $\Sch^d$ which is $2d$. However this embedding will not be trivial. For example consider an element $\bm{x} \in \R^d \subseteq \Sch^d$. In $\Sch_d$ there is only one unital element, while in $\Sch^d$ there are $d$ of them, because $\Sch^d = \bigoplus_{i=1}^d \Sch(i)$, therefore it is impossible to ``match`` the unital elements of $\Sch_d$ and $\Sch^d$ in a trivial way. Therefore it is better to think about $\Sch^d$ and $\Sch_d$ as being two different spaces that however ``share'' planes $\Sch(i)$.
\end{remark}

From now on we will study the functions $f: \Sch^d \rightarrow \Sch_d$. A function $f$ could be explicitly expressed as a function of several hypercomplex variables $f(z_1, ..., z_d) \equiv f(x_1 + e_1 y_1, ..., x_d + e_d y_d)$ with each $z_k \in \Sch(k)$. 

\begin{definition}
A function $f: \Sch^d \rightarrow \Sch_d$ is called {\it hypercomplex differentiable in the variable $z_k$} if the following limit exists
\begin{equation}
\frac{\partial f}{\partial z_k} (z_1, ..., z_0, ..., z_d) = \lim_{z \rightarrow z_0} \frac{f(z_1,...,z, ..., z_d)-f(z_1,...,z_0, ..., z_d)}{z - z_0}.
\end{equation}
\end{definition}

\begin{definition}
	If $f: \Omega \rightarrow \Sch_d$, defined on some open subset $\Omega \subseteq \Sch^d$, is hypercomplex differentiable in each variable separately at every point $z_0 \in \Omega$, then $f$ is called {\it holomorphic} on $\Omega$.
\end{definition}

One obtains the generalized Cauchy-Riemann equations using the same argument as in complex analysis. We give an example for a function $f: \Sch^2 \rightarrow \Sch_2$, which is easy to generalize for any $d>2$.
\begin{example}
	Let us consider a function $f: \Sch^2 \rightarrow \Sch_2$. It depends on $4$ real variables $(x_1,y_1,x_2,y_2)$ because $f(z_1, z_2) \equiv f(x_1+e_1 y_1, x_2 + e_2 y_2)$. Moreover in $\Sch_2$ we can expand the value of $f$ componentwise
	\begin{align}
	\begin{split}
	f \left( z_1, z_2 \right) \equiv f \left( x_1 + e_1 y_1, x_2 + e_2 y_2 \right) = \ &f_0 \left( x_1 + e_1 y_1,  x_2 + e_2 y_2 \right) \\
	&+ e_1 f_1 (x_1 + e_1 y_1,  x_2 + e_2 y_2) \\
	&+ e_2 f_2 (x_1 + e_1 y_1,  x_2 + e_2 y_2) \\
	&+ e_1 e_2 f_{12} (x_1 + e_1 y_1,  x_2 + e_2 y_2).
	\end{split}
	\end{align}
	First we calculate the derivative in the ``real'' direction of the plane $\Sch(1)$ as
	\begin{align}
	\begin{split}
	\lim_{\Delta x \rightarrow 0} \frac{f(x_1 + \Delta x + e_1 y_1, z_2)-f(z_1, z_2)}{\Delta x} = 
	\frac{\partial f_0}{\partial x_1} + e_1 \frac{\partial f_1}{\partial x_1} + e_2 \frac{\partial f_2}{\partial x_1} + e_1 e_2 \frac{\partial f_{12}}{\partial x_1}.
	\end{split}
	\end{align}
	Then calculate the derivative in the orthogonal direction $e_1$, note that $1/e_1 = e_1/e^2_1 = -e_1$,
	\begin{align}
	\begin{split}
	\lim_{\Delta y \rightarrow 0} \frac{f(x_1 +  e_1 y_1 + e_1 \Delta y, z_2)-f(z_1, z_2)}{ e_1 \Delta y} = 
	-e_1 \frac{\partial f_0}{\partial y_1} + \frac{\partial f_1}{\partial y_1} - e_1 e_2 \frac{\partial f_2}{\partial y_1} + e_2 \frac{\partial f_{12}}{\partial y_1}.
	\end{split}
	\end{align}
\end{example}
Equating the corresponding components gives us the Cauchy-Riemann equations for the variable $z_1$
\begin{align}
\label{eq:scheffers equations 2d z1}
\begin{split}
&\frac{\partial f_0}{\partial x_1} = \frac{\partial f_1}{\partial y_1}, \\
&\frac{\partial f_1}{\partial x_1} = -\frac{\partial f_0}{\partial y_1},\\
&\frac{\partial f_2}{\partial x_1} = \frac{\partial f_{12}}{\partial y_1},\\
&\frac{\partial f_{12}}{\partial x_1} = -\frac{\partial f_2}{\partial y_1}. \\
\end{split}
\end{align}
Similarly we get $4$ equations for $z_2$
\begin{align}
\label{eq:scheffers equations 2d z2}
\begin{split}
&\frac{\partial f_0}{\partial x_2} = \frac{\partial f_1}{\partial y_2}, \\
&\frac{\partial f_1}{\partial x_2} = -\frac{\partial f_0}{\partial y_2},\\
&\frac{\partial f_2}{\partial x_2} = \frac{\partial f_{12}}{\partial y_2},\\
&\frac{\partial f_{12}}{\partial x_2} = -\frac{\partial f_2}{\partial y_2}. \\
\end{split}
\end{align}
These $8$ equations provide necessary and sufficient conditions for a function $f(z_1, z_2)$ to be holomorphic. It comes immediately after one observes that all three available directions $e_0, e_1, e_2$ are mutually orthogonal in the Euclidean representation of $\Sch_2$ and by linearity of the derivative operator.

Let us now express the conditions on holomorphic function using the derivative with respect to conjugated variable, where the conjugation of $z_j = a + e_j b$ is defined by $\bar{z}_j = a - e_j b$, $j = 1, ...,d$. Derivative operators are defined similar to \eqref{eq:complex differentiantion z and z bar} by

\begin{align}
\label{eq:hypercomplex differentiantion z and z bar}
\begin{split}
\frac{\partial}{\partial z_j} = \frac{1}{2} \left( \frac{\partial}{\partial x_j} - e_j \frac{\partial}{\partial y_j} \right), \\
\frac{\partial}{\partial \bar{z}_j} = \frac{1}{2} \left( \frac{\partial}{\partial x_j} + e_j \frac{\partial}{\partial y_j} \right).
\end{split}
\end{align}

\begin{definition}
	A function $f: \Omega \rightarrow \Sch_d$, defined on some open subset $\Omega \subseteq \Sch^d$, is  {\it holomorphic} on $\Omega$ if
	\begin{equation}
	\label{eq:generalized Cauchy-Riemann equations}
	\frac{\partial f}{\partial \bar{z}_j} = 0, \ \text{  for } j=1,...,d,
	\end{equation}
	at every point $\bm{z}_0 \in \Omega$.
\end{definition}
\noindent Again we give an example for $f: \Sch^2 \rightarrow \Sch_2$.
\begin{example}
	The derivative with respect to $\bar{z}_1$ of $f: \Sch^2 \rightarrow \Sch_2$ is given by
	\begin{align}
	\begin{split}
	\frac{\partial f}{\partial \bar{z}_1} = \frac{1}{2} \left( \frac{\partial f}{\partial x_1} + e_1 \frac{\partial f}{\partial y_1} \right) = &\frac{1}{2}  \left( \frac{\partial f_0}{\partial x_1} + e_1 \frac{\partial f_1}{\partial x_1} +  e_2 \frac{\partial f_2}{\partial x_1} +  e_1 e_2 \frac{\partial f_{12}}{\partial x_1} \right. \\
	& \left. + e_1 \frac{\partial f_0}{\partial y_1} -  \frac{\partial f_1}{\partial y_1} + e_1 e_2 \frac{\partial f_2}{\partial y_1} - e_2 \frac{\partial f_{12}}{\partial y_1} \right).
	\end{split}
	\end{align}
	\noindent Equating the components in front of each $e_i$ to zero we get equations \eqref{eq:scheffers equations 2d z1}.
\end{example}

We are ready to introduce the Cauchy integral formula for the functions $f: \Sch^d \rightarrow \Sch_d$. The concept of Riemann contour integration as well as Cauchy's integral theorem are well defined in the {\it unital real, commutative and associative} algebras \cite{pedersen1997cauchy}. Some care will be taken to define Cauchy integral formula because some elements of $\Sch_d$ are not invertible. To introduce the concept of integral in the space $\Sch_d$ it is natural to assume (cf. \eqref{eq:hypercomplex variable}) that
\begin{equation}
\diff w = \sum_{\beta = 0}^{2^d-1} e_\beta \diff w_\beta.
\end{equation}
Therefore we understand the integral $\int f(w) \, \diff w$ of $\Sch_d$-valued function $f(w)$ of the variable $w \in \Sch_d$ as integral of differential form $\sum_{\beta} f(w) e_\beta \diff w_\beta$ along a curve in $\R^{2^d}$. As it was pointed out in the Remark \ref{remark:zero divisors}, presence of the inverse function in Cauchy formula implies that inverse of the function $(z-\zeta)$ does not pass through zero divisors in $\Sch_d$. It is true that $\Sch_d$ contains some zero divisors. For example $(e_1-e_2)(e_1+e_2) = 0$. However if we restrict our attention to the functions $\Sch(i) \rightarrow \Sch_d$ then it is easy to see that each nonzero $z \in \Sch(i), \ i = 1, ... ,d$ is invertible simply because each $\Sch(i)$ is a field. Cauchy formula for $f: \Sch(i) \rightarrow \Sch_d$ for a disk $\D(i) \subset \Sch(i)$ will have the form
\begin{align}
\label{eq:Cauchy formula in e_i}
\frac{1}{2 \pi e_i} \int\limits_{\partial \D(i)} \frac{f(\zeta)}{\zeta - z} \, \diff \zeta_i = \begin{cases}
f(z) & \text{if } z \in \D(i), \\
0 & \text{if } z \in \Sch(i) \setminus \overline{\D}(i). 
\end{cases} 
\end{align}

Therefore we can also construct the multidimensional Cauchy integral formula for holomorphic functions $f: \Sch^d \rightarrow \Sch_d$ without any problems. General Cauchy formula for a simple open polydisk $\D_{\bm{j}} \subset \Sch^d$ is thus given by
\begin{align}
\label{eq:Cauchy formula general}
\frac{1}{(2 \pi)^{|\bm{j}|} e_{\bm{j}}} \bm{\int}\limits_{\partial \D_{\bm{j}}} \frac{f \left( \zeta_{\bm{j}} \right)}{ \left( \bm{\zeta} - \bm{z} \right)^{\bm{j}}} \, \diff \bm{\zeta^j} = 
\begin{cases}
f(z) & \text{if } z \in \D_{\bm{j}}, \\
0 & \text{if } z \in \Sch^d \setminus \overline{\D}_{\bm{j}}.
\end{cases} 
\end{align}
This motivates the following equivalent definition of a holomorphic function.

\begin{definition}
	Let function $f: \Omega \rightarrow \Sch_d$, defined on some open subset  $\Omega \subseteq \Sch^d,$ be continuous in each variable separately and locally bounded. The function $f$ is said to be {\it holomorphic} in $\Omega$ if for each $\bm{w} \in \Omega$ there is an $r = r(\bm{w})>0$ such that $\overline{\D^d} (\bm{w},r) \subseteq \Omega$ and
	\begin{equation}
	\label{eq:cauchy formula hypercomplex - definition}
	f(z_1, ..., z_d) = \frac{1}{ \left( 2 \pi \right)^d e_1 \cdot \dotsc \cdot e_d} \oint\limits_{|\zeta_1 - w_1| = r} \dotsi \oint\limits_{|\zeta_d - w_d| = r} \frac{f \left( \zeta_1, \dotsc, \zeta_d \right)}{\left( \zeta_1 - z_1 \right)\cdot \dotsc \cdot \left( \zeta_d - z_d \right)} \, \diff \zeta_1 \dotsc \diff \zeta_d
	\end{equation}
	for all $\bm{z} \in \D^{d}(\bm{w},r)$.
\end{definition}

\noindent Analytic hypercomplex function therefore will be defined as following.

\begin{definition}
	A function $f: \Omega \rightarrow \Sch_d$ is called {\it analytic} (holomorphic) in some open domain $\Omega \subseteq \Sch^d$ if for each $\bm{w} \in \Omega$ there is an $r = r(\bm{w}) >0$ such that $\overline{\D^d} (\bm{w}, r) \subseteq \Omega$ and $f$ can be written as an absolutely and uniformly convergent power series for all $\bm{z} \in \D^d(\bm{w},r)$
	\begin{equation}
	f(\bm{z}) = \sum_{i_1, \dotsc, i_d = 0}^{\infty} a_{i_1, \dotsc, i_d} \left( z_1 - w_1 \right)^{i_1} \cdot \dotsc \cdot \left( z_d - w_d \right)^{i_d}
	\end{equation}
	with coefficients
	\begin{equation}
	a_{i_1, ..., i_d} = \frac{1}{i_1! \cdot \dotsc \cdot i_d!} \left( \frac{\partial }{\partial z_1} \right)^{i_1} \dotsi \left( \frac{\partial }{\partial z_d} \right)^{i_d} f(z).
	\end{equation}
\end{definition}

In case of a single complex variable Hilbert transform relates the conjugated parts of a holomorphic function on the boundary of a unit disk. Unit disk then can be mapped to the upper complex half-plane and we obtain the usual definition of the Hilbert transform on the real line. In case of polydisk in a hypercomplex space $\Sch^d$ we proceed similarly. First we define Hilbert transform on the boundary of unit polydisk and then consider all biholomorphic mappings from the polydisk to the upper space $\Sch^d_+$.

\begin{definition}
	Hilbert transform $\mathring{H}_{\bm{j}} \left[ f \right]$ of a function $f: \partial \D_{\bm{j}} \rightarrow \R$ is given by
	\begin{align}
	\label{eq:hilbert transform on polydisk}
	\mathring{H}_{\bm{j}} \left[ f \right] \left(  e^{e_1 \theta_1}, ..., e^{e_d \theta_d} \right) = \frac{1}{ \left( 4 \pi \right)^{|\bm{j}|}} \bm{\int\limits_{-\pi}^{\pi}} f \left(e^{e_i t_i} | i \in \ind{\bm{j}} \right) \prod_{i \in \ind{\bm{j}}}\cot \left( \frac{\theta_i - t_i}{2}\right) \, \diff \bm{t^j},
	\end{align}
	where integration of $f$ is only over the variables $z_i \equiv e^{e_i t_i}$ indicated by the binary vector $\bm{j} \in \{0,1\}^d$.
\end{definition}
To construct the holomorphic hypercomplex function from its real part, which is defined on the boundary of polydisk, we can find the corresponding hypercomplex conjugated components by $f_{\bm{j}} = \mathring{H}_{\bm{j}} [f]$ and sum the up. The relationship with one-variable Hilbert transform is the following. In complex analysis each holomorphic function on the boundary of the unit disk has the form $f_a = f + i H[f]$, i.e. Hilbert transform relates two real-valued functions. In case of several hypercomplex variables the situation is slightly different. The $\Sch_d$-valued function may be written for each $j=1,\dotsc,d$ in the form $f = f_1 + e_j f_2$, however now $f_1$ and $f_2$ are not real-valued. Functions $f_1$ and $f_2$ have values lying in the $ \spanspan \left\{e_\beta | \beta \not\owns j \right\}$. Therefore Hilbert transform in the variable $z_j$ relates hypercomplex conjugates within the plane $\Sch(j)$ by the expression $f_2 = H_j [f_1]$.

Next we simply map the boundary of a polydisk to the boundary of Scheffers upper space. All biholomorphic mappings \cite{gong2007concise} from the unit polydisk to the upper Scheffers space, i.e. $\D^d \rightarrow \Sch_+^d$, have the form
	\begin{equation}
	\label{eq:polydisk to half-plane}
	\left(w_1, ..., w_d \right) \mapsto \left( \frac{\bar{a}_1 w_1 - e^{e_1 \theta_1} a_1}{w_1 - e^{e_1 \theta_1}}, \dotsc, \frac{\bar{a}_d w_d - e^{e_d \theta_d} a_d}{w_d - e^{e_d \theta_d}} \right), \text{ where } a_i \in \D(i), \ \theta_i \in \R.
	\end{equation}
By picking up one such mapping we easily define the Hilbert transform $H_{\bm{j}}$ on the boundary of $\Sch^d_+$ and get \eqref{eq:Hj}, similarly as it is described in Appendix \ref{appendix:Hilbert transform in 1D}. Hypercomplex analytic signal is defined to be any holomorphic function $f: \Sch_+^d \rightarrow \Sch_d$ on the boundary $\partial \Sch_+^d \simeq \R^d$.
\begin{definition}
	The {\it Scheffers hypercomplex analytic signal} $f_S : \R^d \rightarrow \Sch_d$ is defined on the boundary of upper Scheffers space $\partial \Sch_+^d \simeq \R^d$ from its real part $f: \R^d \rightarrow \R$ by
	\begin{equation}
	f_S \left( \bm{x} \right) = \sum_{\bm{j} \in \{0,1\}^d}  e_{\bm{j}} H_{\bm{j}} \left[ f \right] \left( \bm{x} \right).
	\end{equation}
\end{definition}

In one varialbe complex analysis the Riemann mapping theorem states that for any simply connected open subset $U \subseteq \C$ there exists biholomorphic mapping of $U$ to the open unit disk in $\C$. However in case of several complex variables $d \geq 2$ this result does not hold anymore \cite{krantz2001function}. Poincar{\'e} proved that in any dimension $d \geq 2$ in case of $d$ complex variables the ball is not biholomorphic to the polydisk. Even though the proof of Poincar{\'e} theorem in case of mappings $\Sch^d \rightarrow \Sch_d$ is out of scope of this paper this conjecture has important consequences. We rely mainly on the $d$-polydisk $\D^d \subset \Sch^d$ as a domain of the hypercomplex holomorphic function. If there is no biholomorphic mapping from polydisk in $\Sch^d$ to the ball in $\Sch^d$ then on their boundaries holomorphic functions will be quite different as well, i.e. it will be impossible to map holomorphic function defined as a limit on the boundary of polydisk to the holomorphic function defined on the boundary of a ball. There is one simple argument on why probably there is no biholomorphic mapping from boundary of open polydisk to the boundary of a ball. The boundary of polydisk is given by torus, while the boundary of a ball is given by hypersphere. Torus and sphere are not homeomorphic for $d \geq 2$. Therefore it is hard to expect existence of a biholomorphic mapping between the two topologically different domains.

Analytic signal was defined as extension of some holomorphic function inside polydisk to its boundary that is torus. Extension of the mapping \eqref{eq:polydisk to half-plane} to the boundary relates points on torus $T^d$ with points on $\R^d$. There are other mappings that one can use to relate points on a compact shape with the points of $\R^d$. On the other hand, shape that one chooses to represent $\R^d$ could also be used for convenient parametrization of analytic signal. For example, in one variable complex analysis, phase of analytic signal is given by the angle on the boundary of unit disk. Here we advocate the use of torus $T^d$ as a natural domain not only to define analytic signal but also to parametrize analytic signal's phase. 

Suppose instead we defined analytic signal on a unit sphere in $\Sch^d$. In this case if there is no biholomorphic mapping between ball and polydisk in $\Sch^d$ the definition of analytic signal as a limiting case of holomorphic function on the boundary of the ball will differ from the polydisk case. Very probable that we will not be able to employ partial Hilbert transforms $H_{\bm{j}}$ as simple relationships between conjugated components because there are no ``selected'' directions on the sphere. On the other hand, parametrization of analytic signal by a point on a sphere in $\Sch^d$ could provide an alternative definition for the phase of a signal in $\R^d$. Sphere and torus both locally look similar to $\R^d$. However torus is given by the product $T^d = S^1 \times \dotsc \times S^1$ and $\R^d$ is given by the product $\R^d = \R^1 \times \dotsc \times \R^1$, therefore there is a natural way to assign circle to each direction in $\R^d$. In contrast hypersphere is a simple object that cannot be decomposed. Later in Observation \ref{observation:torus parametrization} we use parametrization on torus to describe certain class of analytic signals.

 \section{Commutative hypercomplex Fourier transform}
 \label{section:commutative hypercomplex Fourier transform}

 In this section we will define the Fourier transform that naturally arises for functions $f: \R^d \rightarrow \Sch_d$. We start by assuming that we are working in some appropriately defined Schwartz space $\mathcal{S} \left( \R^d,\Sch_d \right)$ of rapidly decreasing $\Sch_d$-valued functions \cite{strichartz2003guide}. Fourier transform $F: \mathcal{S} \rightarrow \mathcal{S}$ is an automorphism on $\mathcal{S}$. After defining the Fourier transform we show that analytic signal is supported only on positive quadrant of frequency space, the property that is frequently desired in applications. In other words we are able to recover the phase-shifted functions $f_{\bm{j}}$ by restricting spectrum of the Fourier transform of a real-valued function to the positive quadrant in frequency space. After establishing basic facts about Fourier transform we proceed to the Bedrosian's theorem that allows one to easily construct the hypercomplex analytic signal in the Euler form from the real-valued oscillating function. 
 
 \begin{definition}[Schwartz space]
 	The space $\mathcal{S}_d \left( \R^d,\Sch_d \right)$ of rapidly decreasing functions $f: \R^d \rightarrow \Sch_d$ is 
 	\begin{equation}
 	\mathcal{S} \left( \mathbb{R}^d, \Sch_d \right) = \left \{ f \in C^\infty \left(\mathbb{R}^d, \Sch_d \right) :  \|f\|_{\alpha,\beta} < \infty \ \ \forall \alpha, \beta \in\mathbb{N}^d \right \},
 	\end{equation}
 	where $\alpha, \beta$ are multi-indices, $C^\infty \left( \R^d, \Sch_d \right)$ is the set of smooth functions from $\R^d$ to $\Sch_d$, and
 	\begin{equation}
 	\|f\|_{\alpha,\beta} = \sup_{\bm{x} \in \mathbb{R}^d} \left| \bm{x}^\alpha \partial^\beta f(\bm{x}) \right|.
 	\end{equation}
 	To put it simply, Schwartz class contains all smooth functions for which all the derivatives go to zero at infinity faster than any polynomial with inverse powers.
 \end{definition}

 We define hypercomplex Fourier transform in terms of phase-shifted harmonics. Although the definition is a bit lengthy, we still think that this form allows one to better focus on the harmonic ''pieces`` of Fourier transform as well as to see how hypercomplex numbers are working and where the possible non-commutativity of various hypercomplex algebras could cause troubles.
 \begin{definition}
 	The {\it Fourier transform} $\hat{f} = F \left[ f \right]$, with $\hat{f}: \R^d \rightarrow \Sch_d$, of a function $f: \R^d \rightarrow \Sch_d$ is defined by
 	\begin{equation}
 	 	\label{eq:hyper Fourier direct transform}
 	\begin{aligned}
 	\hat{f}(\bm{\omega}) = \ &\alpha^{0 \dots 0} (\bm{\omega}) - \sum_{i=1}^{d} e_i \alpha^{0 \dots 1(i) \dots 0} (\bm{\omega}) \\&+ \sum_{i<j} e_i e_j \alpha^{0 \dots 1(i,j) \dots 0}(\bm{\omega}) - \sum_{i<j<k} e_i e_j e_k \alpha^{0 \dots 1(i,j,k) \dots 0}(\bm{\omega}) + \dotsc,
 	\end{aligned}
 	\end{equation}
 	where $0 \dots 1(i,j,k) \dots 0$ means that there are $1$-s on the $i$-th, $j$-th and $k$-th positions of binary string. The {\it inverse Fourier transform} is given by
 	\begin{equation}
 	\label{eq:hyper Fourier inverse transform}
 	\begin{aligned}
 	f(\bm{x}) = \ &\hat{\alpha}^{0 \dots 0} (\bm{x}) + \sum_{i=1}^{d} e_i \hat{\alpha}^{0 \dots 1(i) \dots 0} (\bm{x}) \\ &+ \sum_{i<j} e_i e_j \hat{\alpha}^{0 \dots 1(i,j) \dots 0} (\bm{x}) + \sum_{i<j<k} e_i e_j e_k \hat{\alpha}^{0 \dots 1(i,j,k) \dots 0} (\bm{x}) + \dotsc.
 	\end{aligned}
 	\end{equation}
 \end{definition}
 This definition is equivalent (by applying Euler's formula) to the canonical form of hypercomplex Fourier transform
 \begin{align}
 \label{eq:usual hyper Fourier}
 \hat{f}(\omega_1,\dots,\omega_d) = \int\limits_{-\infty}^{\infty} \dotsi \int\limits_{-\infty}^{\infty} f(x_1,x_2,\dotsc,x_d) e^{ -e_1 \omega_1 x_1} \cdot \dotsc \cdot e^{ -e_d \omega_d x_d} \, \diff x_1 \dots \diff x_d, \\
 f(x_1,x_2,\dotsc,x_d) = \frac{1}{ \left( 2\pi \right)^d} \int\limits_{-\infty}^{\infty} \dotsi \int\limits_{-\infty}^{\infty} \hat{f}(\omega_1,\dots,\omega_d) e^{e_1 \omega_1 x_1} \cdot \dotsc \cdot e^{e_d \omega_d x_d} \, \diff x_1 \dotsc \diff x_d.
 \end{align}
 The inverse Fourier transform \eqref{eq:hyper Fourier inverse transform} may be rewritten using \eqref{eq:bracket notation multidimensional} as
 \begin{align}
 \label{eq:hypercomplex inverse with brackets}
 f(\bm{x}) = \left\langle \hat{f}(\bm{\omega}), \alpha_{0\dots0} \right\rangle + \sum_{i} e_i \left\langle \hat{f}(\bm{\omega}), \alpha_{0 \dots 1(i) \dots 0} \right\rangle + \sum_{i<j} e_i e_j \left\langle \hat{f}(\bm{\omega}), \alpha_{0 \dots 1(i,j) \dots 0} \right\rangle + \dotsc.
 \end{align}
 Next we observe that phase-shifted functions $f_{\bm{j}}$ are easily recovered from the restriction of Fourier transform to only positive frequencies.
 \begin{theorem}
 	The function $f_h : \R^d \rightarrow \Sch_d$ defined by the real-valued function $f:\R^d \rightarrow \R$ as 
 	\begin{align}
 	\label{eq:fh definition}
 	f_h(\bm{x}) = \left\langle \hat{f}(\bm{\omega}), \alpha_{0\dots0} \right\rangle_+ + \sum_{i} e_i \left\langle \hat{f}(\bm{\omega}), \alpha_{0 \dots 1(i) \dots 0} \right\rangle_+ + \sum_{i<j} e_i e_j \left\langle \hat{f}(\bm{\omega}), \alpha_{0 \dots 1(i,j) \dots 0} \right\rangle_+ + \dots
 	\end{align}
 	has as components the corresponding phase-shifted functions $f_{\bm{j}}$, i.e.
 	\begin{equation}
 	\label{eq:fh with components fj}
 	\begin{aligned}
 	f_h(\bm{x}) = \ &f(\bm{x}) + \sum_{i} e_i f_{0 \dots 1(i) \dots 0} (\bm{x})\\  &+ \sum_{i<j} e_i e_j f_{0 \dots 1(i,j) \dots 0} (\bm{x}) + \sum_{i<j<k} e_i e_j e_k f_{0 \dots 1(i,j,k) \dots 0}(\bm{x}) + \dots,
 	\end{aligned}
 	\end{equation}
 	and therefore $f_h \equiv f_S$.
 	\begin{proof}
 		For a proof we simply put each term from the sum in \eqref{eq:hyper Fourier direct transform} inside \eqref{eq:fh definition} and observe that components in \eqref{eq:fh with components fj} are defined by \eqref{eq:fj and alphas}.
 	\end{proof}
 \end{theorem}
 \begin{observation}
 	The Fourier transform of $f_S$ is given by
 	\begin{equation}
 	\label{eq:fh spectrum}
 	\hat{f}_S(\bm{\omega}) = \prod_{i=1}^{d} \left[ 1 + \sign (\omega_i) \right] \hat{f} (\omega_1,\dots,\omega_d).
 	\end{equation}
 \end{observation}
 
 In the following we proceed to the Bedrosian's theorem telling us that Hilbert transform of the product of a low-pass and a high-pass functions with non-overlapping spectra is given by the product of the low-pass function by the Hilbert transform of the high-pass function. We rely on the original work \cite{bedrosian1962product}. In the following $\ind(\bm{j}) = \{i| j_i=1 \}$, the set of positions of $1$-s in the vector $\bm{j} \in \{0,1\}^d$ and $\overline{\ind}(\bm{j}) = \{i| j_i=0 \}$ is the complementary set.
 \begin{lemma}
 	\label{lemma:Hjh = ei sign(omega)}
 	For a product of exponentials for a given $\bm{j} \in \{0,1\}^d$ 
 	\begin{equation}
 	h(\bm{x}) = \prod_{i \in \ind(\bm{j})} e^{e_i \omega_i x_i},
 	\end{equation}
 	we have the identity
 	\begin{equation}
 	H_{\bm{j}} \left[ h \right] (\bm{x}) = \prod_{i \in \ind (\bm{j})} e_i \sign(\omega_i) \, e^{e_i \omega_i x_i}.
 	\end{equation}
 	\begin{proof}
 		The result follows directly from the 1-dimensional case by succesive application of Hilbert transform.
 	\end{proof}
 \end{lemma}
 \begin{theorem}[Bedrosian]
 	Let $f,g: \R^d \rightarrow \Sch_d$ and let us given some $\bm{j} \in \{0,1\}^d$. Suppose that for each $k \in \ind(\bm{j})$ we have corresponding $a_k>0$ such that
 	\begin{equation}
 	 	\label{eq:narrowband f}
 	\begin{split}
 	\hat{f}_k \left( \omega_k; x_1,\dotsc, x_{k-1}, x_{k+1}, \dotsc x_d \right) = \int\limits_{\R} f \left( x_1, \dotsc, x_k, \dotsc, x_d \right) &e^{-e_k \omega_k x_k} \, \diff x_k = 0 \ \\ & \ \text{ for all } |\omega_k| > a_k
 	\end{split}
 	\end{equation}
 	and
 	\begin{equation}
 	\label{eq:narrowband g}
 	\begin{split}
 	\hat{g}_k \left( \omega_k; x_1,\dots, x_{k-1}, x_{k+1}, \dots x_d \right) = \int\limits_{\R} g \left( x_1, \dots, x_k, \dots, x_d \right) &e^{-e_k \omega_k x_k} \, \diff x_k = 0 \ \\ & \ \text{for all } |\omega_k| \leq a_k.
 	\end{split}
 	\end{equation}
 	Then we have the identity
 	\begin{equation}
 	\label{eq:bedrosian theorem many dim}
 	H_{\bm{j}} \left[ f \cdot g \right] = f \cdot H_{\bm{j}} \left[ g \right] .
 	\end{equation}
 	\begin{proof}
 		First we give the proof for $1$-dimensional case and then extend it to our case. 
 		We can write the product of two functions $a(x)$ and $b(x)$ in Fourier domain as
 		\begin{equation}
 		a(x) b(x) = \frac{1}{(2 \pi)^2} \int\limits_{\R} \int\limits_{\R} \hat{a}(u) \hat{b}(v) e^{i(u+v)x} \, \diff u \diff v
 		\end{equation}
 		and
 		\begin{equation}
 		\label{eq:Hab 1D}
 		H [a \cdot b] (x) = \frac{1}{(2 \pi)^2} \int\limits_{\R} \int\limits_{\R} \hat{a}(u) \hat{b}(v) i \sign(u+v) e^{i(u+v)x} \, \diff u \diff v.
 		\end{equation}
 		If we assume that for some $a>0$ we have $\hat{a}(\omega) = 0$ for $|\omega| > a$ and $\hat{b} (\omega) = 0$ for $|\omega| \leq a$, then the product $\hat{a}(\omega) \hat{b}(\omega)$ will be non-vanishing only on two semi-infinite stripes on the plane $\left\{ (u,v): |u|<a, |v|>a \right\}$. So for this integration region the value of the integral \eqref{eq:Hab 1D} will not change if we replace
 		\begin{equation*}
 		\sign (u+v) \rightarrow \sign (v).
 		\end{equation*}
 		Then we will have
 		\begin{align}
 		\label{eq:Hab=a Int b}
 		\begin{split}
 		 H [a \cdot b] (x) &= \frac{1}{(2 \pi)^2} \int\limits_{\R} \int\limits_{\R} \hat{a}(u) \hat{b}(v) i \sign(v) e^{i(u+v)x} \, \diff u \diff v \\
 		 & = a(x) \frac{1}{2\pi} \int\limits_{\R} \hat{b}(v) i \sign (v) e^{ivx} \, \diff v.
 		 \end{split}
 		\end{align}
 		But we know from Lemma \ref{lemma:Hjh = ei sign(omega)} that
 		\begin{equation}
 		\label{eq:Hb = Int b i sign v exp}
 		H [b](x) = \frac{1}{2 \pi} \int\limits_{\R} \hat{b} (v) H \left[ e^{ivx'} \right] \left( x \right) \, \diff v = \frac{1}{2 \pi} \int\limits_{\R} \hat{b} (v) i \sign(v) e^{i v x} \, \diff v.
 		\end{equation}
 		So finally we have 
 		\begin{equation}
 		\label{eq:bedrosian 1d}
 		H[a \cdot b] = a \cdot H[b].
 		\end{equation}
 		To prove the final result \eqref{eq:bedrosian theorem many dim}, we succesively apply $1$-dimensional steps taken in \eqref{eq:Hab=a Int b}, \eqref{eq:Hb = Int b i sign v exp} and the result of Lemma \ref{lemma:Hjh = ei sign(omega)}
 		\begin{align}
 		\begin{split}
 		H_{\bm{j}} [f \cdot g] (\bm{x}) = \ &\frac{1}{(2 \pi)^{2 |\bm{j}|}}  \bm{\int}\limits_{\R^{|\bm{j}|}}\bm{\int}\limits_{\R^{|\bm{j}|}} \hat{f}_{\bm{j}} \left( \bm{u}_{\ind(\bm{j})}; \bm{x}_{\overline{\ind}(\bm{j})} \right) \hat{g}_{\bm{j}} \left( \bm{v}_{\ind(\bm{j})}; \bm{x}_{\overline{\ind}(\bm{j})} \right) \\
 		& \cdot \prod_{i \in \ind(\bm{j})} e_i \sign(u_i + v_i) e^{e_i (u_i + v_i) x_i} \, \diff \bm{u^j} \diff \bm{v^j},
 		\end{split}
 		\end{align}
 		where we denoted by $\hat{f}_{\bm{j}} \left( \bm{u}_{\ind(\bm{j})}; \bm{x}_{\overline{\ind}(\bm{j})} \right)$ and $\hat{g}_{\bm{j}} \left( \bm{v}_{\ind(\bm{j})}; \bm{x}_{\overline{\ind}(\bm{j})} \right)$ the Fourier transforms of $f$ and $g$ over the directions indicated by $\bm{j}$. Thus $\hat{f}_{\bm{j}} \left( \bm{u}_{\ind(\bm{j})}; \bm{x}_{\overline{\ind}(\bm{j})} \right)$ has as its arguments frequency variables indexed by $\ind(\bm{j})$ and the remaining spatial variables indexed by the complementary set $\overline{\ind}(\bm{j})$. Finally by repeating the steps taken in \eqref{eq:Hab=a Int b}, \eqref{eq:Hb = Int b i sign v exp} we conclude that $H_{\bm{j}} [f \cdot g]  = f \cdot H_{\bm{j}} [g] $.
 	\end{proof}
 \end{theorem}
 
 \begin{observation}
 	\label{observation:torus parametrization}
 Suppose we have two real-valued functions $A, B: \R^d \rightarrow \R$ that satisfy conditions \eqref{eq:narrowband f} and \eqref{eq:narrowband g} respectively for $\bm{j} = \bm{1}$. Additionally suppose that $B(\bm{x}) = \alpha_{\bm{0}}(\bm{x},\bm{\omega_0})$ for some choice of frequencies $\bm{\omega_0}$. Then we can write the analytic signal $f_S$ of the product function $C(\bm{x}) = A(\bm{x}) \cdot B(\bm{x})$ in compact form
 \begin{equation}
 C_S(\bm{x}) = A(\bm{x}) e^{e_1 \omega_1 x_1} \cdot \dotsc \cdot e^{e_d \omega_d x_d}.
 \end{equation}
 \end{observation}
 \noindent If we suppose more generally that $B(\bm{x})$ is of the form
 \begin{equation}
  B(\bm{x})= \prod_{l=1}^{d} \cos \left( \phi_l(\bm{x}) \right)
 \end{equation}
 and satisfies condition \eqref{eq:narrowband g}, we can write its analytic signal $C_S(\bm{x})$ then as
 \begin{equation}
 \label{eq:narrowband analytic signal}
 C_S(\bm{x}) = A(\bm{x}) e^{e_1 \phi_1(\bm{x})} \cdot \dotsc \cdot e^{e_d \phi_d(\bm{x})}.
 \end{equation}
 For a $1$-dimensional complex-valued function we interpret its multiplication by a complex number as rotation and scaling in the complex plane. Here we have similar geometric interpretation of narrowband hypercomplex analytic signal having the form \eqref{eq:narrowband analytic signal}. First of all, each exponential $e^{e_i \phi_i}$ represent phase rotations in the corresponding hypercomplex subspace $\Sch(i)$ by the angle $\phi_i$. Therefore for $C_S(\bm{x})$ in \eqref{eq:narrowband analytic signal} we have a combination of $d$ separate rotations, i.e. the overall rotation will correspond to a point on the product of $d$ circles $S^1 \times \dotsc \times S^1$. The product of circles is a $d$-dimensional torus $T^d$ (see also discussion in the end of Section \ref{subsection: Holomorphic (analytic) functions of hypercomplex variable}). The total phase of some hypercomplex analytic signal \eqref{eq:narrowband analytic signal} therefore corresponds to some continuous map on torus $\gamma: \R^d \rightarrow T^d$. The amplitude function $A(\bm{x})$ is attached to the points on the torus $T^d$. One last observation is that even though we know that the target domain $\Sch_d$ of analytic signal is not a division ring (e.g. $e_1 + e_2$ is not invertible), we can claim that the value of narrow-band analytic signal of the form \eqref{eq:narrowband analytic signal} is always invertible thus leaving aside all the troubles concerning multiplicative inverse.

\section{Analytic extension into upper space $\Sch^d_+$}
\label{section:analytic extension into upper space}

One can extend hypercomplex analytic signal $f_a : \R^d \rightarrow \Sch_d$ to the holomorphic function in the upper hypercomplex space $\Sch^d_+$ by using Poisson kernel (see Appendix \ref{appendix:Hilbert transform in 1D}) or by using holomorphic Fourier transform. The inverse holomorphic Fourier transform is given simply by extending the domain of usual inverse Fourier transform \eqref{eq:usual hyper Fourier} to $\Sch_+^d$ by
\begin{equation}
\label{eq:holomorphic inverse hypercomplex Fourier transform}
f(\zeta_1, \zeta_2, ... , \zeta_d) = F^{-1}\left[ \hat{f} \right] = \frac{1}{(2\pi)^d} \int\limits_{-\infty}^{\infty} \dots \int\limits_{-\infty}^{\infty} \hat{f}(\omega_1,\dots,\omega_d) e^{e_1 \omega_1 \zeta_1} \cdot ... \cdot e^{e_d \omega_d \zeta_d} \, \diff x_1 \dots \diff x_d,
\end{equation}
where each $\zeta_i \in \Sch_+(i)$.

Next we wish to understand under what conditions the above integral is well behaved. In case the integral in \eqref{eq:holomorphic inverse hypercomplex Fourier transform} is a nice integral, we could use for example Leibniz's rule to check whether generalized Cauchy-Riemann conditions \eqref{eq:generalized Cauchy-Riemann equations} are satisfied. We could fall into troubles in some cases, for example, when $\hat{f}$ is supported on the whole $\R^d$. It will make the modulus of exponentials $e^{e_i \omega_i \zeta_i}$ grow exponentially fast with $\omega_i \rightarrow - \infty$, thus making the integral undefined. Fortunately we already know from the previous section that frequency support of our analytic signal $\hat{f}_S$ is $[0, \infty )^d$. 

The necessary and sufficient conditions for extension of a function into some (hyper-)complex domain (not necessarily upper space) are given by various Paley-Wiener theorems \cite{strichartz2003guide}. In this work we are mainly interested in the extension of analytic signal into upper space $\Sch^d_+$. We will focus on this case only, while other Paley-Wiener type theorems may be constructed similarly to those given in Chapter $7.2$ of \cite{strichartz2003guide}.

We will need several technical definitions. First of all let us define the space of square integrable functions $L^2$ and Hardy space $H^2$ of integrable holomorphic functions as following.

\begin{definition}
	Let $f: \R^d \rightarrow \Sch_d$. We say $f \in L^2 \left( \R^d, \Sch_d \right)$ if
	\begin{equation}
	\bm{\int\limits_{-\infty}^{\infty}} \left| f(x_1, ..., x_d) \right|^2 \, \diff x_1 ... \diff x_d < \infty,
	\end{equation}
	where norm $| \cdot |$ in the space $\Sch_d$ is given by \eqref{eq:norm on Scheffers space}. We say $f \in L^2 \left( \R^d_+, \Sch_d \right)$ when $f$ is supported on $\left[ 0, \infty \right)^d$.
\end{definition}

\begin{definition}[Hardy space]
	\label{definition: Hardy space}
	Let $f: \Sch_+^d \rightarrow \Sch_d$ be holomorphic in $\Sch^d_+$. We say that $f \in H^2 \left( \Sch^d_+, \Sch_d \right)$ if
	\begin{equation}
	\sup_{\substack{y_i > 0 \\ i = 1,...,d}} \ \bm{\int\limits_{-\infty}^{\infty}} \left| f(x_1 + e_1 y_1, ..., x_d + e_d y_d) \right|^2 \, \diff x_1 ... \diff x_d < \infty.
	\end{equation}
\end{definition}
In general case, Hardy spaces $H^p$ are defined separately for open disk and upper half-plane  (see Chapters $13.3$ and $13.4$ of \cite{greene2006function} and \cite{ricci2004hardy}). For an arbitrary $1 < p \leq \infty$, M{\"o}bius transformation is not an isomorphism between Hardy spaces on disk and upper half-plane. Situation is different for $p=2$, where M{\"o}bius transformation of the form \eqref{eq:polydisk to half-plane} is an isometric isomorphism between Hardy spaces for upper space $\Sch^d_+$ and polydisk $\D^d$. Morevoer we can assume that $H^2$ is sufficiently large class of functions because for the case of one complex variable Lemma $13.3.2$ in \cite{greene2006function} tells us that if $0 < p_1 < p_2 < \infty$, then $H^{p_2} \left( \D, \C \right) \subsetneq H^{p_1} \left( \D, \C \right)$. After these observations we let us work in space $H^2 \left( \Sch^d_+, \Sch_d \right)$. Finally we proceed to the central theorem that allows extension of hypercomplex analytic signal into upper space $\Sch^d_+$ by means of holomorphic Fourier transform.

\begin{theorem}[Paley-Wiener]
	\label{theorem:paley-wiener theorem}
	Let $\hat{f}: \R^d \rightarrow \Sch_d$ with $\hat{f} \in L^2 \left( \R^d_+ , \Sch_d \right)$. Then $f (\bm{\zeta}) = F^{-1}\left[ \hat{f} (\bm{\omega}) \right] \left(\bm{\zeta} \right)$ is in $H^2 \left( \Sch^d_+, \Sch_d \right)$.
	
	Furthermore the usual inverse Fourier transform $f(x_1,\dotsc ,x_d)$ of $\hat{f}(\omega_1, \dotsc,\omega_d)$ is the limit of $f(\zeta_1,\dotsc, \zeta_d) \equiv f(x_1 + e_1 y_1,\dotsc, x_d + e_d y_d)$ as each $y_i \rightarrow 0^+$ for $i = 1,\dotsc,d$ in the following sense
	\begin{equation}
	\lim_{(y_1,\dotsc,y_d) \rightarrow (0^+,\dotsc,0^+)} \bm{\int\limits_{-\infty}^\infty} \left| f(x_1 + e_1 y_1,\dotsc, x_d + e_d y_d) - f(x_1, \dotsc, x_d) \right|^2 \, \diff x_1 \dotsc \diff x_d = 0.
	\end{equation}
	
	Conversely, suppose $f \in  H^2 \left( \Sch^d_+, \Sch_d \right)$ then there exist $g \in L^2 \left( \R^d_+, \Sch_d \right)$ such that $g$ is an inverse holomorphic Fourier transform of $f$, i.e. $f = F^{-1}[g]$.
	\begin{proof}
		Proof is essentially the same as in case of one complex variable. Discussion may be found in \cite{strichartz2003guide}, see Theorem $7.2.4$, or for example see Theorem $3.2$ in \cite{ricci2004hardy}.
	\end{proof}
\end{theorem}

In one-variable complex analysis the holomorphic Fourier transform is an operator that is isometric isomorphism due to the Plancherel theorem. If we denote by $\C_+$ and $\C_-$ the upper and lower complex half-planes respectively, Paley-Wiener theorem provides us with decomposition of the space of squarely integrable functions $L^2 \left( \R, \C \right) = H^2 \left( \C_+, \C \right) \oplus H^2 \left( \C_-, \C \right)$. Similar result could easily be obtained for functions $\R^d \rightarrow \Sch_d$, however as we could expect now we will have all possible $2^d$ factors of mixed spaces.
\begin{remark}
	We have the following decomposition
	\begin{equation}
	L^2 \left( \R^d, \Sch_d \right) = \bigoplus_{\bm{j} \in \{0,1\}^d} \Sch_{+ \bm{j}},
	\end{equation}
	where spaces $\Sch_{+ \bm{j}}$ were defined in \eqref{eq: mixed Scheffers space}. This is a concise version of Paley-Wiener theorem \ref{theorem:paley-wiener theorem}.
\end{remark}
 
\section{Why non-commutative Fourier transforms does not fit for d>2?}
\label{section:why clifford- transform doesnt fit}
There were propositions to apply Fourier transforms based on non-commutative algebras, like Clifford and Cayley-Dickson construction, to study instantaneous amplitude, phase and frequency. The aim of this section is to show that {\it there is no} non-commutative Fourier transform in the canonical form \eqref{eq:usual hyper Fourier}, which lead us to the phase-shifted functions $f_{\bm{j}}$ after restriction of the support of Fourier transform only to the positive quadrant. It is interesting to note that B{\"u}low and Sommer in the early work \cite{bulow2001hypercomplex} mention slightly the commutative hypercomplex numbers and associated Fourier transform, however later studies were mainly focused on the non-commutative hypercomplex systems.

We make all the derivations for the case of Clifford algebra-valued Fourier transform and then find that result is also valid for any non-commutative algebra. First we make a brief introduction to the Clifford algebras. A Clifford algebra is a unital associative algebra that is generated by a vector space $V$ over some field $K$, where $V$ is equipped with a quadratic form $Q: V \rightarrow K$. If the dimension of $V$ over $K$ is $d$ and $\left\{e_1, \dots, e_d \right\}$ is an orthogonal basis of $\left( V,Q \right)$, then $\mathcal{C}l(V,Q)$ is a free vector space with a basis
\begin{equation}
\left\{ e_{i_1} e_{i_2} \dots e_{i_k} \ \left| \ 1 \leq i_1 < i_2< \dots < i_k \leq d \text{ and } 0 \leq k \leq d \right. \right\}.
\end{equation}
Element $e_0$ is defined as the multiplicative identity element. Due to the fact that $V$ is equipped with quadratic form we have an orthogonal basis
\begin{equation}
	\left( e_i, e_j \right) = 0 \text{ for } i \neq j, \text{ and } \left( e_i, e_i \right) = Q \left( e_i \right),
\end{equation}
where $\left( \cdot, \cdot \right)$ denotes the symmetric bilinear form associated to $Q$. The fundamental Clifford identity implies that Clifford algebra is anticommutative
\begin{equation}
e_i e_j = - e_j e_i \text{ for } i \neq j, \text{ and } \left( e_i, e_i \right) = Q \left( e_i \right),
\end{equation}
which makes multiplication of the elements of Clifford algebra quite simple, we can put elements of a Clifford algebra in standard order simply by a number of swaps of neighbouring elements.
Every nondegenerate form $Q$ can be written in standard diagonal form:
\begin{equation}
Q(v) = v_1^2 + \dots + v_p^2 - v_{p+1}^2 - \dots - v_{p+q}^2,
\end{equation}
with $d = p+q$. The pair of integers $\left( p,q \right)$ is called the signature of the quadratic form. The corresponding Clifford algebra is then denoted as $\mathcal{C}l_{p,q}(\R)$. For our purposes of analytic signal construction (see discussion on elliptic units in Section \ref{section:Commutative hypercomplex algebra}) we will consider just Clifford algebras $\mathcal{C}l_{0,q}(\R)$ over the field of real numbers $\R$, with the quadratic form 
\begin{equation}
Q(e_i) = -1 \text{ for } 1 \leq i \leq d.
\end{equation}

Each element $x \in \mathcal{C}l_{0,d} \left( \R \right)$ of Clifford algebra may be thought as a linear combination of its basis elements
\begin{equation}
x = x_0 1 + \sum_i x_i e_i + \sum_{i<j} x_{ij} e_i e_j + \sum_{i<j<k} x_{ijk} e_i e_j e_k + \dots
\end{equation}
with real coefficients and for which we also have identities $e_i e_j = - e_j e_i \text{ for } i \neq j, \text{ and } e_i^2 = -1$.

First for our proof it will be handy to define the set-valued function $\Pi$ that returns all permuted products of its arguments. For example for 3 arguments it acts as
\begin{align}
\label{eq:def product function}
\Pi \left( a,b,c \right) = \left\{ a \cdot b \cdot c, \ a \cdot c \cdot b, \ b \cdot a \cdot c, \ c \cdot a \cdot b, \ b \cdot c \cdot a, \ c \cdot b \cdot a \right\}.
\end{align}
For $d$ arguments $\Pi$ acts analogously. A particular Clifford-Fourier transform is then defined as
\begin{equation}
\label{eq:clifford-fourier general direct}
	\hat{f}^{\mathcal{C}l} (\omega_1,\dots,\omega_d) = \int\limits_{-\infty}^{\infty} \dots \int\limits_{-\infty}^{\infty} \rho\left( f(x_1,x_2,\dots,x_d), e^{ -e_1 \omega_1 x_1}, \dots, e^{ -e_d \omega_d x_d} \right)  \, \diff x_1 \dots \diff x_d,
\end{equation}
where $$\rho\left( f(x_1,x_2,\dots,x_d), e^{ -e_1 \omega_1 x_1}, \dots, e^{ -e_d \omega_d x_d} \right) \in \Pi \left( f(x_1,x_2,\dots,x_d), e^{ -e_1 \omega_1 x_1}, \dots, e^{ -e_d \omega_d x_d} \right).$$

A Clifford-Fourier transform \eqref{eq:clifford-fourier general direct} has inverse transform given by
\begin{equation}
\label{eq:clifford-fourier general inverse}
f \left( x_1,\dots,x_d \right) = \frac{1}{ \left( 2 \pi \right)^d} \int\limits_{-\infty}^{\infty} \dots \int\limits_{-\infty}^{\infty} \rho'\left( \hat{f}^{\mathcal{C}l}(\omega_1,\omega_2,...,\omega_d), e^{ e_1 \omega_1 x_1}, \dots, e^{ e_d \omega_d x_d} \right)  \, \diff x_1 \dots \diff x_d,
\end{equation}
for some $$\rho'\left( \hat{f}^{\mathcal{C}l}(\omega_1,\omega_2,...,\omega_d), e^{ e_1 \omega_1 x_1}, \dots, e^{ e_d \omega_d x_d} \right) \in \Pi \left( \hat{f}^{\mathcal{C}l}(\omega_1,\omega_2,...,\omega_d), e^{ e_1 \omega_1 x_1}, \dots, e^{ e_d \omega_d x_d} \right).$$
We can expand the exponentials in \eqref{eq:clifford-fourier general direct} and \eqref{eq:clifford-fourier general inverse} to write down the general Clifford-Fourier transform in terms of functions $\alpha^{\bm{k}}$ and $\hat{\alpha}^{\bm{k}}$
\begin{align}
&\hat{f}^{\mathcal{C}l} = \rho_0 - \sum_{i=1}^d \rho_1(i) + \sum_{i<j} \rho_2(i,j) - \sum_{i<j<k} \rho_3(i,j,k) + \dots, \\
&f = \rho'_0 + \sum_{i=1}^d \rho'_1(i) + \sum_{i<j} \rho'_2(i,j) + \sum_{i<j<k} \rho'_3(i,j,k) + \dots,
\end{align}
for some $\rho_k$ and $\rho'_k$, where
\begin{align}
\begin{split}
&\rho_0 = \alpha^{0\dots0}, \\
&\rho_1(i) \in \Pi\left(e_i, \alpha^{0\dots 1(i) \dots 0}\right), \\
&\rho_2(i,j) \in \Pi\left(e_i, e_j, \alpha^{0\dots 1(i,j) \dots 0}\right), \\
&\rho_3(i,j,k) \in \Pi\left(e_i, e_j, e_k, \alpha^{0\dots 1(i,j,k) \dots 0}\right), \\
&\dots
\end{split}
\end{align}
and
\begin{align}
\begin{split}
&\rho'_0 = \hat{\alpha}^{0\dots0}, \\
&\rho'_1(i) \in \Pi\left(e_i, \hat{\alpha}^{0\dots 1(i) \dots 0}\right), \\
&\rho'_2(i,j) \in \Pi\left(e_i, e_j, \hat{\alpha}^{0\dots 1(i,j) \dots 0}\right), \\
&\rho'_3(i,j,k) \in \Pi\left(e_i, e_j, e_k, \hat{\alpha}^{0\dots 1(i,j,k) \dots 0}\right), \\
&\dots.
\end{split}
\end{align}

We now check whether Clifford-Fourier ``analytic signal'' given by the restriction of Clifford based Fourier transform only to positive frequencies works for defining the instantaneous amplitude and phase. As we did before we use bracket notation \eqref{eq:f0 with brackets}, \eqref{eq:f1 with brackets} and check whether Fourier transform written in the form similar to  \eqref{eq:fh definition} will work. Positive frequency restricted Clifford ``analytic signal'' of the real-valued function $f: \R^d \rightarrow \R$ is given by
\begin{align}
\label{eq:fC definition}
f^{\mathcal{C}l} = \rho'_{+0} + \sum_{i=1}^d \rho'_{+1}(i) + \sum_{i<j} \rho'_{+2}(i,j) + \sum_{i<j<k} \rho'_{+3}(i,j,k) + \dots,
\end{align}
	
with some $\rho'_{+i}$:
\begin{align}
\begin{split}
&\rho'_{+0} = \hat{\alpha}_{\mathcal{C}l+}^{0\dots0}, \\
&\rho'_{+1}(i) \in \Pi\left(e_i, \hat{\alpha}_{\mathcal{C}l+}^{0\dots 1(i) \dots 0}\right), \\
&\rho'_{+2}(i,j) \in \Pi\left(e_i, e_j, \hat{\alpha}_{\mathcal{C}l+}^{0\dots 1(i,j) \dots 0}\right), \\
&\rho'_{+3}(i,j,k) \in \Pi\left(e_i, e_j, e_k, \hat{\alpha}_{\mathcal{C}l+}^{0\dots 1(i,j,k) \dots 0}\right), \\
&\dots,
\end{split}
\end{align}
where $\hat{\alpha}_{\mathcal{C}l+}^{\bm{j}} (\bm{x}) = \left\langle f^{\mathcal{C}l} \left( \bm{\omega} \right), \alpha_{\bm{j}} \left( \bm{x}, \bm{\omega} \right) \right\rangle_+$.
	
\begin{theorem}[Non-existence result]
	\label{theorem:non-existence result}
	The Clifford algebra valued ``analytic signal'' $f^{\mathcal{C}l} : \R^d \rightarrow \mathcal{C}l_{0,d}(\R) $ of the real-valued function $f: \R^d \rightarrow \R$, where $f^{\mathcal{C}l}$ is given by \eqref{eq:fC definition}, in general	\textbf{\textit{does not  have}} as components the corresponding phase-shifted functions $f_{\bm{j}}$, defined as \eqref{eq:fj definition}, i.e. if we have
	\begin{equation}
	\label{eq:fC with components fj}
	f^{\mathcal{C}l}(\bm{x}) = f(\bm{x}) + \sum_{i} e_i \tilde{f}_{0 \dots 1(i) \dots 0} (\bm{x})  + \sum_{i<j} e_i e_j \tilde{f}_{0 \dots 1(i,j) \dots 0} (\bm{x}) + \sum_{i<j<k} e_i e_j e_k \tilde{f}_{0 \dots 1(i,j,k) \dots 0}(\bm{x}) + \dots
	\end{equation}
	then there exist $\bm{j} \in \{0,1\}^d$ such that $|\tilde{f}_{\bm{j}}| \ne |f_{\bm{j}}|$ for any $d>2$.
	Therefore there is no Clifford algebra based Fourier transform of the form \eqref{eq:clifford-fourier general direct} and \eqref{eq:clifford-fourier general inverse} for $d>2$ that will lead us to the phase-shifts $f_{\bm{j}}$ as components of $e_{\bm{j}}$.
	\begin{proof}
		To prove this theorem it will suffice to demonstrate that we are not able to choose the proper ordering of multiplicative terms in \eqref{eq:clifford-fourier general direct} and \eqref{eq:clifford-fourier general inverse} such that $\tilde{f}_{\bm{j}} = f_{\bm{j}}$ for any $\bm{j} \in \{0,1\}^d$.
	
		Next we simply show that already for the elements of degree $1$, i.e. in front of corresponding $e_i$, we already must have $|\tilde{f}_{0\dots 1(i) \dots 0}| \ne |f_{0\dots 1(i) \dots 0}|$ for some $i$. 
		First let us write the Clifford-Fourier transform for elements of degree up to $1$
		\begin{equation}
		\label{eq:fC sum of first 3 e_i}
		\hat{f}^{\mathcal{C}l}(\bm{\omega})= \alpha^{0 \dots 0} (\bm{\omega}) - e_1 \alpha^{10 \dots 0}(\bm{\omega}) - e_2 \alpha^{010 \dots 0}(\bm{\omega}) - e_3 \alpha^{0010 \dots 0}(\bm{\omega}) - \dotsc - e_d \alpha^{0 \dots 01}(\bm{\omega}) + \dots.
		\end{equation}
		For the elements of degree up to $1$ in \eqref{eq:fC sum of first 3 e_i} we may not care about the order of multiplication because $f(\bm{x})$ is real valued. Then we show that there is no ordering of multiplicative terms in the inverse formula \eqref{eq:clifford-fourier general inverse} to obtain the right signs of phase-shifted components in accordance with \eqref{eq:fj and alphas}. For simplicity, but without loss of generality, we consider just the combinations of components corresponing to the first three elements $e_1, e_2$ and $e_3$.
		In this case potentially we have the four following candidates for the inverse Fourier transform.
		\begin{enumerate}
			\item Zero elements are flipped: 
			\begin{align}
			\label{eq:c0}
			\begin{split}
			c_0(\bm{x}) = \ &\hat{\alpha}^{0\dots0}(\bm{x}) + e_1 \left\langle \hat{f}^{\mathcal{C}l}(\bm{\omega}), \alpha_{10\dots0}(\bm{x}, \bm{\omega}) \right\rangle \\&+ e_2 \left\langle \hat{f}^{\mathcal{C}l}(\bm{\omega}), \alpha_{010\dots0}(\bm{x}, \bm{\omega}) \right\rangle + e_3 \left\langle \hat{f}^{\mathcal{C}l}(\bm{\omega}), \alpha_{0010\dots0}(\bm{x}, \bm{\omega}) \right\rangle + \dots.
			\end{split}
			\end{align}
			\item One element is flipped: 
			\begin{align}
			\label{eq:c1}
			\begin{split}
			c_1(\bm{x}) = \ &\hat{\alpha}^{0\dots0}(\bm{x}) + \left\langle \hat{f}^{\mathcal{C}l}(\bm{\omega}), \alpha_{10\dots0}(\bm{x}, \bm{\omega}) \right\rangle e_1  \\&+ e_2 \left\langle \hat{f}^{\mathcal{C}l}(\bm{\omega}), \alpha_{010\dots0}(\bm{x}, \bm{\omega}) \right\rangle + e_3 \left\langle \hat{f}^{\mathcal{C}l}(\bm{\omega}), \alpha_{0010\dots0}(\bm{x}, \bm{\omega}) \right\rangle + \dots.
			\end{split}
			\end{align}
			\item Two elements are flipped:
			\begin{align}
			\label{eq:c2}
			\begin{split}
			c_2(\bm{x}) = \ &\hat{\alpha}^{0\dots0}(\bm{x}) + \left\langle \hat{f}^{\mathcal{C}l}(\bm{\omega}), \alpha_{10\dots0}(\bm{x}, \bm{\omega}) \right\rangle e_1 \\&+ \left\langle \hat{f}^{\mathcal{C}l}(\bm{\omega}), \alpha_{010\dots0}(\bm{x}, \bm{\omega}) \right\rangle e_2 + e_3 \left\langle \hat{f}^{\mathcal{C}l}(\bm{\omega}), \alpha_{0010\dots0}(\bm{x}, \bm{\omega}) \right\rangle + \dots.
			\end{split}
			\end{align}
			\item Three elements are flipped:
			\begin{align}
			\label{eq:c3}
			\begin{split}
			c_3(\bm{x}) = \ &\hat{\alpha}^{0\dots0}(\bm{x}) + \left\langle \hat{f}^{\mathcal{C}l}(\bm{\omega}), \alpha_{10\dots0}(\bm{x}, \bm{\omega}) \right\rangle e_1 \\ &+ \left\langle \hat{f}^{\mathcal{C}l}(\bm{\omega}, \alpha_{010\dots0}(\bm{x}, \bm{\omega}) \right\rangle e_2 + \left\langle \hat{f}^{\mathcal{C}l}(\bm{\omega}), \alpha_{0010\dots0}(\bm{x}, \bm{\omega}) \right\rangle e_3 + \dots.
			\end{split}
			\end{align}
		\end{enumerate}
		
		Therefore for $c_i$ serving as inverse we will obtain the components of degree $2$ of $f^{\mathcal{C}l}$ by simple subtitution of each term of $\hat{f}^{\mathcal{C}l}(\bm{\omega})$ from \eqref{eq:fC sum of first 3 e_i} into \eqref{eq:c0}, \eqref{eq:c1}, \eqref{eq:c2} and \eqref{eq:c3}. That will give us the following hypercomplex components after we restrict the integration only to positive frequencies. 
		\begin{enumerate}
			\item For $c_0$ we have
			\begin{align*}
			\begin{split}
			\tilde{f}^0_{\bm{j}:|\bm{j}|=2} = &-e_2 e_1 \left\langle \alpha^{100 \dots 0} , \alpha_{010 \dots 0} \right\rangle_+ - e_1 e_2 \left\langle \alpha^{010 \dots 0} , \alpha_{100 \dots 0} \right\rangle_+ \\
			&-e_3 e_1 \left\langle \alpha^{100 \dots 0} , \alpha_{0010 \dots 0} \right\rangle_+ - e_1 e_3 \left\langle \alpha^{0010 \dots 0} , \alpha_{100 \dots 0} \right\rangle_+ \\
			&-e_3 e_2 \left\langle \alpha^{010 \dots 0} , \alpha_{0010 \dots 0} \right\rangle_+ - e_2 e_3 \left\langle \alpha^{0010 \dots 0} , \alpha_{010 \dots 0} \right\rangle_+ + \dots.
			\end{split}
			\end{align*}
			
			\item For $c_1$ we have
			\begin{align*}
			\begin{split}
			\tilde{f}^1_{\bm{j}:|\bm{j}|=2} = &-e_2 e_1 \left\langle \alpha^{100 \dots 0} , \alpha_{010 \dots 0} \right\rangle_+ - e_2 e_1 \left\langle \alpha^{010 \dots 0} , \alpha_{100 \dots 0} \right\rangle_+ \\
			&-e_3 e_1 \left\langle \alpha^{100 \dots 0} , \alpha_{0010 \dots 0} \right\rangle_+ - e_3 e_1 \left\langle \alpha^{0010 \dots 0} , \alpha_{100 \dots 0} \right\rangle_+ \\
			&-e_3 e_2 \left\langle \alpha^{010 \dots 0} , \alpha_{0010 \dots 0} \right\rangle_+ - e_2 e_3 \left\langle \alpha^{0010 \dots 0} , \alpha_{010 \dots 0} \right\rangle_+ + \dots.
			\end{split}
			\end{align*}
			
			\item For $c_2$ we have
			\begin{align*}
			\begin{split}
			\tilde{f}^2_{\bm{j}:|\bm{j}|=2} = &-e_1 e_2 \left\langle \alpha^{100 \dots 0} , \alpha_{010 \dots 0} \right\rangle_+ - e_2 e_1 \left\langle \alpha^{010 \dots 0} , \alpha_{100 \dots 0} \right\rangle_+ \\
			&-e_3 e_1 \left\langle \alpha^{100 \dots 0} , \alpha_{0010 \dots 0} \right\rangle_+ - e_3 e_1 \left\langle \alpha^{0010 \dots 0} , \alpha_{100 \dots 0} \right\rangle_+ \\
			&-e_3 e_2 \left\langle \alpha^{010 \dots 0} , \alpha_{0010 \dots 0} \right\rangle_+ - e_3 e_2 \left\langle \alpha^{0010 \dots 0} , \alpha_{010 \dots 0} \right\rangle_+ + \dots.
			\end{split}
			\end{align*}
			
			\item For $c_3$ we have
						\begin{align*}
			\begin{split}
			\tilde{f}^3_{\bm{j}:|\bm{j}|=2} = &-e_1 e_2 \left\langle \alpha^{100 \dots 0} , \alpha_{010 \dots 0} \right\rangle_+ - e_2 e_1 \left\langle \alpha^{010 \dots 0} , \alpha_{100 \dots 0} \right\rangle_+ \\
			&-e_1 e_3 \left\langle \alpha^{100 \dots 0} , \alpha_{0010 \dots 0} \right\rangle_+ - e_3 e_1 \left\langle \alpha^{0010 \dots 0} , \alpha_{100 \dots 0} \right\rangle_+ \\
			&-e_2 e_3 \left\langle \alpha^{010 \dots 0} , \alpha_{0010 \dots 0} \right\rangle_+ - e_3 e_2 \left\langle \alpha^{0010 \dots 0} , \alpha_{010 \dots 0} \right\rangle_+ + \dots.
			\end{split}
			\end{align*}
		\end{enumerate}
	From the rule in \eqref{eq:fj and alphas} we know that with each flipping of $1$ in the upper subscript to $0$ at the same position in the lower subscript the sign of bracket changes. If we look at $\tilde{f}^0_{\bm{j}:|\bm{j}|=2}$, we see that terms that are components of $e_1 e_2$, i.e. $\left\langle \alpha^{100 \dots 0} , \alpha_{010 \dots 0} \right\rangle_+$ and $\left\langle \alpha^{010 \dots 0} , \alpha_{100 \dots 0} \right\rangle_+$ sum up with opposite signs. The same is true for the components of $e_1 e_3$ and $e_2 e_3$ -- they come all with opposite signs. Therefore case of $c_0$-type inverse formula is not in accordance with the rule \eqref{eq:fj and alphas}. For the case $c_1$ we see that the components of $e_1 e_2$ have the same sign as well as components of $e_1 e_3$, however components of $e_2 e_3$ have opposite signs. Thus $c_1$ is not in accordance with \eqref{eq:fj and alphas}. After checking the rule for $c_2$ and $c_3$ we see that we will always have different signs for some component. Therefore we are not able to order properly the terms in Clifford algebra based Fourier transform and its inverse Fourier transform to correctly restore all the phase-shifted functions.
	\end{proof}
\end{theorem}

\begin{corollary}
	From the proof of Theorem \ref{theorem:non-existence result} it follows that {\it essentially  only} commutative hypercomplex algebra ($e_i e_j = e_j e_i$) is a good candidate to provide us with the phase-shifts \eqref{eq:fj and alphas} by using positive frequency restriction for Fourier transforms of the type \eqref{eq:clifford-fourier general direct} and \eqref{eq:clifford-fourier general inverse}. The word ``essentially'' above means that algebras with two elliptic hypercomplex units that anticommute are still allowed when $d=2$.
\end{corollary}

\begin{remark}[Octonions, Sedenions and Cayley-Dickson construction]
	There were propositions to use Cayley-Dickson algebras to define the Fourier transform \eqref{eq:clifford-fourier general direct},  \eqref{eq:clifford-fourier general inverse} and corresponding hypercomplex ``analytic signal'' \cite{hahn2016complex}, however as far as elements $\{e_i\}$ of these algebras are non-commutative we will not reconstruct the phase-shifts correctly.
\end{remark}

 \begin{example}
 	\label{example:quaternionic example}
 	We may wonder what happens in case $d=2$. Is there any Clifford algebra based Fourier transform that provides us with the correct phase-shifted components? It seems that the answer should be - yes. And indeed the symmetric quaternionic based Fourier transform does the job. 
 	Let $\mathbb{Q}$ be the ring of quaternions with the set of basis elements $\left\{ 1,i,j,k \right\}$. The ring of quaternions coincides with the Clifford algebra $\mathcal{C}l_{0,2} \left( \R \right)$ that in turn has the set of basis elements $\left\{1, e_1, e_2, e_1 e_2 \right\}$ where $i = e_1, \ j = e_2, \ k = e_1 e_2$. The symmetric quaternionic Fourier transform is defined as
 	\begin{align}
 	&\hat{f}^q \left( \omega_1, \omega_2 \right) = \int\limits_{\R}\int\limits_{\R} e^{-e_1 \omega_1 x_1} f \left( x_1,x_2 \right) e^{-e_2 \omega_2 x_2} \, \diff x_1 \diff x_2, \\
 	&f \left( x_1, x_2 \right) = \frac{1}{(2\pi)^2} \int\limits_{\R}\int\limits_{\R} e^{e_1 \omega_1 x_1} \hat{f}^q \left( \omega_1,\omega_2 \right) e^{e_2 \omega_2 x_2} \, \diff x_1 \diff x_2,
 	\end{align}
 	which we may rewrite using the notation from \eqref{eq:notation alpha^j many dim} as
 	\begin{align}
 	&\hat{f}^q(\omega_1, \omega_2) = \alpha^{00} - e_1 \alpha^{10} - e_2 \alpha^{01} + e_1 \alpha^{11} e_2, \\
 	&f(x_1, x_2) = \left\langle \hat{f}^q, \alpha_{00} \right\rangle + e_1 \left\langle \hat{f}^q, \alpha_{10} \right\rangle + \left\langle \hat{f}^q, \alpha_{01} \right\rangle e_2 + e_1 \left\langle \hat{f}^q, \alpha_{11} \right\rangle e_2.
 	\end{align}
 	The quaternionic ``analytic signal'' is obtained by restriction of the quaternion Fourier transform to the positive frequencies as we did before
 	\begin{align}
 	\begin{split}
 	f^q_{a} = &\left\langle \alpha^{00}, \alpha_{00} \right\rangle_+ + \left\langle \alpha^{10}, \alpha_{10} \right\rangle_+ + \left\langle \alpha^{01}, \alpha_{01} \right\rangle_+ + \left\langle \alpha^{11}, \alpha_{11} \right\rangle_+ \\
 	&- e_1 \left\langle \alpha^{10}, \alpha_{00} \right\rangle_+ + e_1 \left\langle \alpha^{00}, \alpha_{10} \right\rangle_+ - e_1 \left\langle \alpha^{11}, \alpha_{01} \right\rangle_+ + e_1 \left\langle \alpha^{01}, \alpha_{11} \right\rangle_+ \\
 	& - \left\langle \alpha^{01}, \alpha_{00} \right\rangle_+ e_2 + \left\langle \alpha^{00}, \alpha_{01} \right\rangle_+ e_2 - \left\langle \alpha^{11}, \alpha_{10} \right\rangle_+ e_2 + \left\langle \alpha^{10}, \alpha_{11} \right\rangle_+ e_2 \\
 	& + e_1 \left\langle \alpha^{11}, \alpha_{00} \right\rangle_+ e_2 + e_1 \left\langle \alpha^{00}, \alpha_{11} \right\rangle_+ e_2 - e_1 \left\langle \alpha^{10}, \alpha_{01} \right\rangle_+ e_2 - e_1 \left\langle \alpha^{01}, \alpha_{10} \right\rangle_+ e_2.
 	\end{split}
 	\end{align}
 	As we see, the components of quaternionic analytic signal indeed provide us with the correct phase-shifts given by \eqref{eq:fj and alphas}. 
 \end{example}

\section{Examples and limitations}
\label{section: application examples}

\subsection{Simple example in 3-D}

\begin{figure}[t]
	\begin{minipage}[b]{.48\linewidth}
		\centering
		\centerline{\includegraphics[width=4.0cm]{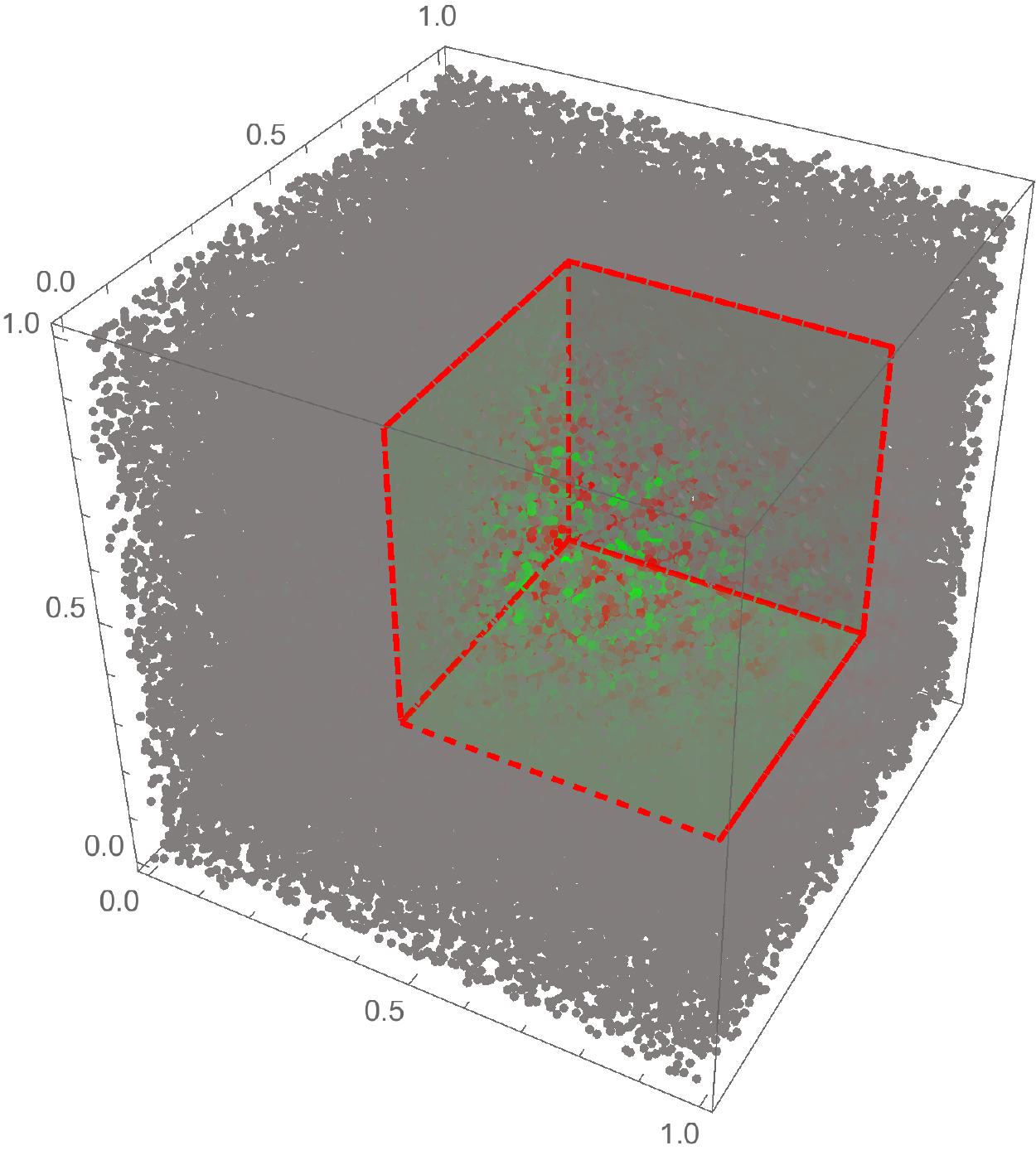}}
		\centerline{\footnotesize (a) Oscillations inside the point-cloud cube}\medskip
	\end{minipage}
	\hfill
	\begin{minipage}[b]{0.48\linewidth}
		\centering
		\centerline{\includegraphics[width=4.0cm]{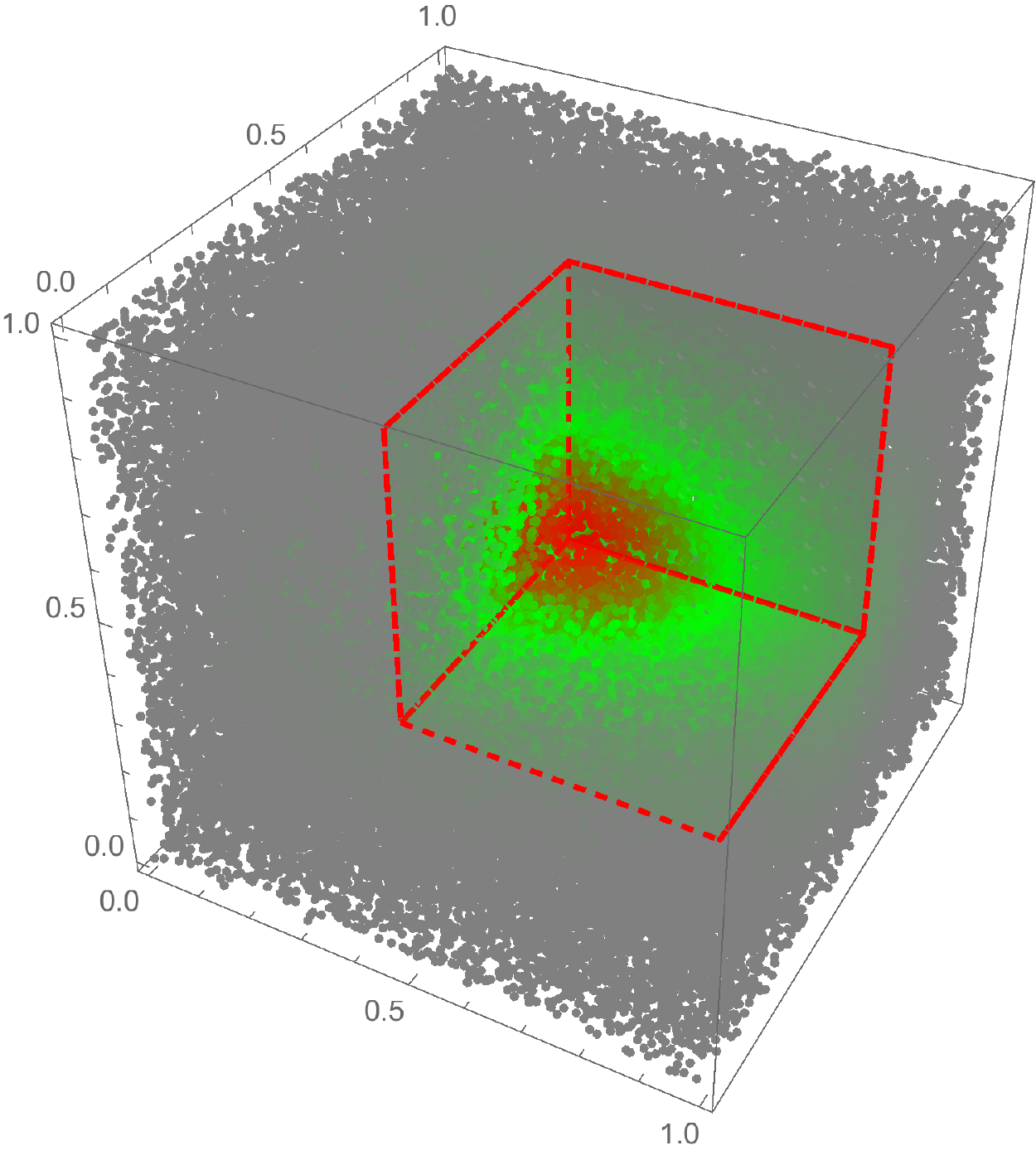}}
		\centerline{\footnotesize (b) Instantaneous amplitude}\medskip
	\end{minipage}
	\vspace{-0.4cm}
	\caption{Point cloud of a cube with a deleted octant is shown. (a) Illustration of oscillating process $f(x,y,z)$ that takes place inside the cube; (b) Instantaneous amplitude of the oscillating process.}
	\label{fig:3d cube}
\end{figure}
As a direct illustration of how the construction of envelope works, we provide an example in 3-D. The oscillating function that we analyse is given by the cosine wave modulated by a Gaussian window:
\begin{align}
\label{eq:signal cube}
\begin{split}
f(x,y,z) & = e^{-10x^2 -20y^2 -20z^2} \cos \left( 50 x \right) \cos \left( 40 y \right) \cos \left( 60 z \right) \\
& = f_x(x) f_y(y) f_z(z).
\end{split}
\end{align}
This particular toy example allows us to separate the variables when calculating the phase-shifted components $f_{\bm{j}}$ from \eqref{eq:fj and alphas}. In three dimensions we will have in total $\left| \{ 0,1 \}^3 \right| = 8$ shifted functions.

Calculation of some $f_{\bm{j}}$ using \eqref{eq:fj definition} reduces to the calculation of direct and inverse cosine and sine transforms. For example for the $f_{100}$ shifted component, we will have
\begin{align}
\label{eq:fj signal cube}
\begin{split}
f_{100}(x,y,z)& = \frac{f_y(y) f_z(z)}{\pi} \int_{0}^{\infty} \int_{-\infty}^{\infty} f_x(x) \sin \left( \omega_1 (x-x') \right) \, \diff x' \diff \omega_1.
\end{split}
\end{align}
Taking into account that $f$ is even in each variable, after expanding the sine of difference, we see that we have to calculate only forward cosine and inverse sine transforms. This double integral can be calculated in semi-analytic form using $\operatorname{erf}(x) = \frac{2}{\sqrt{\pi}} \int_0^x e^{-t^2} \diff t$ by
\begin{align}
\begin{split}
&\frac{2}{\pi} \int\limits_{0}^\infty \sin \left( \omega x \right) \int\limits_0^\infty e^{-\alpha \tilde{x}^2} \cos \left( \omega_x \tilde{x} \right) \cos \left( \omega \tilde{x} \right) \, \diff \tilde{x} \diff \omega \\
& = \frac{1}{2} i e^{-x(\omega_x i + \alpha x)} \left[ \operatorname{erf} \left( \frac{\omega_x/2 - i \alpha x}{\sqrt{\alpha}} \right) - e^{2 \omega_x i x} \operatorname{erf} \left( \frac{\omega_x/2 + i \alpha x}{\sqrt{\alpha}} \right) \right].
\end{split}
\end{align}

To illustrate the resulting envelope of function \eqref{eq:signal cube}, in Fig.~\ref{fig:3d cube} 
we vizualize a point cloud in a cube $[0,1]^3$ with removed octant where each point is colour coded according to the value of $f$ (the origin with respect to function \eqref{eq:signal cube} is shifted by $0.5$ in each direction). This method allows visualization of a function ``inside'' the cube by observing it over the inner faces of the deleted octant. In Fig.~\ref{fig:3d cube} (a) the original $f(x,y,z)$ is plotted, while in Fig.~\ref{fig:3d cube} (b) we plot the envelope function $a(x,y,z)$ obtained from \eqref{eq:envelope in many dim} that was computed on the $8$ phase-shifted functions similarly to (\ref{eq:fj signal cube}).

\subsection{Amplitude on a smooth graph}
\label{sec:typestyle}

Objective of this section is to illustrate amplitude construction for an oscillating function defined over the vertices of a smooth graph or point cloud. The topology of a graph can be captured by a variety of operators like adjacency operator or Laplacian \cite{chung1997spectral}. In this example we start by considering a smooth point cloud patch of curved surface, then we construct the corresponding $k$ nearest neighbors graph on top of the points. Finally we embed the graph in  $\R^2$ and calculate amplitude in the transformed space where the points are flattened. As an embedding of a graph in $\R^2$ we rely on the diffusion maps embedding \cite{coifman2006diffusion, coifman2005geometric}.

\begin{figure}[t!]
	\begin{minipage}[b]{.48\linewidth}
		\centering
		\centerline{\includegraphics[width=4.0cm]{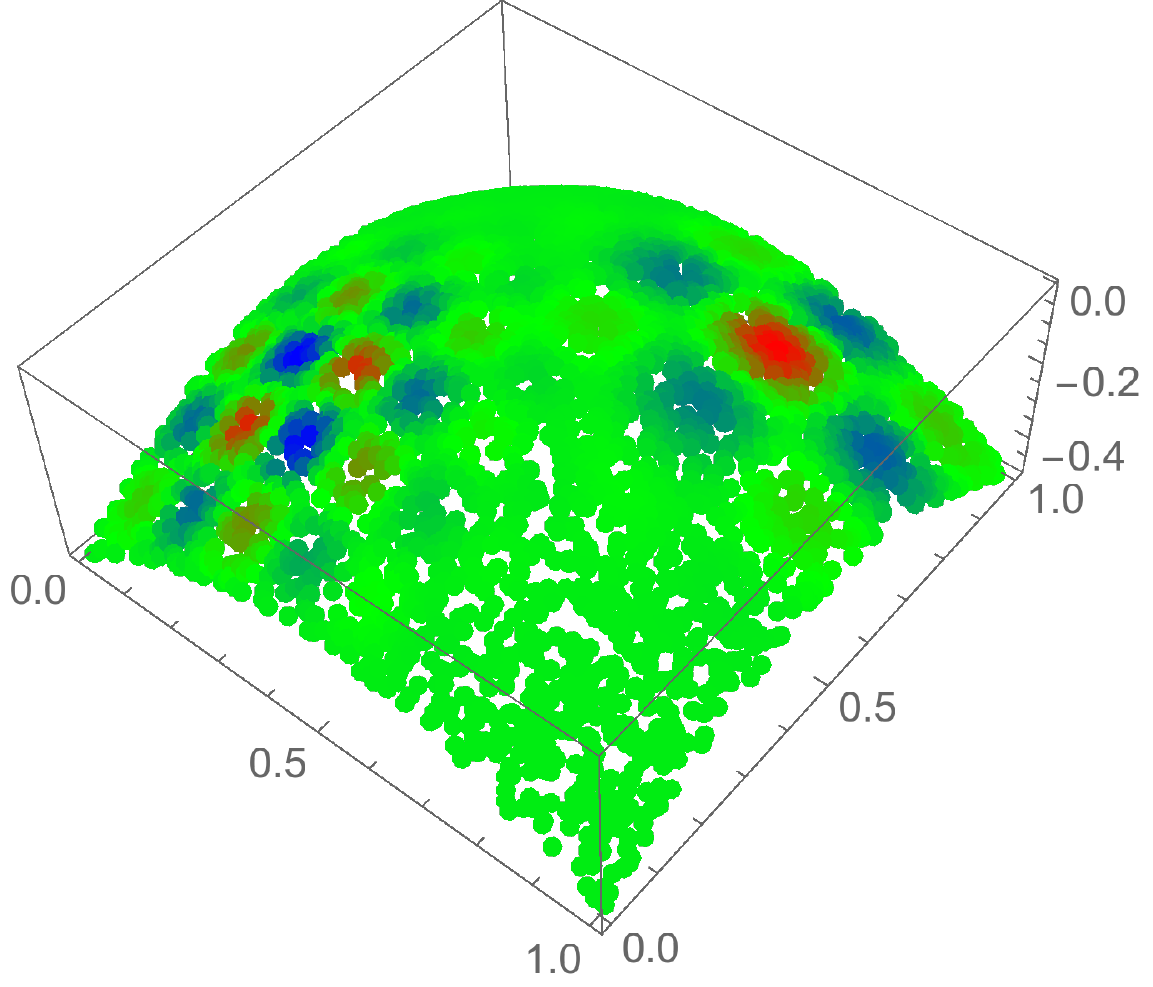}}
		\centerline{\footnotesize (a) Oscillations over a patch of sphere}\medskip
	\end{minipage}
	\hfill
	\begin{minipage}[b]{0.48\linewidth}
		\centering
		\centerline{\includegraphics[width=4.0cm]{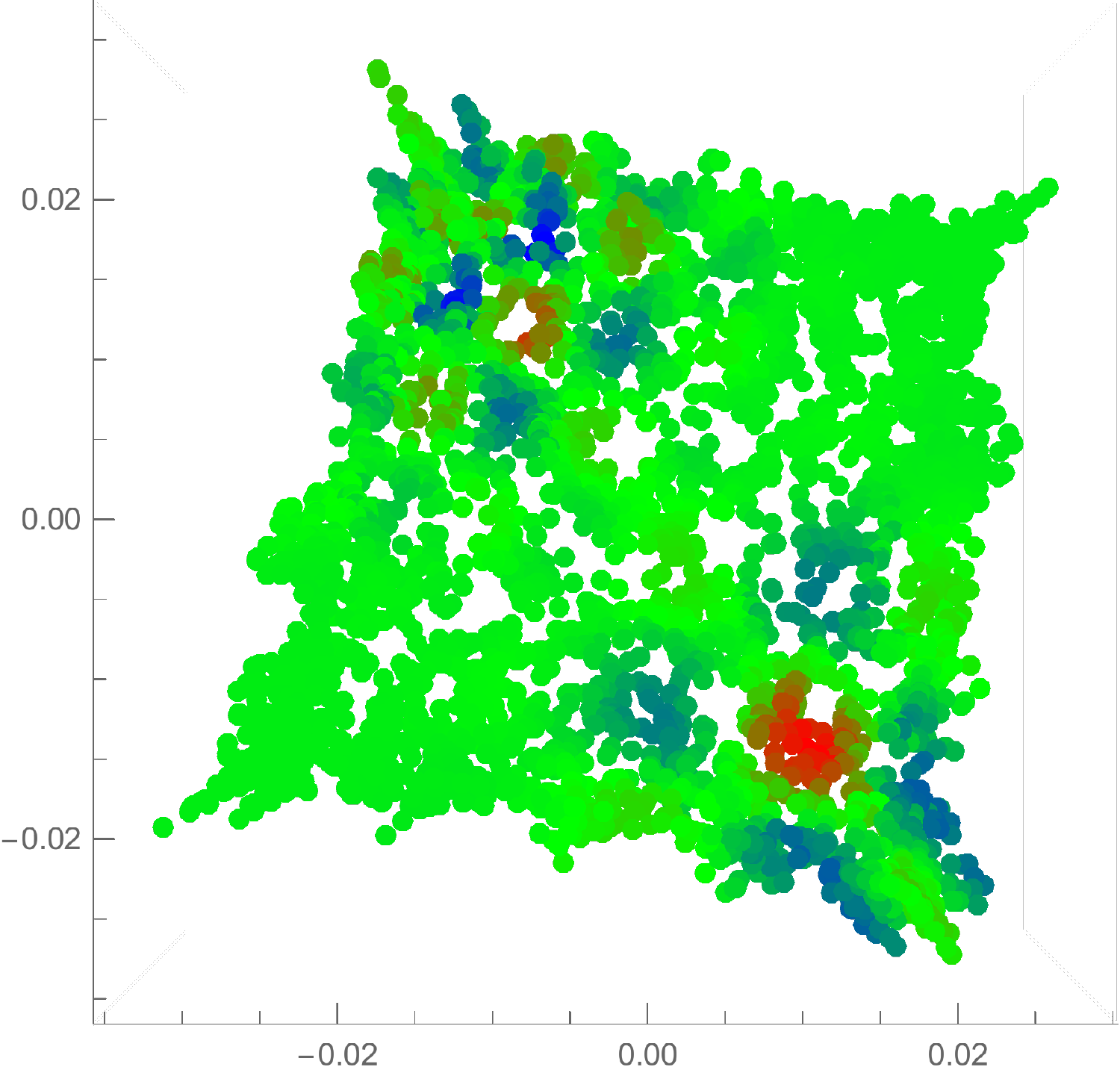}}
		\centerline{\footnotesize (b) Oscillating function in diffusion space}\medskip
	\end{minipage}
	\begin{minipage}[b]{.48\linewidth}
		\centering
		\centerline{\includegraphics[width=4.0cm]{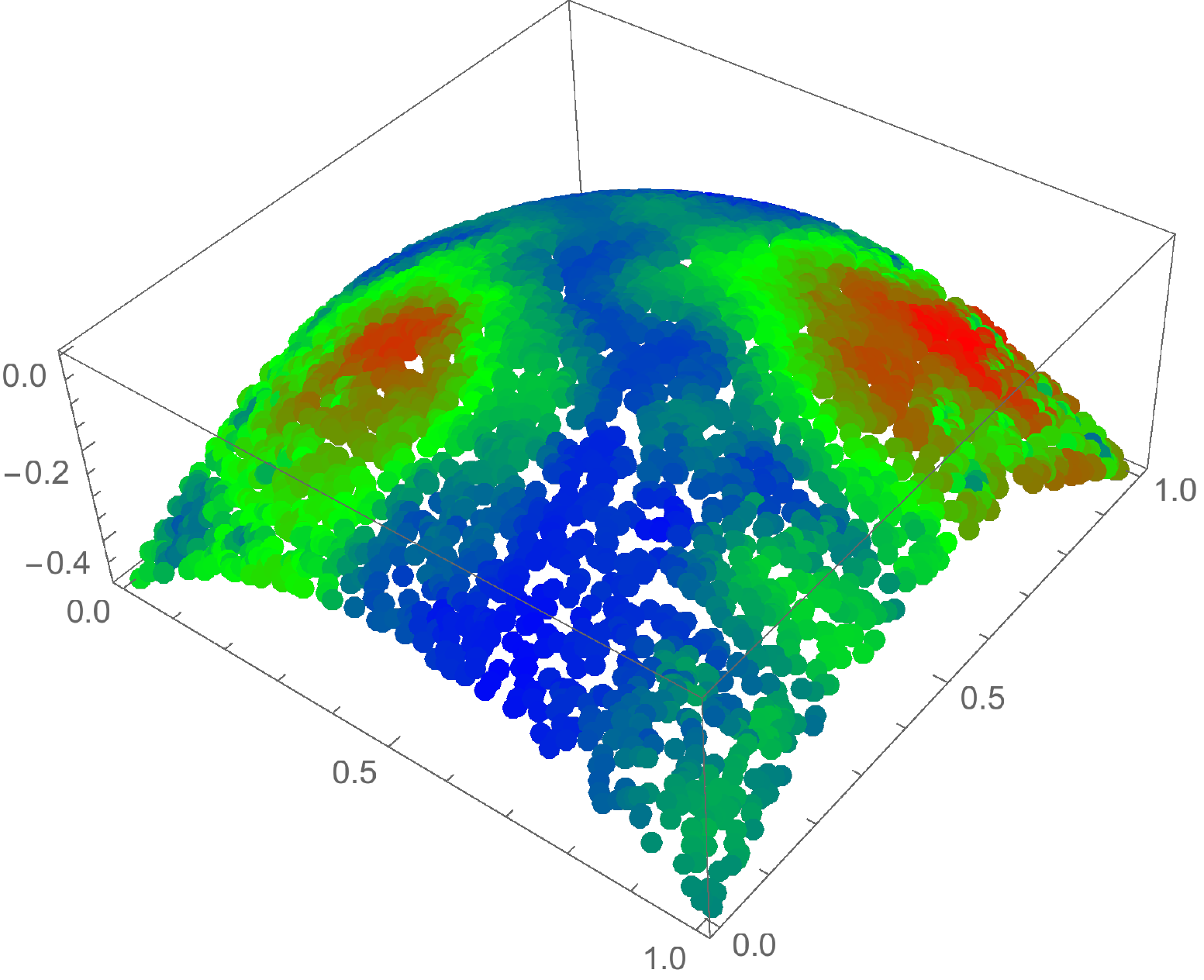}}
		\centerline{\footnotesize (c) Resulting envelope over a patch of sphere}\medskip
	\end{minipage}
	\hfill
	\begin{minipage}[b]{0.48\linewidth}
		\centering
		\centerline{\includegraphics[width=4.0cm]{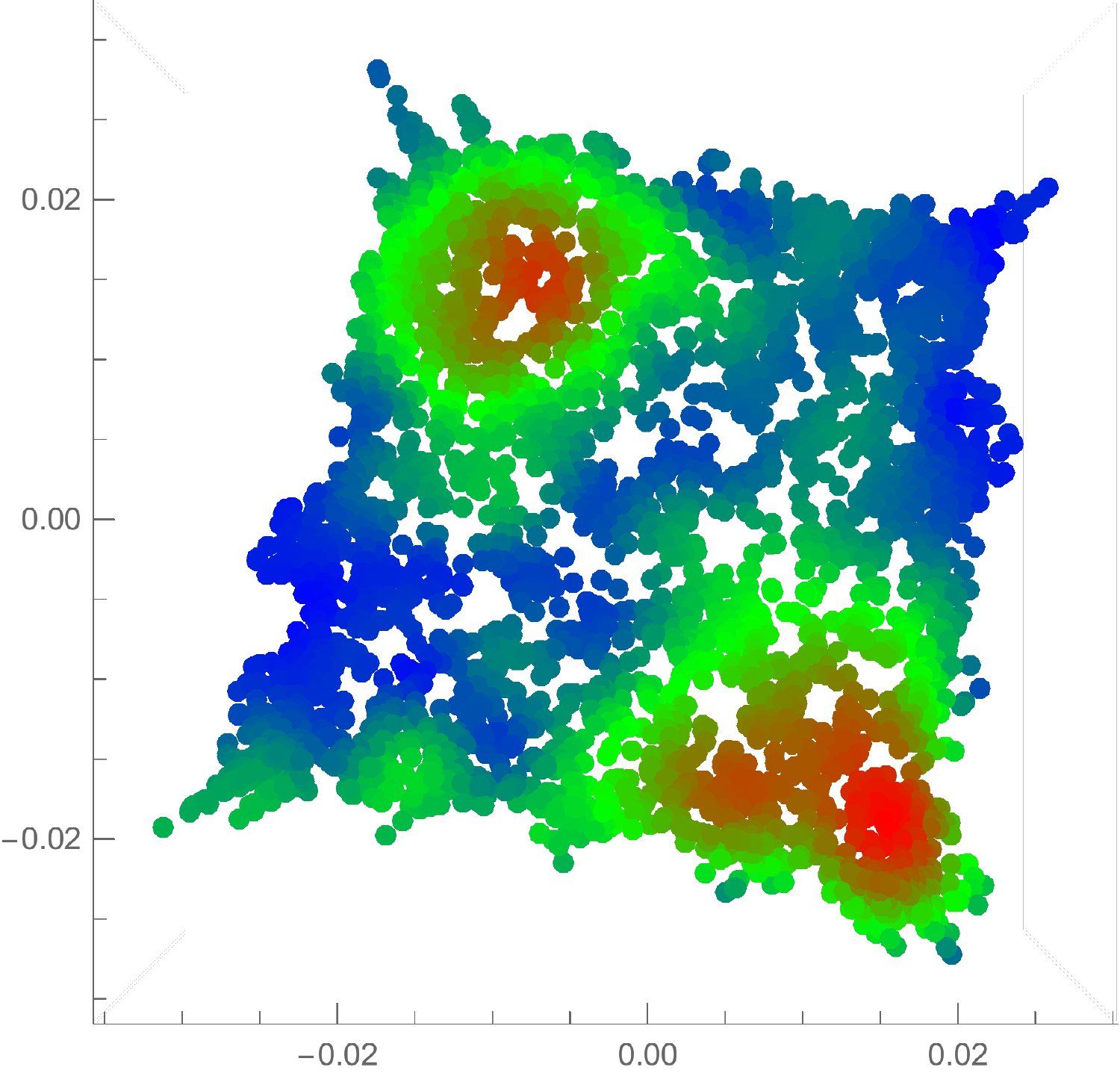}}	
		\centerline{\footnotesize (d) Amplitude in diffusion space}\medskip
	\end{minipage}
	\vspace{-0.3cm}
	\caption{Point cloud and diffusion embedding of the corresponding knn-graph. Oscillation process is given by the sum of modulated gaussians: (a) original point cloud with signal over it; (b) embedding of the point cloud by the first two eigenvectors of normalized Laplacian; (c) envelope of the original signal; (d) envelope of the original signal in diffusion space.}
	\label{fig:point cloud signal}
\end{figure}

The graph in this example is constructed as follows. We take a sampled patch on the sphere consisting of $4096$ randomly picked points as illustrated by Fig.~\ref{fig:point cloud signal} (a). The resulting point cloud is inherently $2$-dimensional although it is embedded in $3$ dimensions. As an oscillating function we consider a sum of two Gaussians modulated by different frequencies that lie on a curved surface. Then, to obtain an associated graph, we consider the $k=20$ nearest neighbours graph to mesh the point cloud. Then we use the diffusion maps embedding, which is given by the first two eigenvectors of the normalised Laplacian, to embed the patch in $\R^2$. The resulting embedding of the corresponding $k$ nearest neighbors graph with the same oscillating process is shown in Fig.~\ref{fig:point cloud signal} (b).

\begin{figure}[t!]
	\begin{minipage}[b]{.48\linewidth}
		\centering
		\centerline{\includegraphics[width=4.0cm]{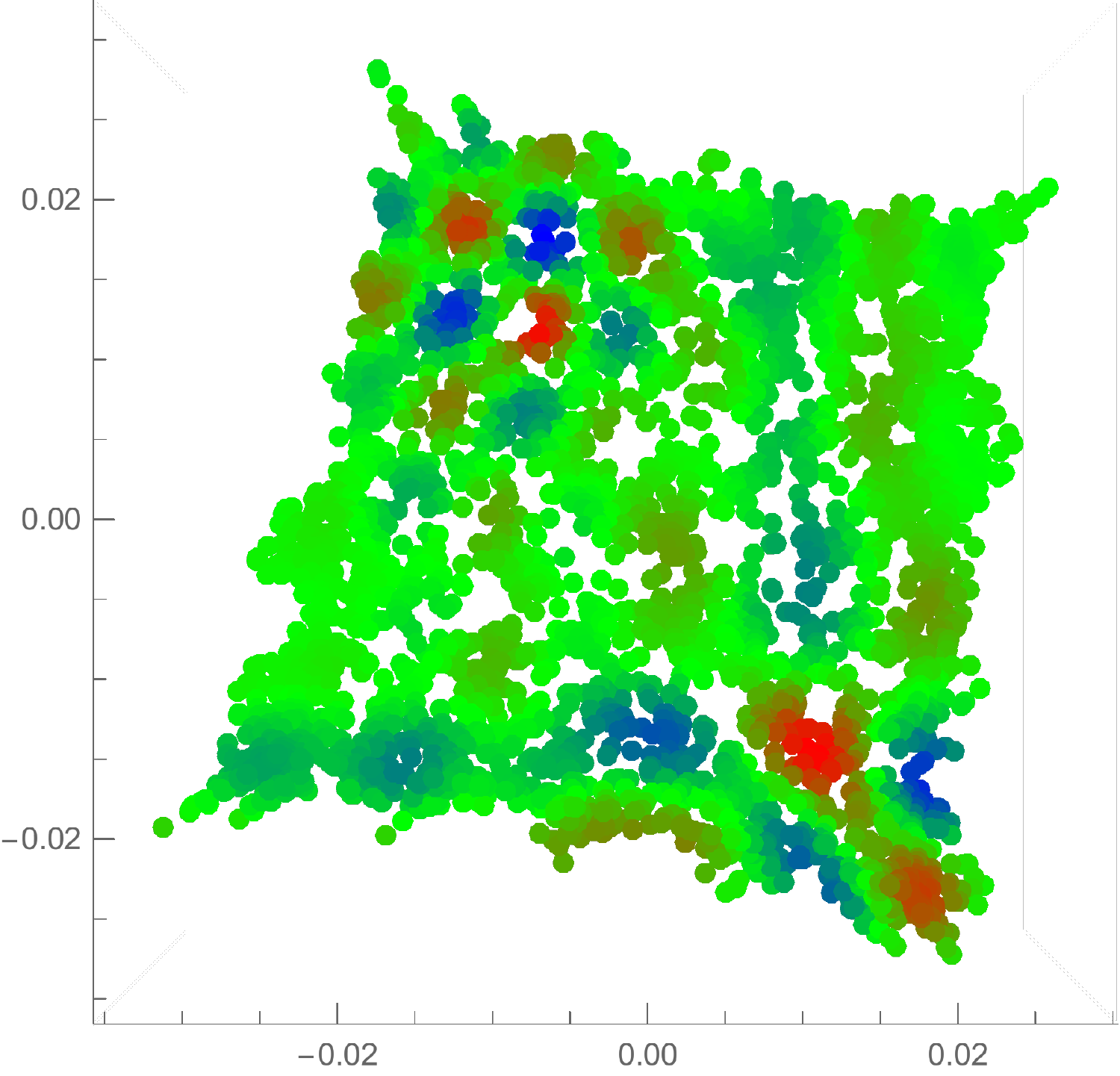}}
		\centerline{\footnotesize (a) $f_{00}$ shifted component}\medskip
	\end{minipage}
	\hfill
	\begin{minipage}[b]{0.48\linewidth}
		\centering
		\centerline{\includegraphics[width=4.0cm]{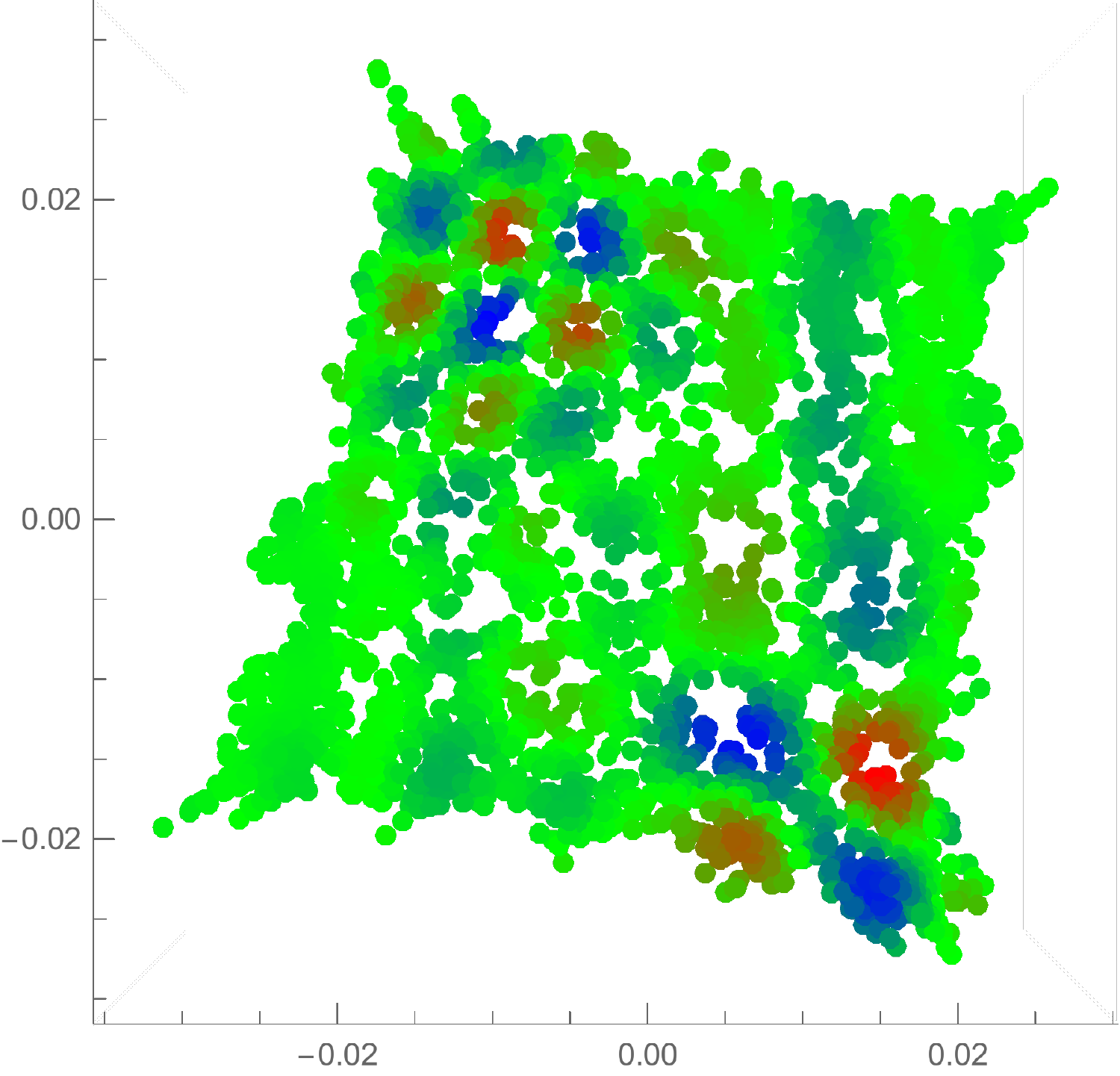}}
		\centerline{\footnotesize (b) $f_{10}$ shifted component}\medskip
	\end{minipage}
	
	\begin{minipage}[b]{.48\linewidth}
		\centering
		\centerline{\includegraphics[width=4.0cm]{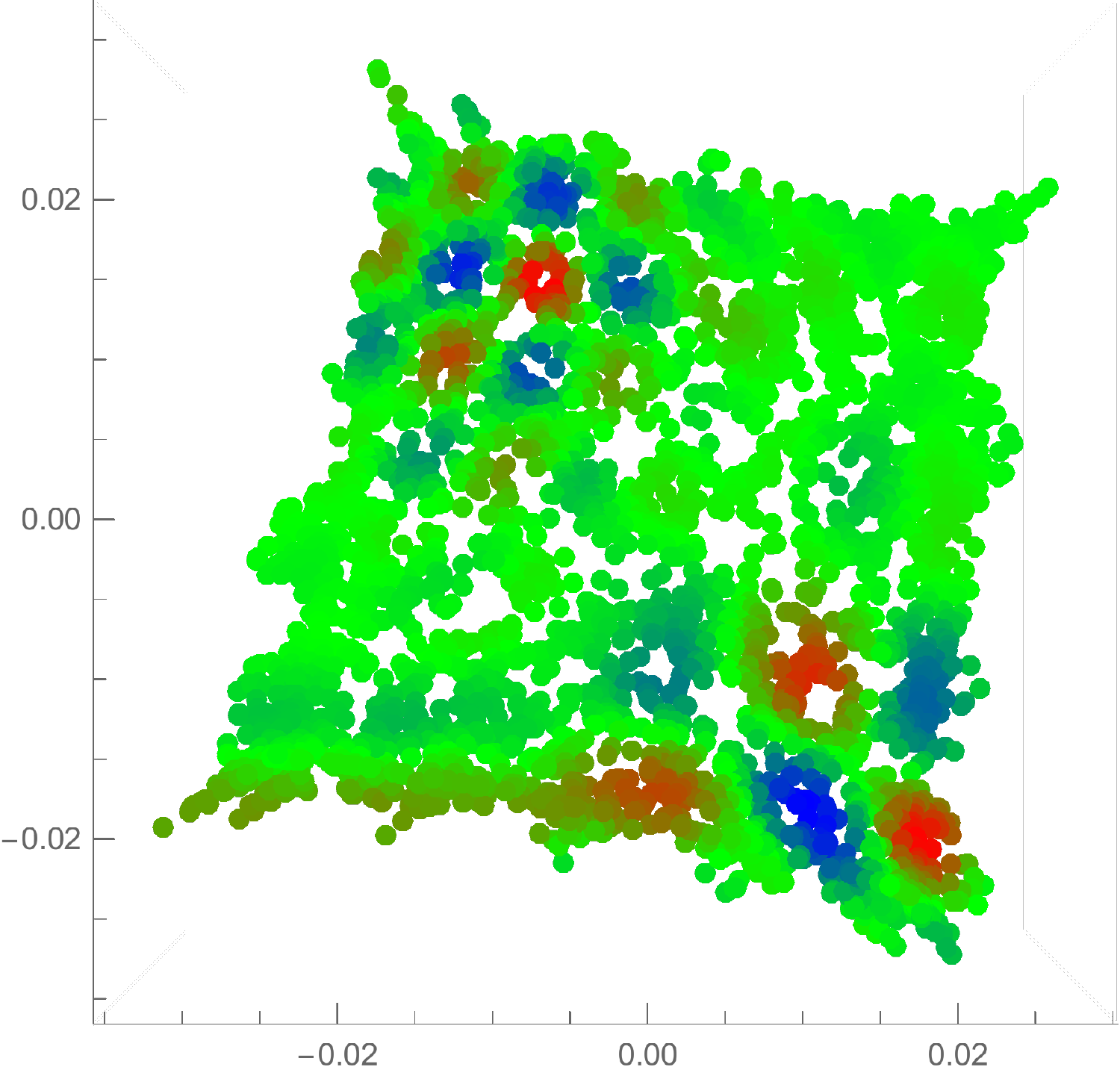}}
		\centerline{\footnotesize (a) $f_{01}$ shifted component}\medskip
	\end{minipage}
	\hfill
	\begin{minipage}[b]{0.48\linewidth}
		\centering
		\centerline{\includegraphics[width=4.0cm]{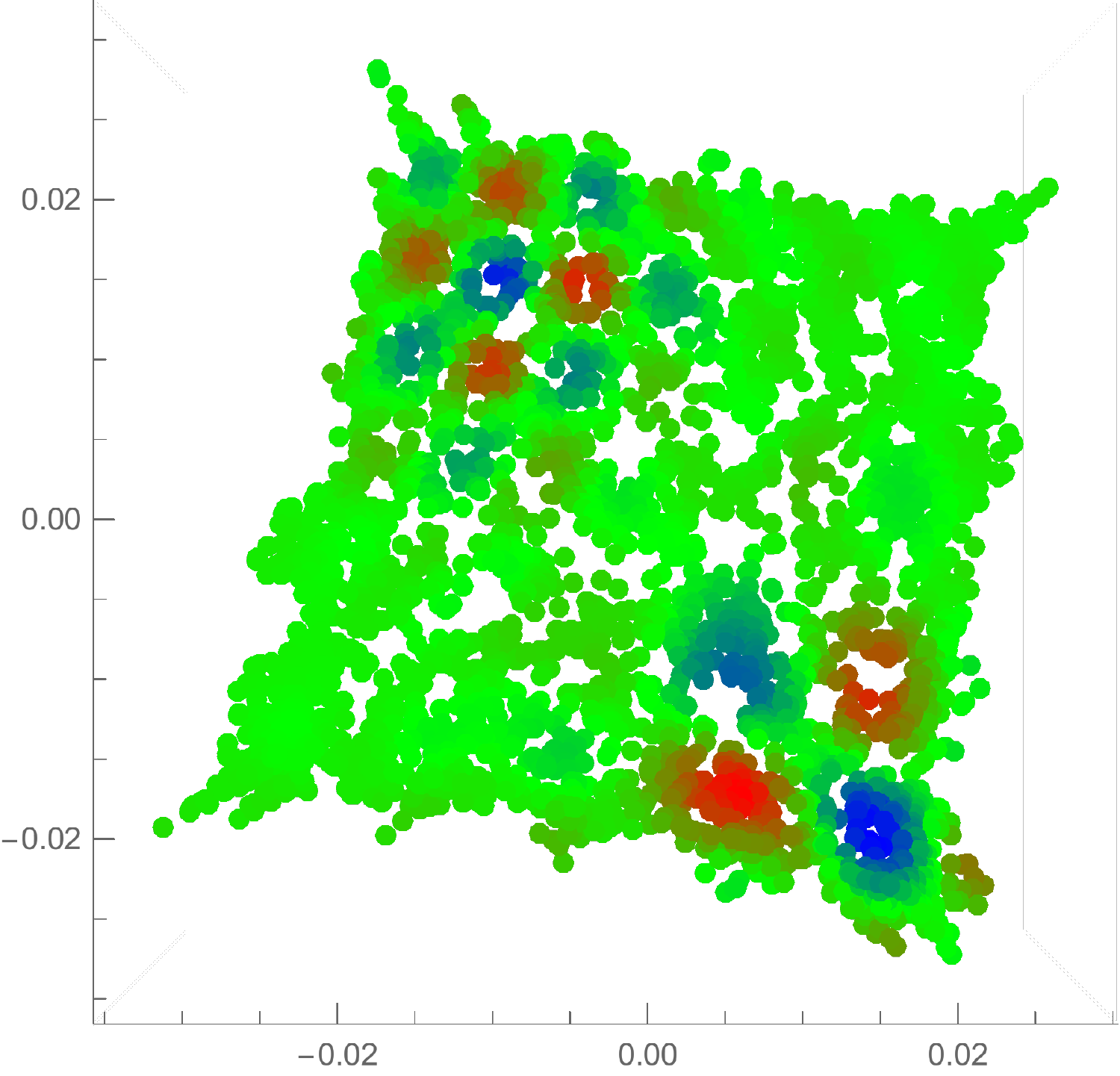}}
		\centerline{\footnotesize (b) $f_{11}$ shifted component}\medskip
	\end{minipage}
	\vspace{-0.3cm}
	\caption{ Phase-shifted components in the diffusion space.}
	\label{fig:phase shifts}
\end{figure}

To calculate the discrete version of hypercomplex Fourier transform $\hat{f}(\bm{\omega})$, we applied the transformation \eqref{eq:usual hyper Fourier} to the discrete oscillating function given in Fig.~\ref{fig:point cloud signal} (b) (axes are rescaled by a factor of $10$). Then corresponding phase-shifted components were constructed by first restricting the spectrum $\hat{f}(\bm{\omega})$ to positive frequencies according to \eqref{eq:fh spectrum} and  applying inverse Fourier transform (\ref{eq:usual hyper Fourier}). All computations were performed in the diffusion space where the initial point cloud is flattened. Discrete version of the Fourier transform \eqref{eq:usual hyper Fourier} was calculated for the first $50$ frequencies with step $1$. Then we calculate the hypercomplex analytic signal by restricting the spectrum to positive frequencies according \eqref{eq:fh spectrum} prior to applying the inverse Fourier transform. In Fig.~\ref{fig:point cloud signal} (c) and (d), the envelope function of Eq.~\eqref{eq:envelope in many dim} is shown over the original point cloud and in the diffusion space, respectively. In Fig.~\ref{fig:phase shifts} the phase-shifted components $f_{\bm{j}}(\bm{x})$ for the four directions $\bm{j}=00,\,10,\,01$ and $11$ in the diffusion space are shown. Even though the directions of oscillations are not perfectly aligned to the basis vectors in embedding space (see Fig.~\ref{fig:point cloud signal} (b)), the resulting amplitude still looks fine. In the following example we clarify this point by looking on how amplitude transforms under rotational transformation of the underlying oscillating process.

\subsection{Amplitude of rotated process}
\label{subsection: Amplitude of rotated process}

In this section we will look how amplitude transforms under rotational transformation of underlying oscillating process. Let us start by considering the simplest $2$-dimensional oscillating process
\begin{equation}
f (x, y ) = \cos (x) \cos (y).
\end{equation}
Intuitively it is clear that such process should have constant amplitude. And indeed using definition \eqref{def:instantaneous amplitude, phase and frequency}, we see that
\begin{equation}
a(x,y) = \sqrt{\cos (x)^2 \cos (y)^2 + \sin (x)^2 \cos (y)^2 + \sin (x)^2 \cos (y)^2 + \sin (x)^2 \sin (y)^2} = 1.
\end{equation}
Let us consider the rotation by $\pi/4$ applied to $f$. In this case the coordinates are transformed according to 

\begin{align}
\label{eq: rotation transforms}
\begin{split}
x' = \frac{1}{\sqrt{2}} x - \frac{1}{\sqrt{2}} y, \ \ \
y' = \frac{1}{\sqrt{2}} x + \frac{1}{\sqrt{2}} y.
\end{split}
\end{align}
By expanding $f'(x', y') \equiv f(x', y')$ we get
\begin{align}
f'(x, y) = \cos \left( \frac{1}{\sqrt{2}} x - \frac{1}{\sqrt{2}} y \right) \cos \left( \frac{1}{\sqrt{2}} x + \frac{1}{\sqrt{2}} y \right) = 
\frac{1}{2} \cos \left( \sqrt{2} x \right) + \frac{1}{2} \cos \left( \sqrt{2} y \right).
\end{align}

The phase-shifted components for $f'$ are (recall that Hilbert transform of constant function is zero)
\begin{align}
\begin{split}
f'_{00} (x, y) = &\frac{1}{2} \cos \left( \sqrt{2} x \right) + \frac{1}{2} \cos \left( \sqrt{2} y \right), \\
f'_{10} (x, y) = & \frac{1}{2} \sin \left( \sqrt{2} x \right), \\
f'_{01} (x, y) = & \frac{1}{2} \sin \left( \sqrt{2} y \right), \\
f'_{11} (x, y) = & 0.
\end{split}
\end{align}
Combining the terms together we get $a'(x,y)$
\begin{align}
\begin{split}
a'(x,y)^2 =  &f'_{00} (x, y)^2 + f'_{10} (x, y)^2 + f'_{01} (x, y)^2 + f'_{11} (x, y)^2 \\
= & \frac{1}{4} \cos \left( \sqrt{2} x \right)^2 + \frac{1}{4} \cos \left( \sqrt{2} y \right)^2 + \frac{1}{2} \cos \left( \sqrt{2} x \right) \cos \left( \sqrt{2} y \right) \\ 
+ & \frac{1}{4} \sin \left( \sqrt{2} x \right)^2  +  \frac{1}{4} \sin \left( \sqrt{2} y \right)^2 \\
= & \frac{1}{2} \left[ 1 + \cos \left( \sqrt{2} x \right) \cos \left( \sqrt{2} y \right) \right],
\end{split}
\end{align}
which is not constant. Therefore definition of amplitude \eqref{def:instantaneous amplitude, phase and frequency} is only valid for oscillations aligned with the chosen orthogonal directions in $\R^d$. 

\subsection{Amplitude of lower dimensional process}
\label{subsection: Amplitude of low dimensional process}

Next we provide an example of amplitude computation for $1$-dimensional oscillating process in $\R^2$. We start by considering the simplest $1$-dimensional oscillations in $\R^2$, given by
\begin{equation}
f(x,y) = \cos(x).
\end{equation}
The phase-shifted components of $f$ are
\begin{align}
\begin{split}
f_{00} (x, y) = & \cos \left( x \right), \\
f_{10} (x, y) = & \sin \left( x \right), \\
f_{01} (x, y) = & 0, \\
f_{11} (x, y) = & 0,
\end{split}
\end{align}
which gives constant instantaneous amplitude $a(x,y) = 1$ as expected. We can conclude more generally that definition of amplitude \eqref{def:instantaneous amplitude, phase and frequency} works well when oscillations in some directions are degenerate.

Let us consider the rotated version of $f$ and calculate its amplitude. As in the previous section we apply transformation \eqref{eq: rotation transforms} to $f(x,y)$ and write $f'$ as
\begin{equation}
f'(x, y) = \cos \left( \frac{1}{\sqrt{2}} x - \frac{1}{\sqrt{2}} y \right) = \cos \left( \frac{1}{\sqrt{2}} x \right) \cos \left( \frac{1}{\sqrt{2}} y \right) + \sin \left( \frac{1}{\sqrt{2}} x \right) \sin \left( \frac{1}{\sqrt{2}} y \right).
\end{equation}
The phase-shifted components of $f'$ are given by
\begin{align}
\begin{split}
f'_{00} (x, y) = & \cos \left( \frac{1}{\sqrt{2}} x - \frac{1}{\sqrt{2}} y \right), \\
f'_{10} (x, y) = & \sin \left( \frac{1}{\sqrt{2}} x - \frac{1}{\sqrt{2}} y \right), \\
f'_{01} (x, y) = & -\sin \left( \frac{1}{\sqrt{2}} x - \frac{1}{\sqrt{2}} y \right), \\
f'_{11} (x, y) = & \cos \left( \frac{1}{\sqrt{2}} x - \frac{1}{\sqrt{2}} y \right).
\end{split}
\end{align}
Summing up the squares of phase-shifted components we get $a'(x,y) = \sqrt{2}$, which is not the expected envelope equal to $1$. We observe similar problem with amplitude calculation of a rotated process as in the previous example.

\section{Conclusion and future work}
\label{sec:conclusion}

In this work we extended analytic signal theory to describe oscillating functions over $d$-dimensional Euclidean space for the case when oscillations are aligned to $d$ orthogonal directions given by orthonormal basis vectors of $\R^d$. There are several directions in which the presented work could be extended. First of all we need a method for amplitude calculation of a general oscillating process where oscillations are observed in various independent directions, not only in the chosen orthogonal. Second, the extension of hypercomplex analytic signal to manifolds with the properly defined hypercomplex holomorphic structure is needed. Third, the development of multidimensional phase-space or space-frequency analysis tools for the presented hypercomplex analytic signal model, like Wigner-Ville and Segal-Bargmann transforms, will allow to understand better the localization properties of multidimensional oscillating processes. Fourth, development of hypercomplex multidimensional filtering theory for the designed hypercomplex analytic signal will bring us a powerful tool for many practical problems. And finally, as a more distant objective there is a hope to find hypercomplex dynamic equations for multidimensional oscillating processes that arise in nature as well as to develop the associated hypercomplex gauge theory that rely on the definition of instantaneous amplitude and phases. From application point of view we are going to investigate how theory presented in this paper could be applied in the domains where multidimensional oscillations are inherent, e.g. image/video analysis, sound and seismic waves, data oscillations in high-dimensional feature space in machine learning as well as complex oscillating phenomena in quantum field theory.

\begin{appendices}
\section{Hilbert transform for unit disk and upper half-plane}
\label{appendix:Hilbert transform in 1D}
Let us briefly describe the essence of the one dimensional Hilbert transform by following the lines of \cite{krantz2009explorations}. The usual way to introduce Hilbert transform is by using the Cauchy formula. If $f$ is holomorphic on $\D \subset \C$ and continuous on disk $\D$ up to the boundary $\partial \D$, we can get the value of $f(z)$ from the values of $f$ on the boundary $\partial \D$ by Cauchy formula
\begin{equation}
f(z) = \frac{1}{2 \pi i} \int\limits_{\partial \D} \frac{f(\zeta)}{\zeta - z} \, \diff \zeta, \ \  z \in \D.
\end{equation}
We can express the Cauchy kernel, $$\frac{1}{2 \pi i} \cdot \frac{\diff \zeta}{\zeta - z} ,$$ by taking $\zeta = e^{i\psi}$ and $z = r e^{i \theta}$, as following
\begin{align}
\label{eq:Cauchy kernel}
\begin{split}
\frac{1}{2 \pi i} \cdot \frac{\diff \zeta}{\zeta - z} &= \frac{1}{2 \pi} \cdot \frac{- i e^{-i \psi} \cdot i e^{i \psi} \diff \psi}{ e^{-i \psi} \left( e^{i \psi} - r e^{i \theta} \right)} =  \frac{1}{2 \pi} \cdot \frac{\diff \psi}{\left(1 - r e^{i(\theta-\psi)}\right)}  = \frac{1}{2 \pi } \cdot \frac{ 1 - r e^{-i(\theta-\psi)} }{\left| 1 - r e^{i(\theta-\psi)} \right|^2} \diff \psi \\
& = \left( \frac{1}{2 \pi } \cdot \frac{ \left(1 - r \cos \left(\theta-\psi \right) \right)}{\left| 1 - r e^{i(\theta-\psi)} \right|^2} \diff \psi \right) + i \left( \frac{1}{2 \pi } \cdot \frac{ r \sin \left(\theta-\psi \right) }{\left| 1 - r e^{i(\theta-\psi)} \right|^2} \diff \psi \right).
\end{split}
\end{align}
If we subtract $\frac{1}{4 \pi} \diff \psi$ from the real part of the Cauchy kernel, we get the Poisson kernel $P_r(\theta)$
\begin{align}
\label{eq:real part of Cauchy kernel}
\begin{split}
\Re \left( \frac{1}{2 \pi i} \cdot \frac{\diff \zeta}{\zeta - z} \right) - \frac{1}{4\pi} \diff \psi &= \frac{1}{2 \pi } \left( \frac{ 1 - r \cos \left(\theta-\psi \right) }{\left| 1 - r e^{i(\theta-\psi)} \right|^2} - \frac{1}{2}  \right) \diff \psi \\
&= \frac{1}{2 \pi } \left( \frac{\frac{1}{2} - \frac{1}{2}r^2}{1-2r \cos \left( \theta - \psi \right) + r^2} \right) \diff \psi =: \frac{1}{2} P_r \left( \theta - \psi \right).
\end{split}
\end{align}
Thus the real part is, up to a small correction, the Poisson kernel. The kernel that reproduces harmonic functions is the real part of the kernel that reproduces holomorphic functions.

We describe briefly the Poisson integral formula. If $\D = \left\{ z: |z| < 1 \right\}$ is the open unit disc in $\C$ and $\partial \D$ is the boundary of $\D$ and there is continuous $g : \partial \D \rightarrow \R$, then the function $u: \D \rightarrow \R$, given by
\begin{equation}
u \left( r e^{i \theta} \right) = \frac{1}{2 \pi} \int\limits_{-\pi}^{\pi} P_r \left( \theta - \tau \right) g \left( e^{i \tau} \right) \, \diff \tau, \ \ 0 \leq r < 1,
\end{equation}
will be harmonic on $\D$ and has radial limit $r \rightarrow 1^-$ that agrees with $g$ almost everywhere on $\partial \D$.

Suppose now that we are given a real-valued function $f \in L^2 \left( \partial \D \right)$. Then we can use the Poisson integral formula to produce a function $u$ on $\D$ such that $u = f$ on $\partial \D$. Then we can find a harmonic conjugate $u^\dagger$ of $u$, such that $u^\dagger(0) = 0$ and $u + i u^\dagger$ is holomorphic on $\D$. As a final goal we aim to produce a boundary function $f^\dagger$ for $u^\dagger$ and get the linear operator $f \mapsto f^\dagger$.

If we define a function $h$ on $\D$ as
\begin{equation}
h(z) = \frac{1}{2 \pi i} \int\limits_{\partial \D} \frac{f(\zeta)}{\zeta - z} \, \diff \zeta, \ \  z \in \D,
\end{equation}
then $h$ will be holomorphic in $\D$. From \eqref{eq:real part of Cauchy kernel} we know that the real part of $h$ is up to an additive constant equal to the Poisson integral $u$ of $f$. Therefore $\Re (h)$ is harmonic on $\D$ and $\Im (h)$ is a harmonic conjugate of $\Re (h)$. Thus if $h$ is continuous up to the boundary, then we can take $u^\dagger = \Im(h)$ and $f^\dagger \left( e^{i\theta} \right) = \lim_{r \rightarrow 1^-} u^\dagger \left( r e^{i \theta} \right)$.

For the imaginary part of the Cauchy kernel \eqref{eq:Cauchy kernel} under the limit $r \rightarrow 1^-$ we get
\begin{align*}
\begin{split}
\frac{1}{2 \pi } \cdot \frac{ r \sin \left(\theta-\psi \right) }{\left| 1 - r e^{i(\theta-\psi)} \right|^2}  \xrightarrow{r \rightarrow 1^-} \frac{1}{2 \pi } \cdot  \frac{ \sin \left(\theta-\psi \right) }{\left| 1 - e^{i(\theta-\psi)} \right|^2} = \ & \frac{\sin (\theta - \psi)}{4 \pi \left( 1 - \cos (\theta - \psi)\right)} 
= \frac{2 \sin \left( \frac{\theta - \psi}{2}\right) \cos \left( \frac{\theta - \psi}{2}\right)}{8 \pi \left(\cos \left( \frac{\theta - \psi}{2}\right)\right)^2} \\
= &\frac{1}{4 \pi} \cot \left( \frac{\theta - \psi}{2}\right).
\end{split}
\end{align*}

Therefore we obtain the Hilbert transform $H: f \mapsto f^\dagger$ on the unit disk $\D$ as following
\begin{equation}
\label{eq:hilbert transform cot}
H \left[ f \right] \left( e^{i\theta} \right) = \frac{1}{4 \pi} \int\limits_{-\pi}^{\pi} f \left( e^{it} \right) \cot \left( \frac{\theta - t}{2} \right) \, \diff t.
\end{equation}
Taylor series expansion of the kernel in \eqref{eq:hilbert transform cot} gives us
\begin{align*}
\cot \left( \frac{\theta}{2} \right) = \frac{\cos \left( \frac{\theta}{2} \right)}{\sin \left( \frac{\theta}{2} \right)} = \frac{1 - \frac{(\theta/2)^2}{2!} + \dots }{\frac{\theta}{2} \left( 1 - \frac{(\theta/2)^2}{3!} + \dots \right)}
= \frac{2}{\theta} + E(\theta),
\end{align*}
where $E \left( \theta \right) = O \left( |\theta| \right)$ is a bounded continuous function. Finally for the Hilbert transform on $\partial \D$ we can write

\begin{align}
\label{eq:hilbert transform on disk}
H \left[ f \right] \left( e^{i\theta} \right) = \frac{1}{4 \pi} \int\limits_{-\pi}^{\pi} f \left( e^{it} \right) \cot \left( \frac{\theta - t}{2}\right) \, \diff t = \frac{1}{4 \pi} \int\limits_{-\pi}^{\pi} f \left( e^{it} \right)  \frac{2}{\theta - t} \, \diff t + \frac{1}{4\pi} \int\limits_{-\pi}^{\pi} f \left( e^{it} \right) E \left( \theta - t \right) \, \diff t.
\end{align}
The first integral above is singular at $t = \theta$ and the second is bounded and easy to estimate. Usually we write for the kernel of Hilbert transform
\begin{equation}
\cot \left( \frac{\theta}{2} \right) \approx \frac{2}{\theta - t},
\end{equation}
by simply ignoring the trivial error term. Finally in \eqref{eq:hilbert transform on disk} we have defined the Hilbert transform for the boundary of unit disk. Hilbert transform for a unit disk relates harmonic conjugates of a periodic function $f$ defined over $\partial \D$. In the followign we briefly outline how to obtain harmonic conjugate for a function defined over $\R \subset \C$.

The unit disk $\D$ can be conformally mapped to the upper half-plane $U = \{ \zeta \in \C: \Im(\zeta) > 0 \}$ by the M{\"o}bius map \cite{krantz2009explorations}
\begin{align*}
&c: \D \rightarrow U, \\
& \zeta \mapsto i \cdot \frac{1-\zeta}{1+\zeta}.
\end{align*}
Since the conformal map of a harmonic function is harmonic, we can carry also Poisson kernel to the upper half-plane \cite{titchmarsh1948introduction}. The Poisson integral equation for $U$ will be
\begin{equation}
u \left( x + i y \right) = \int\limits_{-\infty}^{\infty} P \left( x-t,y \right) f(t) \, \diff t, \ \ y>0,
\end{equation}
with the Poisson kernel given by
\begin{equation}
P \left( x,y \right) = \frac{1}{\pi} \cdot \frac{y}{x^2 + y^2}.
\end{equation}
The harmonic conjugate may be obtained from $f$ by taking convolution with the conjugate kernel
\begin{equation}
Q \left( x,y \right) = \frac{1}{\pi} \cdot \frac{x}{x^2 + y^2}.
\end{equation}
The Cauchy kernel is related to $P$ and $Q$ by the relation $\frac{i}{\pi z} = P(x,y) + i Q(x,y)$.
Finally we are able to construct the harmonic conjugate on the boudary $\partial U = \R$
\begin{equation}
u^{\dagger}(x,y) = \frac{1}{\pi} \int\limits_{-\infty}^{\infty} f(t) \frac{x-t}{(x-t)^2 + y^2} \, \diff t \xrightarrow{y \rightarrow 0^+} \frac{1}{\pi} \int\limits_{-\infty}^{\infty} f(t) \frac{1}{x-t} \, \diff t.
\end{equation}

\section{Cauchy formula by Ketchum and Vladimirov}
\label{appendix: Cauchy formula of Ketchum and Vladimirov}
We briefly describe the setting and the resulting Cauchy formula given in \cite{ketchum1928analytic} and \cite{vladimirov1984superanalysis2}.
First let us take a bounded region  $G \subseteq \C$ with piecewise smooth boundary $\partial G$. Second let us consider the mapping of $G$ given by $l: w \mapsto a + b w$ with $a, b \in \Sch_d$ onto the plane $C = \{ z: z = a+bw, w \in \Sch_d(\C) \}  \subset \Sch_d(\C)$, where $\Sch_d \left( \C \right)$ gives a complexification of $\Sch_d$, i.e. we replace real coefficients with complex in \eqref{eq:hypercomplex variable}. $G$ is mapped to $L$ and $\partial G$ is mapped to $\partial L$ correspondingly. Suppose also that the function $f(z)$ is differentiable  \cite{vladimirov1984superanalysis1} in some neighbourhood $O$ of the closure $\bar{L} = L \cup \partial L$ and $b$ is an invertible element of $\Sch_d(\C)$. Then in the plane $C$ we have the following Cauchy formula
\begin{align}
\label{eq:Cauchy hypercomplex formula}
\frac{1}{2 \pi i} \int\limits_{\partial L} \frac{f(\zeta)}{\zeta - z} \, \diff \zeta = \begin{cases}
f(z) & \text{if } z \in L, \\
0 & \text{if } z \in C \setminus \bar{L}. 
\end{cases} 
\end{align}
As it was outlined in Appendix \ref{appendix:Hilbert transform in 1D} the Cauchy formula plays a vital role in the definition of Hilbert transform and one can directly extend the Hilbert transform and define it in the unit open disk (upper half-plane) of $C$.
\begin{remark}[\cite{vladimirov1984superanalysis2}]
	The Cauchy formula \eqref{eq:Cauchy hypercomplex formula} holds for real Banach algebra $\Sch_d$ in which there exists an element $i^2 = -\epsilon$, where $\epsilon$ is the unit element in $\Sch_d$. The plane $C$ consists of the elements
	\begin{equation}
	C = \{x : x = a + u \epsilon + v i, \ u,v \in \R \}, \ a \in \Sch_d.
	\end{equation}
\end{remark}

The Cauchy formula for several hypercomplex variables is written similarly \cite{vladimirov1984superanalysis2}. Now we have the region $L = L_1 \times \dots \times L_d$, where $L_j = \{ z_j = a_j + b_j w_j\}$ is compact in some open region. By the Hartogs's theorem we have that if the function of several variables $f(z_1, \dots, z_d)$ is analytic with respect to each variable in $G_1 \times \dots \times G_d$, then it is analytic in $G_1 \times \dots \times G_d$ and we have the general Cauchy formula
\begin{equation}
f(z_1, ..., z_d) = \frac{1}{(2 \pi i)^d} \int\limits_{\partial L_1} ... \int\limits_{\partial L_d} \frac{f(\zeta_1, ..., \zeta_d)}{(\zeta_1 - z_1) \cdot ... \cdot (\zeta_d - z_d)} \, \diff \zeta_1 \dots \zeta_d
\end{equation}
for all $z \in L$.

\end{appendices}

\bibliographystyle{apalike}
\bibliography{bibliography}

\begin{thebibliography}{}

\bibitem[SKE, 2017]{SKETCHINGHYPERCOMPLEX}
 (2017).
\newblock
  \href{https://web.archive.org/web/20170309084128/http://history.hyperjeff.net/hypercomplex}{``Sketching
  the History of Hypercomplex Numbers``}.

\bibitem[Alfsmann et~al., 2007]{alfsmann2007hypercomplex}
Alfsmann, D., G{\"o}ckler, H.~G., Sangwine, S.~J., and Ell, T.~A. (2007).
\newblock Hypercomplex algebras in digital signal processing: Benefits and
  drawbacks.
\newblock In {\em Signal Processing Conference, 2007 15th European}, pages
  1322--1326. IEEE.

\bibitem[Bahri et~al., 2008]{bahri2008uncertainty}
Bahri, M., Hitzer, E.~S., Hayashi, A., and Ashino, R. (2008).
\newblock An uncertainty principle for quaternion fourier transform.
\newblock {\em Computers \& Mathematics with Applications}, 56(9):2398--2410.

\bibitem[Bedrosian, 1962]{bedrosian1962product}
Bedrosian, E. (1962).
\newblock A product theorem for {Hilbert} transforms.

\bibitem[Bernstein, 2014]{bernstein2014fractional}
Bernstein, S. (2014).
\newblock The fractional monogenic signal.
\newblock In {\em Hypercomplex Analysis: New Perspectives and Applications},
  pages 75--88. Springer.

\bibitem[Bernstein et~al., 2013]{bernstein2013generalized}
Bernstein, S., Bouchot, J.-L., Reinhardt, M., and Heise, B. (2013).
\newblock Generalized analytic signals in image processing: comparison, theory
  and applications.
\newblock In {\em Quaternion and Clifford Fourier Transforms and Wavelets},
  pages 221--246. Springer.

\bibitem[Bihan and Sangwine, 2010]{bihan2010hyperanalytic}
Bihan, N.~L. and Sangwine, S.~J. (2010).
\newblock The hyperanalytic signal.
\newblock {\em arXiv preprint arXiv:1006.4751}.

\bibitem[Brackx et~al., 2006]{brackx2006two}
Brackx, F., De~Schepper, N., and Sommen, F. (2006).
\newblock The two-dimensional {Clifford-Fourier} transform.
\newblock {\em Journal of mathematical Imaging and Vision}, 26(1-2):5--18.

\bibitem[Bridge, 2017]{bridge2017introduction}
Bridge, C.~P. (2017).
\newblock Introduction to the monogenic signal.
\newblock {\em arXiv preprint arXiv:1703.09199}.

\bibitem[B{\"u}low and Sommer, 2001]{bulow2001hypercomplex}
B{\"u}low, T. and Sommer, G. (2001).
\newblock Hypercomplex signals-a novel extension of the analytic signal to the
  multidimensional case.
\newblock {\em IEEE Transactions on signal processing}, 49(11):2844--2852.

\bibitem[Catoni et~al., 2008]{Catoni2008}
Catoni, F., Boccaletti, D., Cannata, R., Catoni, V., Nichelatti, E., and
  Zampetti, P. (2008).
\newblock {\em The mathematics of {Minkowski} space-time: with an introduction
  to commutative hypercomplex numbers}.
\newblock Springer Science \& Business Media.

\bibitem[Chung, 1997]{chung1997spectral}
Chung, F.~R. (1997).
\newblock {\em Spectral graph theory}.
\newblock American Mathematical Soc.

\bibitem[Clifford, 1871]{clifford1871}
Clifford (1871).
\newblock Preliminary sketch of biquaternions.
\newblock {\em Proceedings of the London Mathematical Society},
  s1-4(1):381--395.

\bibitem[Cockle, 1849]{cockle1849iii}
Cockle, J. (1849).
\newblock {III}. on a new imaginary in algebra.
\newblock {\em Philosophical Magazine Series 3}, 34(226):37--47.

\bibitem[Coifman and Lafon, 2006]{coifman2006diffusion}
Coifman, R.~R. and Lafon, S. (2006).
\newblock Diffusion maps.
\newblock {\em Applied and computational harmonic analysis}, 21(1):5--30.

\bibitem[Coifman et~al., 2005]{coifman2005geometric}
Coifman, R.~R., Lafon, S., Lee, A.~B., Maggioni, M., Nadler, B., Warner, F.,
  and Zucker, S.~W. (2005).
\newblock Geometric diffusions as a tool for harmonic analysis and structure
  definition of data: Diffusion maps.
\newblock {\em Proceedings of the National Academy of Sciences of the United
  States of America}, 102(21):7426--7431.

\bibitem[De~Bie, 2012]{de2012clifford}
De~Bie, H. (2012).
\newblock Clifford algebras, {Fourier} transforms, and quantum mechanics.
\newblock {\em Mathematical Methods in the Applied Sciences},
  35(18):2198--2228.

\bibitem[De~Bie et~al., 2011]{de2011class}
De~Bie, H., De~Schepper, N., and Sommen, F. (2011).
\newblock The class of {Clifford-Fourier} transforms.
\newblock {\em Journal of Fourier Analysis and Applications}, 17(6):1198--1231.

\bibitem[Ell, 1992]{ell1992hypercomplex}
Ell, T.~A. (1992).
\newblock Hypercomplex spectral transformations, phd dissertation.

\bibitem[Ell et~al., 2014]{ell2014quaternion}
Ell, T.~A., Le~Bihan, N., and Sangwine, S.~J. (2014).
\newblock {\em Quaternion Fourier transforms for signal and image processing}.
\newblock John Wiley \& Sons.

\bibitem[Ell and Sangwine, 2007]{ell2007hypercomplex}
Ell, T.~A. and Sangwine, S.~J. (2007).
\newblock Hypercomplex fourier transforms of color images.
\newblock {\em IEEE Transactions on image processing}, 16(1):22--35.

\bibitem[Felsberg et~al., 2001]{felsberg2001fast}
Felsberg, M., B{\"u}low, T., Sommer, G., and Chernov, V.~M. (2001).
\newblock Fast algorithms of hypercomplex {Fourier} transforms.
\newblock In {\em Geometric computing with Clifford algebras}, pages 231--254.
  Springer.

\bibitem[Felsberg and Sommer, 2001]{felsberg2001monogenic}
Felsberg, M. and Sommer, G. (2001).
\newblock The monogenic signal.
\newblock {\em IEEE Transactions on Signal Processing}, 49(12):3136--3144.

\bibitem[Flamant et~al., 2017]{flamant2017polarization}
Flamant, J., Chainais, P., and Le~Bihan, N. (2017).
\newblock Polarization spectrogram of bivariate signals.
\newblock In {\em Acoustics, Speech and Signal Processing (ICASSP), 2017 IEEE
  International Conference on}, pages 3989--3993. IEEE.

\bibitem[Flandrin, 1998]{flandrin1998time}
Flandrin, P. (1998).
\newblock {\em Time-frequency/time-scale analysis}, volume~10.
\newblock Academic press.

\bibitem[Gabor, 1946]{gabor1946theory}
Gabor, D. (1946).
\newblock Theory of communication.
\newblock {\em Journal of the Institution of Electrical Engineers-Part III:
  Radio and Communication Engineering}, 93(26):429--441.

\bibitem[Gong and Gong, 2007]{gong2007concise}
Gong, S. and Gong, Y. (2007).
\newblock {\em Concise Complex Analysis: Revised}.
\newblock World Scientific Publishing Company.

\bibitem[Greene and Krantz, 2006]{greene2006function}
Greene, R.~E. and Krantz, S.~G. (2006).
\newblock {\em Function theory of one complex variable}, volume~40.
\newblock American Mathematical Soc.

\bibitem[Hahn and Snopek, 2011]{hahn2011unified}
Hahn, S. and Snopek, K. (2011).
\newblock The unified theory of n-dimensional complex and hypercomplex analytic
  signals.
\newblock {\em Bulletin of the Polish Academy of Sciences: Technical Sciences},
  59(2):167--181.

\bibitem[Hahn and Snopek, 2013]{hahn2013quasi}
Hahn, S. and Snopek, K. (2013).
\newblock Quasi-analytic multidimensional signals.
\newblock {\em Bulletin of the Polish Academy of Sciences: Technical Sciences},
  61(4):1017--1024.

\bibitem[Hahn and Snopek, 2016]{hahn2016complex}
Hahn, S.~L. and Snopek, K.~M. (2016).
\newblock {\em Complex and Hypercomplex Analytic Signals: Theory and
  Applications}.
\newblock Artech House.

\bibitem[Hamilton, 1844]{hamilton1844ii}
Hamilton, W.~R. (1844).
\newblock Ii. on quaternions; or on a new system of imaginaries in algebra.
\newblock {\em Philosophical Magazine Series 3}, 25(163):10--13.

\bibitem[Hilbert, 1912]{hilbert1912grundzuge}
Hilbert, D. (1912).
\newblock {\em {Grundzuge einer allgemeinen Theorie der linearen
  Integralgleichungen}}.

\bibitem[Huo et~al., 2007]{huo2007survey}
Huo, X., Ni, X., and Smith, A.~K. (2007).
\newblock A survey of manifold-based learning methods.
\newblock {\em Recent advances in data mining of enterprise data}, pages
  691--745.

\bibitem[Ketchum, 1928]{ketchum1928analytic}
Ketchum, P. (1928).
\newblock Analytic functions of hypercomplex variables.
\newblock {\em Transactions of the American Mathematical Society},
  30(4):641--667.

\bibitem[Kodaira, 2006]{kodaira2006complex}
Kodaira, K. (2006).
\newblock {\em Complex manifolds and deformation of complex structures}.
\newblock Springer.

\bibitem[Krantz, 2001]{krantz2001function}
Krantz, S.~G. (2001).
\newblock {\em Function theory of several complex variables}, volume 340.
\newblock American Mathematical Soc.

\bibitem[Krantz, 2009]{krantz2009explorations}
Krantz, S.~G. (2009).
\newblock {\em Explorations in harmonic analysis: with applications to complex
  function theory and the Heisenberg group}.
\newblock Springer Science \& Business Media.

\bibitem[Le~Bihan, 2017]{le2017foreword}
Le~Bihan, N. (2017).
\newblock Foreword to the special issue “{Hypercomplex Signal Processing}”.
\newblock {\em Signal Processing}, 136:1--106.

\bibitem[Le~Bihan and Sangwine, 2008]{le2008h}
Le~Bihan, N. and Sangwine, S. (2008).
\newblock The {H}-analytic signal.
\newblock In {\em 16th European Signal Processing Conference (EUSIPCO-2008)},
  pages P8--1.

\bibitem[Le~Bihan et~al., 2014]{le2014instantaneous}
Le~Bihan, N., Sangwine, S.~J., and Ell, T.~A. (2014).
\newblock Instantaneous frequency and amplitude of orthocomplex modulated
  signals based on quaternion fourier transform.
\newblock {\em Signal Processing}, 94:308--318.

\bibitem[Lee, 2001]{lee2001introduction}
Lee, J.~M. (2001).
\newblock Introduction to smooth manifolds.

\bibitem[Mawardi and Hitzer, 2006]{mawardi2006clifford}
Mawardi, B. and Hitzer, E.~M. (2006).
\newblock {Clifford-Fourier} transformation and uncertainty principle for the
  {Clifford} geometric algebra {$Cl_{3,0}$}.
\newblock {\em Advances in applied {Clifford} algebras}, 16(1):41--61.

\bibitem[Munkres, 2018]{munkres2018analysis}
Munkres, J.~R. (2018).
\newblock {\em Analysis on manifolds}.
\newblock CRC Press.

\bibitem[Pandey, 2011]{pandey2011hilbert}
Pandey, J.~N. (2011).
\newblock {\em {The Hilbert transform of Schwartz distributions and
  applications}}, volume~27.
\newblock {John Wiley \& Sons}.

\bibitem[Pedersen, 1997]{pedersen1997cauchy}
Pedersen, P.~S. (1997).
\newblock Cauchy's integral theorem on a finitely generated, real, commutative,
  and associative algebra.
\newblock {\em Advances in mathematics}, 131(2):344--356.

\bibitem[Ricci, 2004]{ricci2004hardy}
Ricci, F. (2004).
\newblock {\em Hardy Spaces in One Complex Variable}.
\newblock Lecture Course, Available at the webpage of {F. Ricci}.

\bibitem[Sangwine and Le~Bihan, 2007]{sangwine2007hypercomplex}
Sangwine, S.~J. and Le~Bihan, N. (2007).
\newblock Hypercomplex analytic signals: extension of the analytic signal
  concept to complex signals.
\newblock In {\em Signal Processing Conference, 2007 15th European}, pages
  621--624. IEEE.

\bibitem[Scheffers, 1893]{scheffers1893generalisation}
Scheffers, G. (1893).
\newblock Sur la g{\'e}n{\'e}ralisation des fonctions analytiques.
\newblock {\em CR Acad. Sc}, 116:1114.

\bibitem[Strichartz, 2003]{strichartz2003guide}
Strichartz, R.~S. (2003).
\newblock {\em A guide to distribution theory and {Fourier} transforms}.
\newblock World Scientific Publishing Company.

\bibitem[Titchmarsh et~al., 1948]{titchmarsh1948introduction}
Titchmarsh, E.~C. et~al. (1948).
\newblock {\em Introduction to the theory of {Fourier} integrals}, volume~2.
\newblock Clarendon Press Oxford.

\bibitem[Vakman and Vainshtein, 1977]{vakman1977amplitude}
Vakman, D. and Vainshtein, L. (1977).
\newblock Amplitude, phase, frequency—fundamental concepts of oscillation
  theory.
\newblock {\em Sov. Phys. Usp. 20(12)}.

\bibitem[Vladimirov and Volovich, 1984a]{vladimirov1984superanalysis1}
Vladimirov, V.~S. and Volovich, I.~V. (1984a).
\newblock Superanalysis. {I. Differential calculus}.
\newblock {\em Theoretical and Mathematical Physics}, 59(1):317--335.

\bibitem[Vladimirov and Volovich, 1984b]{vladimirov1984superanalysis2}
Vladimirov, V.~S. and Volovich, I.~V. (1984b).
\newblock Superanalysis. {II. Integral calculus}.
\newblock {\em Theoretical and Mathematical Physics}, 60(2):743--765.

\bibitem[Yu et~al., 2008]{yu2008bedrosian}
Yu, B., Zhang, H., et~al. (2008).
\newblock The {Bedrosian} identity and homogeneous semi-convolution equations.
\newblock {\em J. Integral Equations Appl}, 20(4):527--568.

\bibitem[Zhang, 2014]{Zhang2014}
Zhang, H. (2014).
\newblock Multidimensional analytic signals and the {Bedrosian} identity.
\newblock {\em Integral Equations and Operator Theory}, 78(3):301--321.

\end{thebibliography}

\end{document}